\documentclass[12pt]{iopart}
 \expandafter\let\csname equation*\endcsname\relax
 \expandafter\let\csname endequation*\endcsname\relax
\usepackage{bm}
\usepackage{enumerate}
\usepackage[top=1in, bottom=1in, left=1in, right=1in]{geometry}
\usepackage{mathrsfs,color}
\usepackage{amsfonts}\usepackage{multirow}
\usepackage{amssymb}
\usepackage{caption,comment}
\usepackage{blkarray}
\usepackage{amsmath}
\usepackage{float}
\usepackage{fixmath}
\usepackage{amssymb,amsthm}
\usepackage[toc,page]{appendix}
\usepackage{graphicx,afterpage}
\usepackage{algorithm}
\usepackage{algorithmic} 
\usepackage{subfig}
\usepackage[dvipdfm,colorlinks,linkcolor=blue,citecolor=blue]{hyperref}
\numberwithin{equation}{section}
\numberwithin{figure}{section}
\def\theequation{\arabic{section}.\arabic{equation}}
\newcommand\tabcaption{\def\@captype{table}\caption}
\def\thefigure{\arabic{section}.\arabic{figure}}

\DeclareMathOperator{\divr}{div}
\DeclareMathOperator{\grad}{grad}

\DeclareMathOperator{\spn}{span}

\newcommand{\argmin}{\operatornamewithlimits{argmin}}

\begin{document}
\title{Analog Forecasting with Dynamics-Adapted Kernels}

\author{Zhizhen Zhao}

\address{Department of Mathematics and Center for Atmosphere and Ocean Science, Courant Institute of Mathematical Sciences, New York University, New York, NY, USA}
\ead{jzhao@cims.nyu.edu}

\author{Dimitrios Giannakis}
\address{Department of Mathematics and Center for Atmosphere and Ocean Science, Courant Institute of Mathematical Sciences, New York University, New York, NY, USA}
\ead{dimitris@cims.nyu.edu}

\vspace{10pt}

\begin{abstract}
Analog forecasting is a nonparametric technique introduced by Lorenz in 1969 which predicts the evolution of states of a dynamical system (or observables defined on the states) by following the evolution of the sample in a historical record of observations which most closely resembles the current initial data. Here, we introduce a suite of forecasting methods which improve traditional analog forecasting by combining ideas from kernel methods developed in harmonic analysis and machine learning and state-space reconstruction for dynamical systems. A key ingredient of our approach is to replace single-analog forecasting with weighted ensembles of analogs constructed using local similarity kernels. The kernels used here employ a number of dynamics-dependent features designed to improve forecast skill, including Takens' delay-coordinate maps (to recover information in the initial data lost through partial observations) and a directional dependence on the dynamical vector field generating the data. Mathematically, our approach is closely related to kernel methods for out-of-sample extension of functions, and we discuss alternative strategies based on the Nystr\"om method and the multiscale Laplacian pyramids technique. We illustrate these techniques in applications to forecasting in a low-order deterministic model for atmospheric dynamics with chaotic metastability, and interannual-scale forecasting in the North Pacific sector of a comprehensive climate model. We find that forecasts based on kernel-weighted ensembles have significantly higher skill than the conventional approach following a single analog.


\end{abstract}

\vspace{2pc}
\noindent{\it Keywords}: Analog forecasting, kernel methods, out-of-sample extension , delay-coordinate maps \\

\section{Introduction}
In many branches of science and engineering, advances in modeling capabilities and the proliferation of experimental and observational data have spurred the development of new mathematical and computational techniques for predicting the dynamical behavior of complex nonlinear systems with imperfect models and partial knowledge of the initial data. The Earth's climate system is a classical example of such a complex and partially observed system coupling physical processes from the atmosphere, ocean, and land over a wide range of spatial and temporal scales. Currently, as well as in the foreseeable future \cite{Fox-KemperEtAl14}, the equations of motion for the full climate system cannot be integrated via direct numerical simulation and must be parameterized to account for unresolved scales, while in some cases the equations of motion are only partially known \cite{BodenschatzEtAl10}. A topic of significant current interest is how to make accurate and reliable predictions of large-scale climate patterns over timescales ranging from weeks to decades, using imperfect models initialized with the incomplete information about the current state of the system provided by observational networks \cite{PalmerHagedorn06,MeehlEtAl09,CortiEtAl12}. 

One way of making initial-value forecasts in complex systems is through large-scale numerical simulation. For example, in climate science, regional weather models and general circulation models (GCMs) are used extensively to make forecasts over lead times spanning a few hours (in weather forecasting) to several decades (in long-term climate projections). Broadly speaking, these numerical models are based on a dynamical core for the atmosphere and/or ocean derived from the Navier-Stokes equations on a rotating sphere, which is coupled to a complex network of physics models describing processes such as radiative energy transfer and phase transitions. In an operational environment, the models are coupled with a data assimilation system providing initial conditions from the (incomplete) data available from sensors such as satellites, weather stations, and buoys. The significant increases in forecast skill for the weather and climate over the past decades is due to advances in both the forward models and data assimilation systems \cite{PalmerHagedorn06}.  
 
Distinct from large-scale numerical models are parametric and nonparametric empirical low-order methods  \cite{KantzSchreiber04,CressieWikle11}. These techniques do not rely on a large-scale forward model in the forecast step. Rather, predictions are made using a model trained on a reference time series (which may be obtained from historical observations of nature, or a large-scale numerical model), and initialized with the currently observed data from nature. The empirical models generally make coarser forecasts and may also underperform large-scale models in terms of innovation, but their strengths lie in their low computational cost and ability to be trained using data from nature. Low computational cost is desirable in scenarios where real-time or near real-time forecasts are required but the cost of frequent large-scale model runs is prohibitive \cite{WheelerWeickmann01}, and in applications where low-order process models are embedded in large-scale models as subgrid-scale parameterization schemes \cite{DengEtAl14}. Equally important, low-order empirical models can be trained using data acquired from nature, therefore avoiding some of the dynamical model errors present in large-scale forward models. For instance, a major barrier in extending the skill of numerical weather forecasts at timescales greater than about a week is the poor dynamical representation of organized tropical convection \cite{MoncrieffEtAl07}. As a result, low-order empirical models frequently outperform their large-scale counterparts in predicting climate phenomena which are the outcome of convective organization,  including the Indian monsoon \cite{XavierGoswami07} and the Madden-Julian oscillation \cite{LauWaliser11}.  

Within empirical low-order models, a dichotomy can be made between parametric and nonparametric techniques (though the dichotomy is loose, as there exist semiparametric techniques combining aspects of both approaches; e.g., \cite{BerryHarlim16}). Parametric techniques require the specification of an explicit model for the dynamics, whose parameters are estimated statistically from past observations of the system. Classical examples of such techniques are linear and nonlinear regression models \cite{KantzSchreiber04} and Bayesian hierarchical models \cite{CressieWikle11}. When available, prior knowledge of physical laws, or physics-based constraints \cite{MajdaHarlim13}, can significantly facilitate the parametric model design with a corresponding improvement of skill (e.g., \cite{Chen2014}).  Yet, in certain applications, prior knowledge of an appropriate parametric model is not available. Such an application, which we will study in Section~\ref{sec:ccsm} ahead, is the low-frequency variability of the North Pacific Ocean. There, even state-of-art parametric models generally provide poorer forecast skill than simple persistence forecasts \cite{ComeauEtAl15}. In such scenarios, nonparametric models are a viable alternative strategy relying on only general conditions on the properties of the dynamics (e.g., ergodicity). 

Nonparametric techniques utilize past observations of a dynamical system to infer its future behavior without assuming an explicit parametric form of the equations of motion. As a result, their main advantage is their applicability in a wide range of forecasting scenarios. At the same time, nonparametric models must compensate for their flexible structure through plentiful samples in phase space (whereas in parametric techniques a partial sampling of the phase space may be sufficient for accurate parameter estimation).  Classical nonparametric forecasting techniques include local linear models \cite{shiftop1,shiftop2,shiftop3} and analog forecasting \cite{Lorenz1969,vandenDoolEtAl03}; the methods developed in this paper can be viewed as a generalization of the latter technique using ideas from kernel methods in harmonic analysis and machine learning \cite{CoifmanLafon06,BerrySauer15,CoifmanLafon06b,CoifmanHirn13,FernandezEtAl14,Rabin2012}. The methods developed here also have connections with a recently developed nonparametric method called diffusion forecasting \cite{BerryEtAl15} that makes use of related kernel techniques, and we will discuss these connections in Section~\ref{sec:diffusion}.  
       

\subsection{\label{sec:introAnalog}Analog forecasting}

The focus of this paper is on nonparametric methods inspired by the analog forecasting technique. This technique was introduced by Lorenz in 1969 \cite{Lorenz1969} as a method for predicting the time evolution of observables in dynamical systems based on a historical record of training data. In the initialization stage of analog forecasting, one identifies an analog, i.e., the state in the historical record which most closely resembles the current initial data. Then, in the forecast step, the historical evolution of that state is followed for the desired lead time, and the observable of interest is predicted based on its value on the analog. Unlike prediction using GCMs and parametric low-order models, the analog approach makes no assumptions about the dynamics of the system being examined. Thus, if the historical record of observations comes from nature, the method avoids the dynamical model errors of parametric models essentially by construction, allowing the deployment of the method to real-world problems where parametric models are known to perform poorly \cite{Toth1989,XavierGoswami07}. Analog forecasting is also useful in situations where the reference time series itself is the output of numerical models. For instance, in \cite{Branstator2012} analogs are used to assess the long-range internal predictability properties of GCMs. Nonlinear forecasting based on libraries of past histories has also been used in ecology~\cite{Sugihara1990}.   

Three major factors influencing the efficacy of analog forecasting are (i) the identification of ``skillful'' analogs in the training data; (ii) the choice of forecast observable (predictand); (iii) model error in the training data. Geometrically, the ability to identify skillful analogs amounts to being able to identify subsets of the training data whose dynamical evolution will shadow the future time evolution for the given initial data. In general, the required shadowing accuracy to achieve a given forecast accuracy depends on both the dynamical system and the observable, as different observables have different predictability limits. For instance, the predictability of the full state vector is limited by the maximal Lyapunov exponent of the dynamical system, and a trivial observable equal to a constant is predictable at arbitrarily long lead times. Indeed, an important problem in data-driven forecasting techniques is to identify observables which are physically meaningful but also predictable. The properties of the prediction observable also affects the reliability of analog forecasts in situations where the training data have model error. In particular, in some situations it may be preferable (e.g., when the time span of the available data from nature is short), or necessary (e.g., when the predictand is not directly observable) to use a training dataset generated by an imperfect model, despite that analog forecasting cleverly avoids dynamical model error if the training data are acquired for nature.      

In real-world applications, the problem of analog identification is compounded by the fact that the initial data are often high-dimensional and incomplete (i.e., insufficient to uniquely determine the future dynamical evolution of the system). Moreover, highly predictable observables are not known a priori. As a concrete example, consider climate forecasting in the North Pacific Ocean given observations of sea surface temperature (SST) over a spatial grid. In this example, which will be studied in Section~\ref{sec:ccsm}, the dimension of the initial data is equal to the number of spatial gridpoints where SST measurements are taken. For the spatial resolutions typical of current-generation observational networks and GCMs ($ \lesssim 1^\circ$) the number of North Pacific gridpoints is  $ O( 10^4 ) $. However, knowledge of SST at a single instance of time is not sufficient to determine its future evolution, as many of the essential degrees of freedom of the atmosphere and the ocean (e.g., the vertically-resolved circulation, temperature, and density fields) are not represented in SST snapshots. As a result, similar SST snapshots may be produced by states lying far apart on the attractor, and identifying analogs solely on the basis of similarity of SST snapshots is likely to lead to poor prediction skill. 

Regarding the choice of prediction observable, considerable efforts are being made by oceanographers to identify patterns which characterize the variability of the ocean on timescales of several years. In the North Pacific, one such prominent patterns are the Pacific decadal oscillation (PDO) \cite{MantuaHare02}, which is conventionally obtained by applying principal components analysis (PCA) to seasonally-detrended and low-pass filtered  SST data. In general, an observable measuring the PDO activity [e.g., the principal component (PC) corresponding to the pattern] is subjective, in the sense that it is defined through a data analysis algorithm applied to a finite dataset. PCA-type algorithms, in particular, are known to have poor skill in extracting dynamically significant patterns in certain nonlinear dynamical systems \cite{AubryEtAl93,CrommelinMajda04}, motivating the development of alternative data-driven approaches for extracting observables for empirical forecasting. 

\subsection{Contributions of this work}

The central theme of this work is that kernel methods, in conjunction with state-space reconstruction methods for dynamical systems, can lead to significant improvements in the traditional formulation of analog forecasting, benefiting both the identification of skillful analogs, as well as the construction of intrinsic slow modes for prediction. To motivate our approach, we note that a generic feature of many dynamical systems of interest in science and engineering is that, after initial transients have decayed, the system  evolves on low-dimensional subsets of phase space (attractors) \cite{KatokHasselblatt97,DymnikovFilatov97}. These nonlinear structures are embedded in the high-dimensional ambient data space in potentially complicated ways, meaning that the best-fitting linear hyperplane is likely to be of significantly higher dimension than the intrinsic dimension of the data. Kernel algorithms \cite{BelkinNiyogi03,CoifmanLafon06,CoifmanLafon06b,TingEtAl10,Rabin2012,BerrySauer15,FernandezEtAl14} are well-suited to handle nonlinear geometrical structures of data such as manifolds. Here, our objective is to improve the traditional analog forecasting using these techniques, adapted to deal with data generated by dynamical systems \cite{GiannakisMajda12a,BerryEtAl13,Giannakis15}.   

For our purposes, a kernel will be a measure of similarity between pairs of samples in data space. Suitably normalized \cite{CoifmanLafon06,BerrySauer15}, these objects naturally lead to local averaging operators acting on observables such that. In particular, given previously unseen initial data, these operators act by averaging the observable values over the samples of the training data which are most similar to the initial data with respect to the kernel. Conceptually,  averaging extends the notion of single-analog identification in conventional analog forecasting to the construction of a weighted ensemble of analogs with weights determined by the averaging operator.   

For suitably constructed kernels, the behavior of the associated averaging operators can be understood through a Riemannian metric on the nonlinear manifold sampled by the data that depends on the kernel \cite{BerrySauer15}. In this context, kernels can therefore be thought of as inducing a Riemannian geometry, sometimes referred to as diffusion geometry, to the training data. Having in mind that the skill of analog forecasting depends on the ability to geometrically identify good analogs in the training data, we propose to use the properties of a geometry  adapted to the dynamical system generating the data to improve the efficacy of this step. In particular, following \cite{GiannakisMajda12a,GiannakisMajda13,BerryEtAl13,Giannakis15}, our approach is to take advantage of a special structure present in data generated by dynamical systems, namely the time ordering of the samples which is the outcome of the dynamics. 

The time-ordering of the data samples has been utilized at least since the early 1980s in state-space reconstruction methods \cite{Takens81,BroomheadKing86,SauerEtAl91}. In these works, it was established that under generic conditions one can recover the topology of the attractor of a dynamical system from time-ordered partial observations by embedding the data into a higher-dimensional space consisting of lagged sequences of observations over a temporal window. At the very least, when the initial data are incomplete, this suggests to perform analog forecasting using Euclidean distances or similarity kernels defined in delay-coordinate space. Besides topological aspects, however, delay-coordinate mapping also influences the geometry of the data irrespective of whether or not the initial data is complete. That is, pairwise distances on the data manifold in delay coordinate space depend not only on ``snapshots'' at single time instances, but on dynamical trajectories (``videos''). In \cite{BerryEtAl13}, it was shown that adding delays increasingly biases the kernel-induced metric towards the most stable subspace of the Lyapunov metric, enhancing the kernel's timescale separation capability. This approach was also used heuristically in earlier work on so-called nonlinear Laplacian spectral analysis (NLSA) algorithms \cite{GiannakisMajda12a,GiannakisMajda12b,GiannakisMajda13},  where kernels in delay-coordinate space were found to have significantly higher skill in recovering patterns analogous to the PDO in North Pacific data from a GCM. Moreover, a set of spatial patterns recovered via NLSA were found to yield Galerkin-reduced models reproducing faithfully the chaotic regime transitions in a low-order model for the atmosphere \cite{CrommelinMajda04} where PCA is known to fail dramatically. The low-order atmospheric model and the high-dimensional North Pacific GCM data will both be employed here as applications of our kernel analog forecasting techniques. Moreover, we will study an example with model error in the training data simulated by treating North Pacific SST output from a GCM operating at a given atmospheric resolution as ``nature'' and the corresponding output obtained from a model with a coarser atmosphere as training data from an imperfect model.

 Besides delay-coordinate mappings, the kernel used in NLSA features an additional dependence on the time-tendency of the data evaluated through finite differences of the data in time. In \cite{Giannakis15}, it was shown that the finite-differenced data provide an empirical approximation of the dynamical vector field on the attractor generating the dynamics. That work also established that by incorporating a suitable directional dependence, biasing the kernel towards states which are mutually aligned with the dynamical flow, one gains additional invariance properties and ability to extract slow dynamical time scales. These so-called cone kernels will be employed here as our preferred similarity kernel to identify skillful analogs in the training data, and we will also use kernel eigenfunctions associated with cone kernels to construct slow observables with high predictability.   

With any kernel, the ability to perform analog forecasts hinges on the ability to evaluate observables at points lying outside the training dataset. Here, we carry out these tasks using two popular out-of-sample extension techniques, namely the Nystr\"om method \cite{Nystrom1930,BengioEtAl03,CoifmanLafon06b} and Laplacian pyramids \cite{Rabin2012,FernandezEtAl14}.  The Nystr\"om method performs particularly well when the prediction observable is itself a kernel eigenfunction (which will be the case in the North Pacific SST application), but may become ill-conditioned and/or biased for more general observables. In those situations, the multiscale procedure for out-of-sample extension in Laplacian pyramids is preferable. Used in conjunction with a bistochastic kernel normalization \cite{CoifmanHirn13}, Laplacian pyramids is also well-suited for analog forecasting with incomplete initial data; a situation encountered frequently in real-world applications. Both the Nystr\"om method and Laplacian pyramids perform out-of-sample extension through one or more kernel integral operators acting on observables of the dynamical system. Thus, analog forecasting with these techniques can be thought of as an ensemble generalization of conventional single-analog forecasting. In applications, we find that these methods have significantly higher forecast skill than what is possible with single analogs.

This paper is organized as follows. In Section~\ref{sec:mathback}, we describe the mathematical framework of kernel analog forecasting, focusing on the properties of analog forecasting with Nystr\"om extension and Laplacian pyramids for general types of kernels. We discuss specific examples of dynamics-adapted kernels and their normalization in Section~\ref{sec:kernel}. In Section~\ref{sec:err}, we discuss forecast error metrics and the uncertainty quantification of our scheme. In Section~\ref{sec:diffusion}, we provide a comparison of kernel analog forecasting with the diffusion forecast method \cite{BerryEtAl15}. In Section~\ref{sec:results}, we present applications of kernel analog forecasting in a low-order model for the atmosphere and the North Pacific sector of comprehensive climate models. We conclude in Section~\ref{sec:conclusions}.  

\section{\label{sec:mathback}Kernel analog forecasting}

In this Section, we present the mathematical framework of our kernel analog forecasting strategies. These techniques are inspired by Lorenz's analog method \cite{Lorenz1969}, so, after establishing our notation and main assumptions in Section~\ref{sec:notation}, we provide an overview of conventional forecasting in Section~\ref{sec:analog}. In Section~\ref{sec:basicKernel}, we introduce a basic version of kernel analog forecasting based on averaging with a single operator. While this would not be our method of choice in practice, it serves as an illustration of the main ideas and challenges in kernel analog forecasting. In Sections~\ref{sec:nystrom} and~\ref{sec:pyramids}, we present refinements of the  basic kernel analog forecasting approach using Nystr\"om out-of-sample extension \cite{CoifmanLafon06b} and Laplacian pyramids \cite{Rabin2012,FernandezEtAl14}. 


\subsection{\label{sec:notation}Notation and preliminaries}
 
Consider time-ordered samples $ \{ x_0, x_1, \ldots, x_{N-1} \} $ of a signal in a space $ X $ taken uniformly at times $ \{ t_0, t_1, \ldots, t_{N-1} \} $ with $ t_i = ( i - 1 ) \tau $. The signal is generated by a dynamical system, and observations of $  y \in X $ at a single instance of time or over a finite time interval will serve as initial data to forecast a quantity of interest (a prediction observable)  $ f $.  The prediction observable is generated by the same dynamical system generating $ \{ x_i \} $, takes values in a space $ B $, and the values $ \{ f_i \} $ at the sampling times $ \{ t_i \} $ are also observed. 

Mathematically, we consider that the dynamical system evolves in an abstract (unobserved) phase space $ A $ and the observed data in $ X $  are the outcome of a mapping $ \pi : A \mapsto X $ sending the system state $ a_i \in A $ at time $ t_i $ to the data sample $ x_i = \pi( a_i ) \in X $. Similarly, the prediction observable is a map $ f : A \mapsto B $ such that $ f_i = f( a_i ) $. For instance, in the ocean application of Section~\ref{sec:ccsm}, we have $ X = \mathbb{ R }^d $ and $ B = \mathbb{ R } $, where $x_i \in X $ is the SST field measured at $ d $ spatial gridpoints in the North Pacific ocean at time $ t_i $, and $ f_i \in B $ is the corresponding value of a low-frequency mode of variability such as the PDO. Moreover, $ A $ is the phase space of the Earth's climate system. In general, we are interested in a partial-observations scenario where $ x_i $ provides incomplete knowledge of the system state $ a_i $, i.e., $ \pi $ is a non-invertible map on its image. We call such initial data incomplete. On the other hand, we say that the initial data are complete if $ \pi $ is invertible on its image. In the applications of Section~\ref{sec:ccsm}, observations of North Pacific SST alone are not sufficient to uniquely determine the full state of the climate system and the future evolution of the PDO, hence the initial data are incomplete.   

Throughout, we assume that $ A $ and $ X $ are measurable spaces with $ \sigma $-algebras $ \mathcal{ A } $ and $ \mathcal{ X } $, respectively, and $ \pi $ is a measurable map from $ ( A, \mathcal{ A } ) $ to $ ( X, \mathcal{ X } ) $. In the case of $ X $, we also assume that it is a metrizable space (though we will not explicitly use its topology), and that $ \mathcal{ X } $ is its Borel $ \sigma $-algebra. We assume that the image $ M = \pi( A ) $ of the phase space under $ \pi $ is measurable, and denote the $ \sigma $-algebra $ \mathcal{ X } \cap M $ by $ \mathcal{ M } $. Also, we denote the finite set of observations $\{ x_i \} \subseteq M $ by $ \hat M $. 

Turning to the prediction observable, we consider that $ B $ is a Banach space with the corresponding Borel $ \sigma $-algebra $ \mathcal{ B } $ and that $ f : ( A, \mathcal{ A } ) \mapsto ( B, \mathcal{ B } ) $ is measurable. In what follows, we develop our methodology for scalar-valued observables, $ B = \mathbb{ R } $, but a similar treatment can be applied for more general classes of vector- and function-valued observables for which ergodic theorems apply \cite{Krengel85}. In general, we distinguish between prediction observables defined objectively for the system, and observables constructed through a data analysis technique. In the low-order atmospheric model studied in Section~\ref{sec:toyModel}, $ f $ will be the components of the state vector that capture metastable regime transitions---these are examples of objectively-defined observables. On the other hand, in Section~\ref{sec:ccsm} the goal will be to predict the time-evolution of the top two low-frequency modes constructed by processing the data via kernel algorithms. Desirable features of such data-driven observables is to evolve on slow intrinsic dynamical timescales, and have high smoothness as functions on the phase space $ A $. In Section~\ref{sec:kernel} ahead, we will see that dynamics-adapted kernels provide a promising route for constructing physically meaningful data-driven observables with favorable predictability properties.  

In this work, we consider that the dynamics on $ ( A, \mathcal{ A } ) $ is governed by a group or semigroup of measure-preserving transformations $ \Phi_t : A \mapsto A $ with time parameter $ t $ and invariant measure $ \alpha $; that is, $ \alpha( \Phi_t^{-1}( S ) ) = \alpha( S ) $ for all $ S \in \mathcal{ A } $. The invariant measure of the dynamics induces a probability measure $ \mu $ on $ ( M, \mathcal{ M } ) $ such that $ \mu( S ) = \alpha( \pi^{-1}( S ) ) $ for $ S \in \mathcal{ M } $.  We also assume that $ \Phi_t $ is ergodic, i.e.,  $ \alpha( S ) $ is equal to either 0 or 1 for every $ \Phi_t $-invariant set $ S $. Let $ T $ be the time domain of the system. For concreteness, we develop our approach in the continuous-time invertible (group) case, $ T = \mathbb{ R }$,  $ \Phi_s \circ \Phi_t = \Phi_{s+t} $, $ \Phi_t^{-1} = \Phi_{-t} $, and we assume that the dynamical system sampled at the discrete times $ \{ \ldots, -2 \tau, -\tau, 0, \tau, 2 \tau, \ldots \} $ is also ergodic. Counterparts of our results for continuous-time sampling follow by replacing discrete time averages $ N^{-1} \sum_{i=0}^{N-1} f_i $ by continuous time integrals $ t^{-1} \int_0^t f( t' ) \, dt' $. Our methods are also applicable in non-invertible (semigroup) systems, $ T = \{ t \geq 0 \} $, with the exception of constructions utilizing backwards delay-coordinate maps (see Section~\ref{sec:delay}).   

At a minimum, we require that the prediction observable $ f $ is an integrable scalar-valued function in $ L^1( A, \alpha ) $, i.e.,
\begin{equation}
  \label{eq:barf} \bar f = \int_A f \, d\alpha < \infty.
\end{equation}  
This requirement will be re sufficient for the most basic formulation of our analog forecasting techniques (see Section~\ref{sec:basicKernel}). In several other instances (e.g., the method based on Nystr\"om extension in Section~\ref{sec:nystrom}), we additionally require that $ f $ is a square integrable-observable in $ L^2( A, \alpha ) \subset L^1( A, \alpha ) $. We denote the inner product and norm of $ L^2( A, \alpha ) $ by $ \langle f, g \rangle = \int_A f g \, d\alpha $ and $ \lVert f \rVert = \langle f, f \rangle^{1/2} $, respectively.  

By the Birkhoff pointwise ergodic theorem, the time average of an observable $ f $ in $ L^1( A, \alpha ) $ along the sampled states $ a_i $, 
\begin{equation}
  \label{eq:birkhoff} \hat f = \frac{ 1 }{ N } \sum_{i=0}^{N-1} f_i, \quad f_i = f( a_i ),
\end{equation}
converges as $ N \to \infty $ to $ \bar f $ for $ \alpha $-almost any starting point $ a_0 $. Intuitively, we think of $ N \to \infty $ as the limit of large data. The time average in~\eqref{eq:birkhoff} can also be expressed as an expectation value with respect to the purely atomic probability measure $ \hat \alpha = N^{-1} \sum_{i=0}^{N-1} \delta_{a_i} $, where the $ \delta_{a_i} $ are Dirac measures on  $ ( A, \mathcal{ A } ) $ centered at  $ a_i $. That is, we have $ \hat f = \int_A f \, d \hat \alpha $. The ability to approximate expectation values such as~\eqref{eq:barf} with the corresponding empirical time averages in~\eqref{eq:birkhoff} is crucial for the analog forecasting techniques developed below. 



\subsection{\label{sec:analog}Conventional analog forecasting}

Broadly speaking, analog forecasting \cite{Lorenz1969} produces a forecast $ \hat F_t( y ) $ of an observable $ f $ of a dynamical system at forecast lead time $ t $ given initial data $ y $ in the space $ X$. Assuming that $ t $ is a non-negative integer multiple $ k $ of the sampling interval $ \tau $, the necessary ingredients needed to carry out this task are (i) the training time series $ \hat M  = \{ x_i \} $ with $ x_i = \pi( a_i ) $ lying in the same space as $ y $; (ii) the corresponding time series of the observable values $ f_i = f( a_i ) $; (iii) a distance function $ D : X \times X \mapsto \mathbb{ R } $. Typically, when $ X = \mathbb{ R }^n $, $ D $ is set to the Euclidean distance $ D( y, x ) = \lVert y - x \rVert $. In a perfect-model scenario, the training samples $ ( x_i, f_ i ) $ are generated by the same dynamical system as the dynamical system being forecast---this would be the case if $ x_i $ are observations of nature. Mathematically, this means that the initial data are the outcome of the same observation map as the training data, i.e., we have
\begin{equation}
  \label{eq:perfectIC}
  y = \pi( b ), \quad b \in A.
\end{equation}  

In an imperfect-model scenario, the $ ( x_i, f_i ) $ have systematic biases compared to the time evolution of these quantities in nature. Moreover, there does not necessarily exist a state $ b $ in the phase space of the imperfect model underlying the data $ y $ at forecast initialization as in~\eqref{eq:perfectIC}. Such situations arise when the training data have been generated by a  numerical model with dynamical errors relative to nature. However, as stated in Section~\ref{sec:introAnalog}, it may be desirable to use a long integration from a numerical model to improve the sampling density of state space despite model errors. Moreover, in some cases, the pairs $ ( x_i, f_i ) $ may not be observationally accessible. A notable example from climate science where this issue is encountered is forecasting and estimation of the interior circulation and thermal structure of the ocean. These variables are not accessible via remote sensing from satellites, and are only sparsely known from in situ sensors such as gliders and floats. For the remainder of this section we will restrict attention to the perfect model scenario where~\eqref{eq:perfectIC} holds.   

Let $ U_t $, $ t \in T$,  be the operator on $ L^1( A, \alpha ) $ whose action is given by composition with the evolution map:  
\begin{equation}
  \label{eq:koopman}
   U_t f  = f \circ \Phi_t, \quad U_tf( a ) = f( \Phi_t( a) ). 
\end{equation}
That is, $ U_t f $ is a time-shifted observable whose value at $ a $ is given by evaluating $ f $ at the point $ \Phi_t( a ) $ lying in the future of $ a $ at time $ t $. Intuitively, $ f_t( a ) $ is the value of a ``perfect''  forecast for $ f $ at lead time $ t $ initialized at the dynamical state $ a \in A $. Note that for $ t = k \tau $ sufficiently small, $ k \in \mathbb{ N } $, and $ a_i \in \{ a_0, \ldots, a_{N-1} \} $, $ U_t f( a_i ) = f_{i+k} $  is given by a $ k $-step forward shift of the $ \{ f_i \} $ time series. In ergodic theory, the operators $ U_t $ are known as Koopman operators \cite{EisnerEtAl15}. The set $ \{ U_t \} $ forms a group (or a semigroup in the non-invertible case) under composition of operators with $ U_s \circ U_t = U_{s+t} $ and (in the group case) $ U_t^{-1} = U_{-t} $. Koopman operators are similarly defined for observables in the Hilbert space $ L^2( A, \alpha ) \subset L^1( A, \alpha ) $. These operators are isometric, $ \lVert U_t f \rVert = \lVert f \rVert $, and are unitary with $ U_t^*  = U_t^{-1} = U_{-t} $ if $ \{ U_t \} $ is a group.     
 
Analog forecasting is a two-step procedure which involves finding the sample $ x_i $ in the training data lying closest to $ y $ with respect to the distance on $ X $, and then time-advancing the observable values $ f_i $  to produce a forecast. That is, one first identifies the analog $ a_i \in A $ underlying the initial data $ y $ via
\begin{equation}
  \label{eq:analogDist}
  i = \argmin_{j \in \{ 1, \ldots, n \} } D( y, x_j ), \quad x_j = \pi( a_i ), 
\end{equation}
and then evaluates the forecast at lead time $ t = k \tau $ via
\begin{equation}
  \label{eq:analogShift}
  \hat F_t( y ) := f_{i+k} = U_t f( a_i ).
\end{equation}
Note that the process of analog identification is implicit in the sense that in order to evaluate~\eqref{eq:analogShift} it suffices to know the timestamp $i$ of $ a_i $ from~\eqref{eq:analogDist}, but explicit knowledge of the state $ a_i \in A $ is not needed. Moreover, in~\eqref{eq:analogDist} we have tacitly assumed that the minimizer of $ D( y, \cdot ) $ is unique in the training dataset $ \hat M $; if multiple minimizers are found (an unlikely situation in practice), then~\eqref{eq:analogShift} can be replaced by the average forecast over those minimizers. 

Conventional analog forecasting also has a measure-theoretic interpretation which will be useful in the development of kernel analog forecasting ahead. In particular, $ \hat F_t( y ) $ is equal to the expectation value of $ U_t f $ with respect to the Dirac measure centered at the analog, i.e., 
\begin{equation}
  \label{eq:analogShift2}
  \hat F_t( y ) = \int_A U_t f \, d \delta_{a_i}.
\end{equation}  
Given the training data $ \hat M $ and the initial data $ y $ from~\eqref{eq:perfectIC}, the forecast error (residual) of conventional analog forecasting is
\begin{equation}
  \label{eq:analogError}
  \hat r_t( b ) =  U_t f( b ) - \hat F_t( y )  =  U_t f( b ) - U_t f( a_i )  =   \int_A U_t f  ( d \delta_b - d \delta_{a_i} ).
\end{equation}
Note that in an imperfect-model scenario $ \hat F_t( y ) $ would generally be evaluated for initial data lying outside of $M $, and~\eqref{eq:perfectIC} and~\eqref{eq:analogError} would not hold. 

It is evident from~\eqref{eq:analogDist}--\eqref{eq:analogShift2} that analog forecasting is a fully nonparametric approach, relying solely on the geometric identification of nearest neighbors in the training data and empirical time shifts of time series. These properties make the method especially attractive in situations where one does not have knowledge of the equations of motion, or an appropriate parametric model is not available. For instance, in the ocean application in Section~\ref{sec:ccsm} construction of parametric regression models for the low-frequency modes is especially hard, and linear autoregressive models have worse skill than a trivial persistence forecast \cite{ComeauEtAl15}. 

Yet, despite its attractive features, conventional analog forecasting suffers from high risk of overfitting the training data. As is evident from~\eqref{eq:analogShift}, the conventional analog forecast  (and hence the corresponding forecast error in~\eqref{eq:analogError}) is a highly non-smooth function of the initial data---as $ y $ varies on $ X $, $ \hat F_t( y ) $ changes discontinuously to the value $ U_t f( a_i ) $ of the observable on the analog that happens to minimize the distance to $ y $, and the forecast error can exhibit large variations under small perturbations of the initial data. Moreover, it follows from~\eqref{eq:analogShift2} that conventional analog forecasting utilizes information from a single state in the training data to predict the future of the observable, but one would expect that other states in the training data also carry useful predictive information. Our objective in Sections~\ref{sec:basicKernel}--\ref{sec:pyramids} ahead is to address these shortcomings using kernel methods.  

\subsection{\label{sec:basicKernel}Basic kernel analog forecasting} 

According to the discussion in Section~\ref{sec:analog}, conventional analog forecasting can be described as a mapping $ y \mapsto \delta_{a_i} $ of the initial data in $ X $ to a Dirac measure $ \delta_{a_i} $ on $ ( A, \mathcal{ A } ) $, followed by computation of the expectation value of the time-shifted observable $ U_t f $ with respect to $ \delta_{a_i} $. In the simplest version of our kernel analog forecasting techniques, we replace $ \delta_{a_i} $ by a probability measure $ \hat \nu_y $ with the properties that (i) expectation values with respect to $ \hat \nu_y $ depend continuously on $ y $; (ii) in the limit of large data, $ N \to \infty $, $ \hat \nu_y $ converges to a probability measure $ \nu_y $ which is absolutely continuous with respect to the invariant measure $ \alpha $ of the dynamics.   

A key ingredient of kernel analog forecasting is a kernel function, $ K : X \times X \mapsto \mathbb{ R } $, mapping pairs of points in the observation space $  X$ to a non-negative real number. Intuitively, we think of $ K( y, x ) $ as a measure of similarity between the data samples $ x $ and $ y $, in the sense that large values of $ K( y, x ) $ indicate that $ x $ and $ y $ are highly similar, and values of $ K( y, x ) $ close to zero indicate that $  x $ and $ y $ are highly dissimilar. In what follows, $ x $ will typically lie in the training dataset $ \hat M $ whereas $ y $ will be a previously unseen sample. Mathematically, we require that $ K $ has the following properties:
\begin{enumerate}
\item $ K $ is non-negative; 
\item The functions $ K( y, \cdot ) $ and $ K( \cdot, x ) $ are measurable for every $ y \in X $ and $ x \in M $;
\item There exists a constant $ C < \infty $ such that $ K \leq C $;
\item There exists a constant $ c > 0 $ such that $ \int_M K( y, \cdot ) \, d\mu \geq c $ for all $ y \in X $.
\item The function $ K( \cdot, x ) $ is continuous for every $ x \in M $. 
\end{enumerate} 
Among these requirements, properties~(iii) and~(iv) are not essential, but are very convenient for the exposition below. These assumptions could be weakened so long as quantities derived from $ K $ (e.g., the density $ \rho $ ahead) remain sufficiently smooth. Property~(v) is also non-essential and can be relaxed if a continuous dependence of the kernel analog forecast on the initial data is not required.  

By properties~(ii) and (iii) and the fact that $ \mu $ is finite, $ K( y, \cdot ) $ and $ K( \cdot, x ) $ are functions in $ L^p( M, \mu ) $, $ 1 \leq p \leq \infty $. Similarly, $ K( y, \pi( \cdot ) ) $ and $ K( \pi( \cdot ), x ) $ are in $ L^p( A, \alpha ) $.  In particular, the ergodic theorem in~\eqref{eq:birkhoff} applies, and the time average $ \hat q( y ) $, where
\begin{equation}
  \label{eq:ksum}
  \hat q( y ) = \int_M K( y, \cdot  ) \, d\hat \mu = \int_A K( y, \pi( \cdot ) ) \, d\hat \alpha = \frac{ 1 }{ N } \sum_{i=0}^{N-1} K(y,x_i),   
\end{equation}
converges to
\begin{displaymath}
  q( y ) = \int_M K( y, \cdot ) \, d\mu = \int_A K( y, \pi( \cdot ) ) \, d\alpha.
\end{displaymath}
By property (iii), $ q( y ) $ is bounded away from zero, and therefore for $ \alpha $-almost every starting point $ a_0 \in A $ and for sufficiently large $ N $, $ \hat q( y ) $ is nonzero---hereafter, we will assume that this is the case. The quantities $ \hat q( y ) $ and $ q( y ) $ will be useful in various kernel normalizations throughout the paper. 

A basic example of a kernel that meets the properties listed above in the case that $ X $ is equipped with distance function $ D $ is the radial Gaussian kernel,
\begin{displaymath}
  \label{eq:KGauss}
  K_\epsilon( y, x ) = e^{-D^2(y,x)/\epsilon},
\end{displaymath}
where $ \epsilon $ is a positive bandwidth parameter. This kernel is continuous with respect to both of its arguments. Radial Gaussian kernels are useful for approximating heat kernels on manifolds, and have been used extensively in harmonic analysis and machine learning \cite{BelkinNiyogi03,CoifmanLafon06,HeinEtAl05,TingEtAl10,CoifmanLafon06,Singer06}. Note that if $ M $ and/or $ X $ are noncompact, the radial Gaussian kernel may fail to meet property~(iv). Such situations can be handled by replacing~\eqref{eq:KGauss} via so-called variable-bandwidth kernels \cite{BerryHarlim15}, where the squared distance in the exponential function is scaled by a bandwidth function becoming large at distant points with small sampling density. In Section~\ref{sec:kernels}, we will discuss anisotropic Gaussian kernels which also take certain properties of the dynamical system generating the data into account. 

To construct a probability measure on $ ( A, \mathcal{ A } ) $, we normalize the kernel and obtain weights $ \hat \rho( y, x_i ) $ such that  $ \hat \rho( y, x_i ) \geq 0 $ and
$ N^{-1}  \sum_{i=0}^{N-1} \hat  \rho( y, x_i ) = 1 $. There are several ways to perform this normalization \cite{CoifmanLafon06,CoifmanHirn13,BerrySauer15}, but arguably the simplest one is to divide $ K( y, x ) $ by the (nonzero) kernel average in~\eqref{eq:ksum}, i.e., 
\begin{equation}
  \label{eq:hatRho}
  \hat \rho( y, x ) = \frac{ K( y, x ) }{ \hat q( y ) }.
\end{equation}
This type of normalization is called left normalization \cite{BerrySauer15} due to the fact that the normalizer $ \hat q( y ) $ is evaluated at the ``left'' argument ($y$) of $ K( y, x ) $. In Section~\ref{sec:normalization} ahead we will also consider alternative normalization strategies designed to control forecast biases. With either of these strategies,  as $ N \to \infty $  the function $ \hat \rho( y, \cdot ) $ converges for $ \alpha $-almost every starting point $ a_0 $ to a density 
\begin{equation}
  \label{eq:rho}
  \rho( y, \cdot ) = \frac{ K( y, x ) }{ q( y ) }
\end{equation}
in $ L^1( A, \alpha ) $ such that $ \int_A \rho( y, \pi(\cdot) ) \, d\alpha = 1 $. Moreover, $ \rho( y, \cdot ) $ is bounded by $ C / c $.   

Using $ \hat \rho $ and $ \rho $, we define the atomic probability measure
\begin{displaymath}
  \hat \nu_y = \frac{ 1 }{ N } \sum_{i=0}^{N-1} \hat \rho( y, x_i ) \delta_{a_i},
\end{displaymath}
 on $ (A,\mathcal{ A }) $ and the measure $ \nu_y $ such that 
\begin{displaymath}
  \nu_y( S ) = \int_S \rho( y, \pi( \cdot ) ) \, d \alpha, \quad S \in \mathcal{ A }.
\end{displaymath}
Under the assumptions on $ K $ stated above, $ \nu_y $ is absolutely continuous with respect to the invariant measure of the dynamics and $ d \nu_y / d \alpha = \rho( y, \cdot ) < \infty $. Moreover, it follows directly from the Birkhoff ergodic theorem that for every observable $  f \in L^1( A, \alpha ) $ the expectation value 
\begin{displaymath}
  \hat F( y ) = \int_A f \, d \hat \nu_y = \frac{ 1 }{ N } \sum_{i=0}^{N-1} \hat \rho( y, x_i )  f_i 
\end{displaymath}
converges to
\begin{displaymath}
  F( y ) = \int_A f \, d\nu_y = \int_A f \rho( y, \cdot ) \, d\alpha
\end{displaymath}
for $ \alpha $-almost every starting point $ a_0 $. 

In basic kernel analog forecasting, the forecast $ \hat F_t( y ) $ of $ f $ at lead time $ t = k \tau $ given initial data $ y $ is obtained by evaluating the expectation of the time-shifted observable $ U_t f $ from~\eqref{eq:koopman}, i.e.,
\begin{equation}
  \label{eq:basicAnalog}
  \hat F_t( y ) = \int_A U_t f \, d \hat \nu_y = \frac{ 1 }{ N } \sum_{i=0}^{N-1}  \hat{\rho}( y, x_i ) f_{i+k}.
\end{equation}
In the limit of large data, $ N \to \infty $, $ \hat F_t( y ) $ converges to
\begin{equation}
  \label{eq:basicAnalogLarge}
  F_t( y ) = \int_A U_t f \, d \nu_y.
\end{equation}
Unlike traditional analog forecasting via~\eqref{eq:analogShift2}, the kernel analog forecast in~\eqref{eq:basicAnalog} depends continuously on the initial data due to the continuity of $ K( \cdot, x ) $. Moreover, the forecast value $ \hat F_t( y ) $ incorporates information from multiple samples in the training data. For this reason, we sometimes refer to kernel analog forecasts as ensemble analog forecasts, distinguishing them from single-analog approaches such as the conventional method in Section~\ref{sec:analog}. As with~\eqref{eq:analogShift2}, $ \hat F_t( y ) $ can be evaluated for arbitrary initial data in $ X $. However, for $ y $ restricted to $ M = \pi( A ) $ there exists a function $ \hat f_t \in L^1(A, \alpha ) $ such that $ \hat f_t( b ) = \hat F_t( \pi( b ) ) $. Conceptually, we think of $ \hat f_t $ as lying ``above'' $ \hat F_t $. Similarly, there exists a function $ f_t \in L^1( A, \alpha ) $ lying above $ F_t $. 


For initial data of the form $ y = \pi( b ) $ with $ b \in A $, the error in forecasts made via~\eqref{eq:basicAnalog} is (cf.~\eqref{eq:analogError})
\begin{displaymath}
  \hat r_t( b ) = U_t f ( b ) - \hat F_t( y ) = \int_A U_t f \, ( d \delta_b - d \hat \nu_y ), 
\end{displaymath}
and in the limit of large data that error becomes
\begin{displaymath}
  r_t( b ) =  U_t f ( b ) - F_t( y ) = \int_A U_t f \, ( d \delta_b - d \nu_y ).
\end{displaymath}
Taking the expectation value of $ r_t $ with respect to the invariant measure of the dynamics, and using the fact that $ \int_A U_t f \, d\alpha = \int_A f \, d \alpha = \bar f $, it follows that the expected forecast error over all initial conditions is given by
\begin{equation}
  \label{eq:barE}
  \bar r_t  = \int_A r_t \, d \alpha = \bar f - \int_A  \int_A \rho( y, x ) U_t f( a ) \, d \alpha( a ) \, d\alpha(b ),
\end{equation}
where $ x = \pi( a ) $. 

We now comment on the behavior of the expected error with respect to time, the mixing properties of the dynamics, and the properties of the kernel, assuming that $ f $ is a square-integrable observable in $ L^2( A, \alpha ) $. 

First, note that $ \bar r_t $ is  generally nonzero, even at $ t = 0 $ and for complete initial data. However, if the dynamical system is mixing, then $ \bar r_t $ approaches zero as $ t \to \infty $ for both complete and incomplete initial data. More specifically, mixing implies that for any two observables $ f, g \in L^2( A, \alpha ) $, 
\begin{displaymath}
  \lim_{t\to\infty} \int_A ( U_t f ) g  \, d \alpha = \int_A f  \, d\alpha \int_A g \, d \alpha,
\end{displaymath}
and for any probability measure $ \nu $, absolutely continuous with respect to $ \alpha $,
\begin{displaymath}
  \lim_{t\to\infty} \int_A U_t f \, d \nu = \int_A f \, d\alpha.
\end{displaymath}
Together, these two properties imply that under mixing dynamics the expected error vanishes at asymptotic times, $ \lim_{t \to \infty} r_t = 0 $. As expected, in this case, the mean squared error (MSE) $ \lVert r_t \rVert^2 $ approaches the ensemble variance of $ f $, namely, 
\begin{displaymath}
   \lim_{t\to\infty} \lVert r_t \rVert^2 = \lim_{t\to\infty} \int_A r^2_t  \, d \alpha = \int_A ( f - \bar f )^2 \, d \alpha.
\end{displaymath}

Analog forecasting via~\eqref{eq:basicAnalog} becomes unbiased, $ \bar r_t = 0 $, at all times and irrespectively of the mixing properties of the dynamics and the completeness of the observations if $ \rho( y, x ) $ is bistochastic, that is 
\begin{displaymath}
  \int_M \rho( y, \cdot ) \, d\mu = \int_M \rho( \cdot, x ) \, d\mu = 1.
\end{displaymath}
Indeed, in this case, the double integral in~\eqref{eq:barE} is equal to $ \bar f $, and $ r_t $ vanishes identically. While $ \rho( y, x ) $ constructed via~\eqref{eq:rho} is generally not bistochastic, this observation motivates the use of density functions with the bistochastic property. In Section~\ref{sec:normalization}, we will discuss two approaches for constructing density functions which become bistochastic in the limit of large data, either via a careful kernel normalization \cite{CoifmanHirn13}, or a combination of normalization and kernel localization \cite{CoifmanLafon06,BerrySauer15}. 

Even with bistochastic density functions, however, basic kernel analog forecasting is overly diffusive and can suffer from large root mean squared error (RMSE). To see this, note that the forecast formula for $ F_t( y ) = f_t( b ) $ can be expressed in terms of the action of an averaging operator $ P : L^2( A, \alpha ) \mapsto L^2( A, \alpha ) $ such that
\begin{displaymath}
  P f( b ) = \int_A p( b, a ) f( a ) \, d\alpha( a ), \quad p( b, a ) = \rho( \pi( b ), \pi( a ) ).
\end{displaymath}
In particular, we have $ f_t( b ) = P U_t f( b ) $. Assume now that $ P $ is a ``good'' averaging operator for the dynamical system in the sense that it is positive-definite and self-adjoint on $ L^2( A, \alpha ) $, and $ \alpha $ is an ergodic invariant measure of the Markov semigroup $ \{ P^n \}_{n \in \mathbb{ N }} $.  Assume also for simplicity that $ f $ has zero mean on $ L^1(A, \alpha) $. Even in this favorable scenario,  the RMSE $ \lVert r_t \rVert = \lVert (I - P) U_t f \rVert $ cannot be less than $ \gamma \lVert f \rVert $, where $ \gamma $ is the spectral gap of $ P $.  In other words, the fact that the forecast in~\eqref{eq:basicAnalog} is determined by averaging $ U_t f $ on $ A $ means that even if the initial data $ y $ are complete and happen to lie exactly on the training dataset, $  F_t( y ) $ will generally have nonzero error. The methods put forward in the following two sections attempt to address these deficiencies while maintaining the desirable features of basic kernel analog forecasting.        

\subsection{\label{sec:nystrom}Kernel analog forecasting with Nystr\"om extension}

Our first refinement of basic kernel analog forecasting is based on the Nystr\"om method for out-of-sample extension in reproducing kernel Hilbert spaces \cite{CoifmanLafon06b}. This method builds a data-driven, finite-dimensional Hilbert space of observables which can be evaluated exactly at arbitrary points on $ A $ using only information from the finite training dataset. If it happens that the prediction observable $ U_t f $ actually lies in that Hilbert space for all lead times of interest, then Nystr\"om extension essentially provides an error-free forecast for arbitrary initial data. Of course, the requirement that $ U_t f $ lies entirely in this space  will rarely hold in practice, and analog forecasting via the Nystr\"om method will accrue errors and biases from the residual of $ U_t f $ lying outside that finite-dimensional space. Nevertheless, the method is likely to perform well with data-driven observables (see Section~\ref{sec:notation}) formed by linear combinations of leading kernel eigenfunctions, such as the patterns of low-frequency SST variability studied in Section~\ref{sec:ccsm}. 

\subsubsection{\label{sec:geomHarm}Nystr\"om extension for partially observed ergodic systems}

The starting point of analog forecasting with Nystr\"om extension is a positive-semidefinite kernel $ \sigma : X \times X  \mapsto \mathbb{ R } $. Similarly to the kernel $ K $ in Section~\ref{sec:basicKernel}, $ \sigma $ provides a measure of pairwise similarity between points in the observation space, but it is required to have somewhat different properties, as follows:
\begin{enumerate}
\item Symmetry: $ \sigma( y, x ) = \sigma( x, y ) $ for all $ x, y \in X $;
\item Non-negativity: $ \sum_{i,j=1}^n c_i c_j \sigma( x_i, x_j ) \geq 0 $ for all $ n \in \mathbb{ N } $, $ x_1, \ldots, x_n \in X $, and $ c_1, \ldots, c_n \in \mathbb{ R } $;
\item Boundedness: $ \sigma \leq C $ for a number $ C > 0 $;
\end{enumerate}
Property~(iii), in conjunction with the fact that $ ( M, \mathcal{ M }, \mu ) $ is a probability space, implies that  $ \sigma( y, \cdot ) $ and $ \sigma( \cdot, x ) $ are square-integrable functions in $ L^2( M, \mu ) $. Similarly, $ \sigma( y, \pi( \cdot ) ) $ and $ \sigma( \pi( \cdot ), x ) $ are square-integrable functions in $ L^2( A, \alpha ) $. As with property~(iii) of $ K $ in Section~\ref{sec:basicKernel}, the boundedness assumption on $ \sigma $ can be relaxed at the expense of increasing the complexity of the exposition below.   

For our purposes, it is natural to work with positive-semidefinite kernels constructed from  non-symmetric kernels $ K $ on $ X \times X $ satisfying the conditions of Section~\ref{sec:basicKernel}. As with kernel normalization, there are several ways of carrying out this procedure, and as concrete examples we consider 
\begin{displaymath}
  \hat \sigma_L( y, x ) = \frac{ 1 }{ N } \sum_{i=0}^{N-1} \hat \rho( x_i,y ) \hat \rho( x_i,x ), \quad \hat \sigma_R( y, x ) = \frac{ 1 }{ N } \sum_{i=0}^{N-1} \hat \rho( y,x_i ) \hat \rho( z,x_i ),
\end{displaymath}
where $ \hat \rho( y, x_i ) $ are the weights determined by left normalization of $ K $ via~\eqref{eq:hatRho}. We call the procedures to obtain $ \hat \sigma_L $ and $ \hat \sigma_R $ ``left'' and ``right'' symmetrizations, respectively. Properties (i) and (ii) follow directly by construction of these kernels, and property (iii) follows from the boundedness of $ \hat \rho $ established in Section~\ref{sec:basicKernel}. In the limit of large data, $ N \to \infty $, $ \hat \sigma_L( y, x ) $ and $ \hat \sigma_R( y, x ) $ respectively converge to the symmetric, positive semidefinite, bounded kernels 
\begin{displaymath}
  \sigma_L( y, x ) = \int_M \rho(  \cdot,y  ) \rho( \cdot, x  ) \, d\mu, \quad \sigma_R( y, x ) = \int_M \rho( y,\cdot  ) \rho( x, \cdot ) \, d\mu. 
\end{displaymath} 
Note that, in general, the symmetric kernels introduced above are neither row- nor column-stochastic with respect to the invariant measure of the dynamics, i.e.,  
\begin{displaymath}
  \int_M \hat \sigma_L( y,  \cdot  ) \, d\mu \neq 1, \quad   \int_M \hat \sigma_L(  \cdot  ,  x ) \, d\mu \neq 1,
\end{displaymath}
and similarly for $ \sigma_L $, $ \hat \sigma_R $, and $ \sigma_R $. In Section~\ref{sec:normalization}, we we will discuss alternative normalization strategies of $ K $ \cite{CoifmanHirn13,BerrySauer15,CoifmanLafon06} leading to bistochastic symmetric kernels in the limit of large data. Hereafter, when convenient we use the symbol $ \hat \sigma $ to represent either of $ \hat \sigma_L $ or $ \hat \sigma_R $, and $ \sigma $ to represent either of $ \sigma_L $ or $ \sigma_R $. Moreover, we use the shorthand notations 
\begin{equation}
  \label{eq:sHat}
  \hat s( b, a ) = \hat \sigma( \pi( b ), \pi( a ) ), \quad  s( a, b ) = \sigma( \pi( b), \pi( a ) ). 
\end{equation}
Because $ \hat s $ and $ s $ meet analogous conditions to (i)--(iii) above,  they are symmetric positive-semidefinite kernels on $ A \times A $. We write $ \hat s_L $ and $ s_L $  ($ \hat s_R $ and $ s_R $) whenever we wish to distinguish between the kernels on $ A \times A $ derived by left (right) symmetrization.

Having constructed a positive-semidefinite kernel with any of the above methods, our method follows closely the geometric harmonics technique of Coifman and Lafon \cite{CoifmanLafon06b}, with the difference that we work with a partially observed system (i.e.,  our goal is to construct spaces of functions on $ A $ as opposed to $ M = \pi( A ) $), and we employ the the kernel $ \hat s $ constructed from finite data (as opposed to $ s $ which is more closely related to the case studied in \cite{CoifmanLafon06b}). In particular, we turn attention to the reproducing kernel Hilbert space (RKHS) $ \hat{\mathcal{ H }} $ on  $ M $ associated with $ \hat s $.  According to the Moore-Aronszajn theorem \cite{Aronszajn50}, this space is the unique Hilbert space of scalar-valued functions on $ M $ with inner product $ \langle \cdot, \cdot \rangle_{\hat{\mathcal{ H }}} $, such that (i) for every $ b \in A $, the function $ \hat s( b, \cdot ) $ is in $ \hat{\mathcal{ H } }$; (ii) for every $ f \in \hat{\mathcal{ H }} $ and $ b \in A $, point evaluation can be expressed in terms of the bounded linear operator  
\begin{displaymath}
  f \mapsto f( b ) = \langle f, \hat s( b, \cdot ) \rangle_{\hat{\mathcal{ H }}}. 
\end{displaymath}
This last equation expresses the reproducing property of $ \hat{\mathcal{ H } }$, and it follows from this property that $ \langle \hat s( b, \cdot ), \hat s( a, \cdot ) \rangle_{\hat{\mathcal{ H }}} = \hat s( b, a ) $. More generally, $ \hat{\mathcal{ H }} $ consists of functions of the form $ f = \sum_{i=1}^\infty c_i \hat s(  \cdot, b_i ) $, with $ \sum_{i,j=1}^\infty c_i c_j \hat s( b_i, b_j ) < \infty $. The inner product between $ f $ and another function $ f' = \sum_{i=1}^\infty c'_i \hat s( \cdot, b'_i  ) $ is $ \langle f, f' \rangle_{\hat{\mathcal{ H }}} =  \sum_{i,j=0}^\infty c_i c'_j \hat s( b_i, b'_j ) $. We define the RKHS $ \mathcal{ H } $ associated with the kernel $ s $ and its inner product $ \langle \cdot, \cdot \rangle_\mathcal{ H } $ in an analogous manner, but note that this space cannot be constructed from finite datasets. Note that due to~\eqref{eq:sHat} all functions in $ \hat{ \mathcal{ H } } $ and $ \mathcal{ H } $ are constant on the pre-images $ \pi^{-1}( x ) $ of points in $ x \in M $. Moreover, $ f = \sum_{i=1}^\infty c_i \hat s(  \cdot, b_i ) $ lies above a unique function $ F = \sum_{i=1}^\infty c_i \hat \sigma( \cdot, \pi( b_i ) ) $ on $ X $ such that $ f( b ) = F( \pi( b ) ) $ for any $ b \in A $. The function $ F $ is in the RKHS on $ X $ associated with $ \hat \sigma $.  
  
Next, consider the Hilbert space  $ L^2( A, \hat \alpha ) $ equipped with the purely atomic measure $ \hat \alpha = N^{-1} \sum_{i=0}^{N-1} \delta_{a_i} $ on the sampled points in $ M $. This space consists of all measurable, square-summable functions $ f : M \mapsto \mathbb{ R } $ such that $ \int_A f^2 \, d\hat\alpha = N^{-1} \sum_{i=0}^{N-1} f^2( a_i ) < \infty $, and we denote the inner product and norm of this space by
\begin{displaymath}
  \langle f, g \rangle_{\hat\alpha} = \int_A f g \, d \hat \alpha = \frac{ 1 }{ N } \sum_{i=0}^{N-1} f( a_i ) g( a_i ), \quad \lVert f \rVert_{\hat \alpha} = \langle f, f \rangle_{\hat\alpha}^{1/2},
\end{displaymath}
respectively. Since in an ergodic system the set of fixed points and periodic orbits has measure zero, the sampled states $ \{ a_0, \ldots, a_N \} $ are all distinct with probability 1. Therefore, $ L^2( A, \hat \alpha ) $ is naturally isomorphic to $ \mathbb{ R }^N$, equipped with the inner product $ \langle f, g \rangle_{\mathbb{ R }^N } = N^{-1} \sum_{i=0}^{N-1} f_i g_i $. Our analysis could equally be performed in this space, but we prefer to work with $ L^2( A, \hat \alpha ) $ to emphasize that elements of  $ L^2( A, \hat \alpha ) $ and  $ L^2( A, \alpha ) $ are functions on the same underlying space. 

Because the set of singularities of any function in $ L^2( A, \alpha ) $ has $ \alpha $-measure zero, it follows that for $ \alpha $-almost any starting point $ a_0 $, every $ f $ in $  L^2( A, \alpha ) $ is also in $ L^2( A, \hat \alpha ) $. In particular, this is true for $ \hat s( b, \cdot ) $ for every $ b \in A $.  On the other hand, that $ f $ is in $ L^2( A, \hat \alpha ) $ does not guarantee that it is also in $ L^2( A, \alpha ) $. Indeed, the set of singularities of $ f \in L^2( A, \hat \alpha ) $ has $ \hat \alpha $ measure zero, but can have nonzero $ \alpha $ measure, and therefore $ \int_A f^2 \, d\alpha $ can be infinite. If, however, $ \lVert f \rVert_{\hat{\mathcal{H}}} $ is finite, it follows that $ \lVert f \rVert $ and $ \lVert f \rVert_{\hat \alpha} $ are both finite. Therefore, $ \hat{\mathcal{H}} $ lies in the intersection of $ L^2(A,\alpha ) $ and $ L^2( A, \hat \alpha ) $. Nystr\"om extension techniques exploit this property to stably map subspaces of $ L^2( A, \hat \alpha ) $ to subspaces of $  L^2( A, \alpha ) $. Due to the reproducing property, functions in these subspaces can be evaluated at arbitrary points on $ A $ without approximation.   
 
To perform Nystr\"om extension, we begin by introducing the integral operator $ \hat S : L^2( A, \hat \alpha ) \mapsto \hat{\mathcal{ H } } $, mapping functions in the Hilbert space with the atomic sampling measure to the RKHS through the formula
\begin{displaymath}
  \hat S f( b ) = \int_A \hat s( b, \cdot ) \, d\hat \alpha = \frac{ 1 }{ N } \sum_{i=0}^{N-1} \hat s( b, a_i ) f( a_i ).
\end{displaymath}
The adjoint, $ \hat S^* : \hat{ \mathcal{ H } } \mapsto L^2( A, \hat \alpha ) $, carries out the reverse operation, and it is a consequence of the reproducing property \cite{CoifmanLafon06b} that for any $ g \in \hat{\mathcal{H}} $ and any $ f \in L^2( A, \hat \alpha ) $,
\begin{displaymath}
  \langle \hat S^* g, f \rangle_{\hat\alpha} = \langle g, \hat S f \rangle_{\hat{\mathcal{H}}} = \langle g, f \rangle_{\hat\alpha}.
\end{displaymath}
This means that for $ \hat \alpha $-almost every $ a $, $ \hat S^* g( a ) = g( a ) $, or, equivalently, $ \hat S^* g( a_i ) = g( a_i ) $ on the sampled states $ a_i $. Due to this property, we interpret $ \hat S^* $ as a restriction operator from functions in $ \hat{\mathcal{ H } } $ to functions in $ L^2( A, \hat \alpha ) $. 

Consider now the positive-semidefinite self-adjoint operator $ \hat G : L^2( A, \hat \alpha ) \mapsto L^2( A, \hat \alpha ) $, $ \hat G = \hat S^* \hat S $.  By the spectral theorem, there exists an orthonormal basis $ \{ \hat \phi_0, \ldots, \hat \phi_{N-1} \} $ of $ L^2( A, \hat \alpha ) $ consisting of eigenfunctions of this operator. Denoting the corresponding, non-negative, eigenvalues by $ \{ \hat \lambda_0, \ldots, \hat \lambda_{N-1} \} $, the eigenfunctions and eigenvalues satisfy the relation
\begin{displaymath}
  \langle g, \hat G \hat \phi_k \rangle_{\hat\alpha} = \langle \hat S g, \hat S \hat \phi_k \rangle_{\hat{\mathcal{H}}} = \hat \lambda_k \langle g, \hat \phi_k \rangle_{\hat \alpha}
\end{displaymath}
for any test function $ g \in L^2( A, \hat \alpha ) $. That is, we have
\begin{equation}
  \label{eq:eig}
  \frac{ 1 }{ N } \sum_{i,j=0}^{N-1} g( a_i ) \hat s( a_i, a_j )  \hat \phi_k( a_j ) = \hat \lambda_k \sum_{i=0}^{N-1} g( a_i ) \hat \phi_k( a_i ).
\end{equation}
Writing
\begin{displaymath}
   \hat s_L( a_i, a_j ) = \frac{ 1 }{ N } \sum_{k=0}^{N-1} \hat p( a_k ,  a_i  ) \hat p(  a_k   a_j ) , \quad \hat s_R( a_i, a_j ) = \frac{ 1 }{ N } \sum_{k=0}^{N-1} \hat p( a_i,  a_k  ) \hat p(  a_j,  a_k ),
\end{displaymath}
where $ \hat p( a_i, a_j ) = \hat \rho( \pi( a_i ), \pi( a_j ) ) $, the solutions of this eigenvalue problem can be found by computing the singular value decomposition of the $ N \times N $ matrix $ \mathbf{ P } = [ \hat p ( a_i, a_j ) / N ] $. In the case of left symmetrization, we have $ ( \hat \phi_k( a_0 ), \ldots, \hat \phi_k( a_{N-1} ) )^\top = \vec v_k $, where $ v_k $ is the $ k $-th right singular vector of $ \mathbf{ P } $, and in the case of right symmetrization $ ( \hat \phi_k( a_0 ), \ldots, \hat \phi_k( a_{N-1} ) )^\top = \vec u_k $, where $ \vec u_k $ is $ \mathbf{ P } $'s  $ k $-th left singular vector. In both cases, $ \hat \lambda_k = \hat \sigma_k^2 $, where $ \hat \sigma_k $ is the $ k $-th singular value of $ \mathbf{ P } $. In fact, in some cases, including $ \hat \rho $ constructed via the methods of Section~\ref{sec:normalization}, $ \mathbf{ P } = \mathbf{ D }^{-1} \mathbf{ S } $ for a diagonal matrix $ \mathbf{ D } $ and a symmetric matrix $ \mathbf{ S } $, and therefore it is a normal matrix. In such cases, the left and right singular vectors of $ \mathbf{ P } $ coincide with its eigenvectors, and its singular values are equal to the absolute values of its eigenvalues.     

Given a function $ f = \sum_{i=0}^{N-1} c_i \hat \phi_i  $ in $ L^2( A, \hat \alpha ) $, it is a consequence of the reproducing property that $ c_i = \langle \hat \phi_i, f \rangle_{\hat \alpha} = \hat \lambda_i  \langle \hat \phi_i, f \rangle_{\hat{\mathcal{ H}}} $. Moreover, given another function $ g = \sum_{i=0}^{N-1} c_i' \hat \phi_i $ we have $ \langle f, g \rangle_{\hat{\mathcal{H}}} = \sum_{i=0}^{N-1} c_i c'_i  / \hat \lambda_i $. It therefore follows that the elements of $  L^2( A, \hat \alpha ) $ which are in $ \hat{\mathcal{H}} $ are all functions   $ f  = \sum_{i=0}^{N-1} c_i \hat \phi_i $, $ \sum_{i=0}^{N-1} c_i^2 < \infty $,  such that $ \lVert f \rVert^2_{\hat{\mathcal{H}}} = \sum_{i=0}^{N-1} c^2_i / \hat \lambda_i $ is finite. Intuitively, the kernel $ \hat s $ induces a measure of roughness of functions in $ L^2(A,\hat\alpha) $ through the corresponding RKHS norm; in particular, functions that do not project strongly on the eigenfunctions corresponding to small eigenvalues have small RKHS norm and are deemed smooth, or weakly oscillatory. 

An important property of the eigenfunctions of $ \hat G $ with nonzero corresponding eigenvalues is that the associated functions 
\begin{equation}
  \label{eq:psi}
  \hat \psi_i = \hat \lambda_i^{-1/2}  \hat S \hat \phi_i, \quad \hat\psi_i \in \hat{\mathcal{H}},
\end{equation}
are orthonormal on $ \hat{\mathcal{ H }} $, 
\begin{displaymath}
  \langle \hat \psi_i, \hat \psi_j \rangle_{\hat{\mathcal{H}}} = ( \hat \lambda_i \hat \lambda_j )^{-1/2}  \langle \hat S \hat \phi_i, \hat S \hat \phi_j \rangle_{\hat{\mathcal{H}}} = \langle \hat \phi_i, \hat \phi_j \rangle_{\hat\alpha} = \delta_{ij},
\end{displaymath}
and moreover these functions are uniquely defined at every point on $A $ (as opposed to $ \hat \phi_i $ whose values are arbitrary on sets of $ \hat \alpha $ measure zero). It is straightforward to verify that $ \hat\psi_i $ is an eigenfunction of the positive-semidefinite, self-adjoint operator $ \hat S \hat S^* : \hat{\mathcal{H}} \mapsto \hat{\mathcal{H}} $ with corresponding eigenvalue $ \hat \lambda_i $. Coifman and Lafon \cite{CoifmanLafon06b} call such functions geometric harmonics. Ordering the eigenvalues in decreasing order, $ \hat \lambda_0 \geq \hat \lambda_1 \geq \cdots \geq \hat \lambda_{N-1} $, the set $ \{ \hat \psi_0, \ldots, \hat \psi_{l-1} \} $ with $ \hat \lambda_{l-1} > 0 $ can be interpreted as the set of $ l $ orthonormal functions in $ \hat{\mathcal{H}} $ whose elements are maximally concentrated on the sampled points, in the sense of extremizing the Rayleigh quotient $ \lVert f \rVert_{\hat{\mathcal{H}}}^2 / \lVert f \rVert^2_{\hat \alpha} $.   

While the $ \hat\psi_i $ are guaranteed to be square-integrable functions  by virtue of the fact that $ \hat{\mathcal{H}} \subset L^2(A,\alpha) $, they are generally non-orthogonal  on $ L^2(A,\alpha ) $. In particular, we have 
\begin{displaymath}
  \tilde w_{ij} = \langle \hat \psi_i, \hat \psi_j \rangle = \frac{ ( \hat \lambda_i \hat \lambda_j )^{-1/2} }{ N^2 } \sum_{k,l=0}^{N-1} \hat \phi_i( a_k ) \tilde s_2( a_k, a_l ) \hat \phi_j( a_l ), \quad \tilde s_2( a_k, a_l ) = \int_A \hat s( a_k, \cdot ) \hat s( a_l, \cdot ) \, d\alpha.
\end{displaymath}
Therefore, $ \langle \hat \psi_i, \hat \psi_j \rangle \neq 0 $ for eigenfunctions corresponding to distinct eigenvalues unless the $ ( \hat \phi_i( a_0 ), \ldots, \hat \phi_i( a_{N-1} ) ) $ happen to be eigenvectors of the matrix $ [ \tilde s_2( a_k, a_l ) ] $. Note that  $ \tilde s_2 $, and therefore $ \tilde w_{ij} $, cannot be computed exactly from finite data. 

Despite their lack of orthogonality at finite $ N$, the geometric harmonics converge to orthogonal functions on $ L^2( A, \alpha ) $ as $ N \to \infty $, in the sense that their counterparts $ \{ \psi_0, \psi_1, \ldots \} $ obtained from the kernel $ s $ are orthogonal on $ L^2( A, \alpha ) $. The procedure to obtain $ \{ \psi_i \} $ in the limit of large data is completely analogous to the finite $ N $ case, and can be carried out essentially by replacing all quantities with overhats in the above with their non-overhat counterparts. That is, instead of $ \hat S $ and $ \hat G $, we have the kernel integral operators $ S : L^2(A,\alpha) \mapsto \mathcal{ H } $ and $ G = S^* S : L^2(A,\alpha) \mapsto L^2(A,\alpha ) $, respectively. In this case, $ G $ is a Hilbert-Schmidt operator, whose eigenfunctions $ \{ \phi_0, \phi_1, \ldots \} $ provide an orthonormal basis of $ L^2( A, \alpha ) $, and the functions $ f = \sum_{i=0}^\infty c_i \phi_i \in L^2(A,\alpha) $ which are also in $ \mathcal{ H } $ satisfy $ \lVert f \rVert_{\mathcal{H}}^2 = \sum_{i=0}^\infty c^2_i / \lambda_i < \infty $. Moreover, the geometric harmonics $ \psi_i = S \phi_i $ are orthonormal eigenfunctions of the operator $ S S^* : \mathcal{ H } \mapsto \mathcal{ H } $, and it follows immediately that 
\begin{equation}
  \label{eq:w}
  w_{ij} = \langle \psi_i, \psi_j \rangle =  ( \lambda_i \lambda_j )^{-1/2} \langle S \phi_i, S \phi_j \rangle  = ( \lambda_i \lambda_j )^{-1/2}  \langle \phi_i, G \phi_j \rangle_\mathcal{H}   = \lambda_i \delta_{ij}.
\end{equation}
In the limit of large data, $ \tilde w_{ij} \to w_{ij} $ and the $ \{ \hat \psi_i \} $ become $ L^2(A,\alpha) $-orthogonal.

The lack of orthogonality of the geometric harmonics on $ L^2(A,\alpha ) $ can also be understood as a form of overfitting the training data. In particular, setting $ \hat w_{ij} = \langle \hat \psi_i, \hat \psi_j \rangle_{\hat{\mathcal{H}}} = \hat \lambda_i \delta_{ij} $ and 
\begin{displaymath}
  \hat s_2( a_k, a_l ) = \frac{ 1 }{ N } \sum_{j=0}^{N-1} \hat s( a_k, a_j ) \hat s( a_l, a_j ),
\end{displaymath}
we write $   \tilde w_{ij} = \hat w_{ij} + \Theta_{ij} $, where 
\begin{displaymath}
  \Theta_{ij} = \frac{ ( \hat \lambda_i \hat \lambda_j )^{-1/2} }{ N^2 } \sum_{k,l=0}^{N-1} \hat \phi_i( a_k ) ( \tilde s_2( a_k, a_l ) - \hat s_2( a_k, a_l ) ) \phi_j( a_l )
\end{displaymath}
are quantities that vanish as $ N \to \infty $. Thus, for a collection $ \{ \hat \psi_0, \ldots, \hat \psi_{l-1} \} $ of geometric harmonics, we can study the matrices $ \tilde{\mathbf{ W }} = [ \tilde w _{ij} ] $, $ \hat{\mathbf{ W }} = [ \hat w_{ij} ] $, and $ \mathbf{ \Theta } = [ \Theta_{ij } ] $. Here, the ratio of the matrix norms $ \lVert \mathbf{ \Theta } \rVert / \lVert \hat{\mathbf{ W }}  \rVert $ measures how strongly the inner product relationships of $  \{ \hat \psi_0, \ldots, \hat \psi_{l-1} \} $ on the Hilbert space $ L^2( A, \hat \alpha ) $ associated with the sampling measure differ from those on the Hilbert space $ L^2( A, \alpha ) $ associated with the invariant measure. Large values of this ratio signify that $ \{ \hat \psi_0, \ldots, \hat \psi_{l-1} \} $ overfits the training data. Another way of characterizing overfitting is through the ratio $ \kappa( \tilde{\mathbf{W}})/ \kappa(\hat{\mathbf{W}}) $ of the condition numbers of $ \tilde{\mathbf{W}}$ and $ \hat{\mathbf{W}} $. Qualitatively, for fixed $ N $, we expect both $ \lVert \mathbf{ \Theta } \rVert / \lVert \hat{\mathbf{ W }}  \rVert $ and $ \kappa(\tilde{\mathbf{W}} )/ \kappa(\hat{\mathbf{W}} )$ to increase with $ l $, so there is a tradeoff between the explanatory power afforded by using large $ l $ and the increased risk of overfitting the training data. We will return to this point in Section~\ref{sec:nystromAnalog} ahead.
    


\subsubsection{\label{sec:nystromAnalog}Analog forecasting} 

We now have the necessary ingredients to perform analog forecasting with Nystr\"om extension. First, setting a bandwidth parameter $ l $, we construct an $ l $-dimensional subspace of $ L^2( A, \alpha ) $ spanned by the first $ l $ geometric harmonics, $ \hat B_l = \spn\{ \hat \psi_0, \ldots, \hat \psi_{l-1} \} $. Intuitively, we think of $ \hat B_l $ as a space of bandlimited observables with respect to the kernel $ \hat s $. Three key properties of an observable $ \eta = \sum_{i=0}^{l-1} c_i  \hat \psi_i \in \hat B_l $ are: 
\begin{enumerate}
\item The expansion coefficients $ c_i $ can be determined without approximation from the values of $ \eta $ on the finite training dataset by evaluating
  \begin{equation}
    \label{eq:cNystrom}
    c_i = \langle \eta, \hat \psi_i \rangle_{\hat{\mathcal{H}}} = \lambda_i^{-1/2} \langle \eta,  \hat \phi_i \rangle_{\hat \alpha} =  \frac{ \lambda_i^{-1/2} }{ N } \sum_{j=0}^{N-1} \eta( a_j ) \hat \phi_i( a_j );
  \end{equation}
\item $ \eta $ can be evaluated at arbitrary points on $ A $ using
  \begin{displaymath}
    \eta(b) =  \sum_{i=0}^{l-1} c_i \hat \psi_i( b ) = \frac{1}{N}\sum_{i=0}^{l-1} \sum_{j=0}^{N-1} \hat s( b, a_j ) c_i \hat \lambda_i^{-1/2} \hat \phi_i( a_j );
  \end{displaymath}
\item $ \eta $ lies above a unique function $ H : X \mapsto \mathbb{ R } $ given by
  \begin{displaymath}
    H( y ) =  \frac{1}{N} \sum_{i=0}^{l-1} \sum_{j=0}^{N-1} \hat \sigma ( y, x_j ) c_i \hat \lambda_i^{-1/2} \hat \phi_i( a_j ),
  \end{displaymath}
  and we have $ \eta( b ) = H( \pi( b ) ) $. 
\end{enumerate}
Note that, restricted to $ M \subseteq X $,  $H $ is a square-function in $ L^2(M,\mu )$. In what follows, we denote the function on $ X $ lying above $ \hat \psi_i $ by $ \hat \Psi_i $. 

Next, consider the prediction observable $ f \in L^2( A, \alpha ) $. Fixing a time interval $ [ 0, t_1 ] $, we compute the expansion coefficients
\begin{equation}
  \label{eq:ctNystrom}
  c_i( t ) = \langle \hat \psi_i, U_tf \rangle_{\hat{\mathcal{H}}} = \frac{ 1 }{ N } \sum_{j=0}^{N-1} U_t f( a_i ) \hat \phi_i( a_j ) =  \frac{ 1 }{ N } \sum_{j=0}^{N-1} f_{i+k} \hat \phi_i( a_j )
\end{equation}
for all $ i \in \{ 0, \ldots, l-1 \} $ and $ t = k \tau \in [ 0, t_1 ] $. Then, given an initial condition $ y \in X $, we define the Nystr\"om analog forecast as
\begin{equation}
  \label{eq:fNystrom}
  \hat F_t( y ) = \sum_{i=0}^{l-1} c_i( t ) \hat \Psi_i( y ).
\end{equation}
For initial data of the form $ y = \pi( b ) $, $ b \in A $, the forecast error with this scheme is
\begin{displaymath}
  \hat r_t( b ) = \hat F_t( y ) - U_t f( b ).
\end{displaymath}

To interpret this forecast, suppose first that $ U_t f $ lies in $ \hat B_l $ for all discrete times $ t = k \tau \in [ 0, t_1 ] $. Then, we have $ U_t f = \sum_{i=0}^{l-1} c_i( t ) \hat \psi_i $, and for initial data of the form $ y = \pi( b ) $ the forecast function $ \hat F_t( y ) $ equals $ U_t f( b ) $ and the forecast error vanishes. Note that there is no approximation in this result despite the fact that the training dataset contains a finite number of samples.

Of course, the assumption that $ U_t f $ is a bandlimited observable in $ \hat B_l $ for all lead times of interest is unlikely to hold in practice. First, in many forecast scenarios the prediction observable is pre-determined, and the chances of it lying entirely in a finite-dimensional data-driven subspace are small. In the language of Section~\ref{sec:notation}, this situation would likely arise for objectively defined observables. Even if $ U_t f $ were in $ \hat B_l $ at a fixed time (say at $ t = 0 $), $ \hat B_l $ is generally not an invariant subspace of the dynamics, and there is no reason to expect $ U_t f $ to lie in this space at other times. 

Given a general prediction observable $ f \in L^2( A, \alpha ) $, then with probability 1  (i.e., excluding cases of $ \alpha $-measure zero where $ f $ is singular on the sampled states $ a_i $) we can write    
\begin{displaymath}
  U_tf = \eta_{t,l} + \hat r_{t,l},
\end{displaymath}
where $ \eta_{t,l} = \sum_{i=0}^{l-1} c_i( t ) \hat \psi_i $ is in $ \hat B_l $, and $ \hat r_{t,l} \in B_l^\perp $ is a residual in the orthogonal complement of $ \hat B_l $ in $ L^2( A, \hat \alpha ) $.  The coefficients $ c_i( t ) $ can be computed without approximation using~\eqref{eq:ctNystrom}, and moreover the refinement $  \eta_{t,l+1}-\eta_{t,l} $ at each $ l $ is $ L^2(A, \hat \alpha ) $-orthogonal to the previous approximation $ \eta_l $. Thus,
\begin{equation}
  \label{eq:resHat}
  \lVert \hat r_{t,l+1} \rVert^2_{\hat\alpha} = \lVert \hat r_{t,l} \rVert^2_{\hat\alpha} - \lVert \eta_{t,l+1} - \eta_{t,l} \rVert^2_{\hat\alpha},
\end{equation} 
and the residual norm on $ L^2( A, \hat \alpha) $ is a non-increasing function of $ l $, i.e., $ \lVert r_{t,l+1} \rVert_{\hat\alpha} \leq \lVert r_{t,l} \rVert_{\hat\alpha} $.

However, this does not mean that the residual norm $ \lVert r_{t,l+1} \rVert $ on $ L^2( A, \alpha ) $, which is equal to the RMSE of the forecast for initial data drawn from the invariant measure, is also a non-increasing function of $ l $. Indeed, since the $ \hat \psi_i $ are non-orthogonal on $ L^2( A, \alpha ) $ (see Section~\ref{sec:geomHarm}), the refinement at each $ l $ is not $ L^2( A,  \alpha ) $-orthogonal to the previous approximation, and instead of~\eqref{eq:resHat} we have
\begin{equation}
  \label{eq:rTL}
  \lVert \hat r_{t,{l+1}} \rVert^2 = \lVert \hat r_{t,l} \rVert^2 - \lVert \eta_{t,l+1} - \eta_{t,l} \rVert^2 + 2 \langle \eta_l, \eta_{l+1} - \eta_l \rangle,
\end{equation}
which is not guaranteed to be non-increasing. An increasing residual norm with $l $ is interpreted as overfitting the training data. According to~\eqref{eq:w}, in the limit of large data the geometric harmonics become $ L^2(A,\alpha) $-orthogonal, and the expected residual becomes a non-increasing function of $ l $.

Equation~\eqref{eq:rTL} can also be expressed in terms of the Gramm matrix elements $ \tilde w_{ij} = \hat w_{ij} + \Theta_{ij} $ introduced in Section~\ref{sec:geomHarm}, namely,
\begin{displaymath}
  \lVert \hat r_{t,l+1} \rVert^2 = \lVert \hat r_{t,l} \rVert^2 - \hat \lambda_{l+1} c_{l+1}^2  + 2 \sum_{i=1}^l c_i c_{l+1} \Theta_{i,l+1}.
\end{displaymath}
From the above, we see that the possibility of increasing expected residual with $ l $ is related to the non-orthogonality of the geometric harmonics with respect to the invariant measure of the dynamics. At fixed sample number $ N $, the coefficients $ \Theta_{ij } $ generally increase in magnitude with $ i $ and $j $, and the risk of increase of $ \lVert r_{t,l} \rVert $ also increases. In other words, we are faced with the usual tradeoff between the ability to predict large classes of observables (large dimension of $ \hat{B}_l $) and risk of overfitting the training data. In the applications of Section~\ref{sec:results}, we will select values of $ l $ achieving a good balance between these competing sources of error by cross-validation against a dataset independent of the training data. A more detailed study of such strategies is beyond the scope of the present paper, but should be an interesting avenue of future research.

Despite the potential shortcomings of the Nystr\"om method when dealing with objectively-defined observables, in several interesting applications the objective is to actually construct data-driven observables that capture the salient features of high-dimensional data and have favorable predictability properties. Eigenfunctions of kernels designed specifically for analyzing dynamical system data are theoretically \cite{BerryEtAl13,Giannakis15} and experimentally \cite{GiannakisMajda12a,GiannakisMajda12b,BushukEtAl14,SzekelyEtAl14} known to perform well in terms of timescale separation and physical interpretability. Therefore, observables lying in the bandlimited spaces associated with these kernels are natural candidates for prediction. An attractive property observables in these spaces is that they have vanishing reconstruction error at forecast initialization (i.e., $ \hat F_0(y) = f(b) $ for all $ y = \pi( b ) $). In Section~\ref{sec:kernels}, we will argue that these spaces may persist (i.e., be approximately invariant) under the dynamical flow, improving the predictability of the corresponding data-driven observables on short to intermediate times.

\subsection{\label{sec:pyramids}Kernel analog forecasting with Laplacian pyramids}

In this Section, we present a kernel analog forecasting technique based on the Laplacian pyramids algorithm for multiscale approximation of functions introduced by Rabin and Coifman \cite{Rabin2012} and further improved by Fern\'andez et al.~\cite{FernandezEtAl14}. This forecasting approach combines aspects of the basic kernel analog forecasting technique in Section~\ref{sec:basicKernel} with RKHS ideas from Section~\ref{sec:nystrom}. In what follows, we begin by presenting the approach in the limit of large data (Section~\ref{sec:pyramidsLarge}), and then discuss its implementation for finite datasets (Section~\ref{sec:pyramidsFinite}). Throughout this Section, we assume that the prediction observable $ f $ and all kernel-derived densities are square-integrable functions in $ L^2( A, \alpha ) $.  


\subsubsection{\label{sec:pyramidsLarge}Laplacian pyramids in the limit of large data}

First, consider the basic kernel analog forecast $ F_t(y ) $ in~\eqref{eq:basicAnalogLarge}, determined  by the expectation of $ U_t f $ with respect to the kernel-dependent probability measure $ \nu_y $. For initial data of the form $ y = \pi( b ) $, $ b \in A $,  this expression can alternatively be expressed in terms of an averaging operator, $ P_0 : L^2( A, \alpha ) \mapsto L^2( A, \alpha ) $, namely
\begin{equation}
  \label{eq:P0}
  f_{t,0}( b ) = P_0 U_tf( b ) = \int_A p_0( b, a ) U_t f( a ) \, d\alpha( a ).
\end{equation}
Here, $ p_0( b, a ) = \rho_0( y, \pi( a) ) $, and $ \rho_0 $ is a density determined by normalization of a kernel $ K $ on $ X \times X $, e.g., via~\eqref{eq:rho} so that $ \int_A p_0( b, \cdot ) \, d\alpha = 1 $ and $ \int_M \rho_0( y, \cdot ) \, d \mu = 1 $ for all $ b \in A $ and $ y \in X $. (The reason for introducing the ``0'' subscripts will become apparent below.)  Moreover, as discussed in Section~\ref{sec:basicKernel}, the error in $ f_{t,0} $ is given by  
\begin{displaymath}
  r_{t,0}( b ) = U_t f( b ) - P_0 U_t f( b ).
\end{displaymath}
For an observable with mean $ \bar f = \int_A f \, d\alpha = \int_A U_t f \, d\alpha $ we can also write  
\begin{displaymath}
  f_{t,0}( b ) = \bar f + P_0 U_t f'( b ), \quad r_{t,0}(b) = U_t f'( b ) - P_0 U_t f'( b ), 
\end{displaymath} 
where $ f' = f - \bar f $. 


 
While the Laplacian pyramids procedure can be operationally carried out for arbitrary averaging operators, in order to place bounds on the RMS forecast error, and ensure stability of the algorithm, we require that $ \lVert ( I - P_0 ) f \rVert \leq \lVert f \rVert $. This condition can be met for arbitrary observables in $ L^2( A, \alpha )$ if $ P_0 $ is positive-semidefinite and self-adjoint on $ L^2(A,\alpha ) $, and $ \alpha $ is an ergodic invariant measure of the Markov semigroup $ \{ P^n_0 \}_{n\in\mathbb{ N }} $ (see also Section~\ref{sec:basicKernel}). That is, we must have  $ \lim_{n\to\infty} \frac{ 1 }{ n } \sum_{k=1}^n P_0^k f = \int_A f \, d \alpha $ for all $ f \in L^2(A,\alpha ) $. The ergodicity of $ P_0 $ is equivalent to the requirement that $ P_0 f = f $ implies that $ f $ is $ \alpha $-almost everywhere constant. Moreover, that $ P_0 $ is self-adjoint on $ L^2(A,\alpha) $ implies that the kernel $ p_0 $ is symmetric and therefore bistochastic, $ \int_A p_0( b, \cdot ) \, d\alpha = \int_A p_0( \cdot, a ) \, d\alpha = 1 $. Such kernels can be constructed from more general similarity kernels via the normalization procedure developed in \cite{CoifmanHirn13}, which we will also employ in Section~\ref{sec:bistochastic}. 

These assumptions lead to the residual bound  
\begin{equation}
  \label{eqRT0}
  \lVert r_{t,0} \rVert = \lVert ( I - P_0 ) U_t f' \rVert \leq \lVert f' \rVert,
\end{equation}
where we have used the fact that $ \lVert U_t f' \rVert = \lVert f' \rVert $.  Moreover, the expected residual $ \int_A r_{t,0} \, d \alpha $ vanishes due to the bistochasticity of $ p_0 $ (see Section~\ref{sec:basicKernel}), and the residual norm $ \lVert r_{t,0} \rVert $ is equal to the RMS forecast error.  Note if $ U_t f' $ is not in the nullspace of $ P_0 $ the inequality in~\eqref{eqRT0} becomes strict, i.e., $ \lVert r_{t,0} \rVert < \lVert f' \rVert $. Furthermore, because $ p_0( b, \cdot ) = \rho_0( \pi( b ), \pi( \cdot ) ) $, the function $ f_{t,0} : A \mapsto \mathbb{ R } $ lies above a function $ F_{t,0} $ on $ X $ given by 
\begin{displaymath}
  F_{t,0}( y ) = \int_A \rho_0( y, \pi( a ) ) U_tf( a) \, d\alpha(a).
\end{displaymath} 
This function agrees with $ f_{t,0}( b ) $ for initial data of the form $ y = \pi( b ) $, and its restriction on $ M = \pi(A ) $ is a square-integrable function in  $L^2(M,\mu)$.  

While basic analog forecasting would stop at~\eqref{eq:P0}, Laplacian pyramids treats $ r_{t, 0} $ as another function which is to be approximated by averaging. In particular, noticing that $ P_0 U_t f $ lies in a RKHS $ \mathcal{ H }_0 \subset L^2( A, \alpha ) $ associated with the kernel $ p_0 $, it adjusts the forecast by adding a correction, $ g_{t,1} $, which lies in a RKHS $ \mathcal{ H }_1 $ associated with a kernel $ p_1 $. 

There are several ways of constructing $ \mathcal{ H }_1 $, and more generally a sequence $ \{ \mathcal{ H }_0, \mathcal{ H }_1, \ldots, \mathcal{ H }_l \} $ of RKHSs such that $ \mathcal{ H }_i \subset L^2(A,\alpha ) $. In the setting of exponentially decaying, bandwidth-dependent kernels used in this work, one can simply fix the functional form of the kernel $ K_\epsilon $ (e.g, use the radial Gaussian family in~\eqref{eq:KGauss}), and set the $ \mathcal{ H }_i $ to the RKHSs associated with the bistochastic kernels $ p_i $ constructed from $ K_{\epsilon_i} $ through the procedure in Section~\ref{sec:bistochastic} for a decreasing sequence $ \epsilon_0 > \epsilon_1 > \cdots > \epsilon_l $. Refs.~\cite{Rabin2012,FernandezEtAl14} use successive subdivisions of the form $ \epsilon_i = \epsilon_0 / 2^i $, which we also employ here. Due to the bistochasticity of the $ p_i $, all of the corresponding averaging operators $ P_i : L^2( A, \alpha ) \mapsto L^2( A, \alpha ) $ are self-adjoint, and the range of $P_i $ is contained in $ \mathcal{ H }_i $.  


Having constructed a nested-RKHS family, we begin by approximating the residual at iteration~1 by acting on it with $ P_1 $:
\begin{displaymath}
  g_{t,1}( b ) = P_1 r_{t,0}( b ) = \int_A p_1( b, a ) r_{t,1}( a ) \, d\alpha( \alpha ). 
\end{displaymath}
The function $ g_{t,1} $ is now in $ \mathcal{ H }_1 $, and the residual $ r_{t,1}( b ) = r_{t,0}-g_{t,1} $ can be bounded in the $ L^2(A,\alpha) $ norm through  
\begin{displaymath}
  \lVert r_{t,1}  \rVert \leq \lVert r_{t,0}  \rVert \leq \lVert f' \rVert.
\end{displaymath}
We then define the first-level approximation of $ U_t f $ as $ f_{t,1} = f_{t,0} + g_{t,1} $, and this function is in  $ \mathcal{ H }_1 $. Since both $ p_0 $ and $ p_1 $ are bistochastic, this approximation has vanishing expected error, $ \int_A r_{t,1} d\alpha = 0 $, and $ \lVert r_{t,1} \rVert $ is equal to the RMS forecast error.

Iterating this procedure $ l $ times yields the $ l $-level Laplacian pyramids approximation of the forecast observable $ f $ at lead time $ t $: 
\begin{equation}
  \label{eq:LapApprox}
  f_{t,l} = f_{t,0} + \sum_{i=1}^l g_{t,i}, \quad f_{t,l} \in \mathcal{ H }_l,
\end{equation} 
where $ f_{t,0} $ is given by~\eqref{eq:P0}, and $ g_{t,i} $ are functions in $ \mathcal{ H }_i $ determined iteratively through the formulas
\begin{equation}
  \label{eq:LapApprox2}
  g_{t,i} = P_i r_{t,i-1}, \quad 
  r_{t,i} = 
  \begin{cases}   
    f_t - f_{t,0}, & i = 0, \\
    r_{t,{i-1}} - g_{t,i}, & i \geq 1.
  \end{cases}
\end{equation}
This approximation has the residual norm 
\begin{equation}
  \label{eq:rTLLP}
  \lVert r_{t,l} \rVert \leq \lVert r_{t,l-1} \rVert \leq \cdots \leq \lVert r_{t,0} \rVert \leq \lVert f' \rVert,
\end{equation}
and the above inequalities become strict if the $ r_{t,i} $ and $ U_tf' $ are not in the nullspaces of the corresponding $ P_i $. Moreover, $ \int_A r_{t,l} \, d\alpha $ vanishes by bistochasticity of the $ p_i $, and $ \lVert r_{t,i} \rVert $ is equal to the RMS forecast error.  As with $ f_{t,0} $, $ f_{t,l} $ lies above a function $ F_{t,l} : X \mapsto \mathbb{ R } $ such that $ F_{t,l}( \pi( b ) ) = f_{t,l}( b ) $, and the restriction of $ F_{t,l} $ on $ M $ is in $ L^2(M,\mu )$. 

In general, the limiting behavior of the algorithm as $ l \to \infty $ depends on  whether  or not the limit space in the sequence $ \{ \mathcal{ H }_l \} $ is the whole $ L^2(A,\alpha)$ space. Since the kernels associated with $ \mathcal{ H }_l $ have the form $ K_{\epsilon_l}( \pi( \cdot ), \pi( \cdot ) ) $, functions in $ \mathcal{ H }_l $ are always  constant on the pre-images $ \pi^{-1}( x) $ of points $x \in M $.  Therefore, if $ \pi $ is not invertible on its image (i.e., the observations are incomplete), $ \mathcal{ H }_l $ will converge to a strict subspace of $ L^2(A,\alpha) $ and $ \lim_{l \to \infty} \lVert r_{t,l} \rVert $ will generally be nonzero. 

Note that while the requirement that the $ P_i $ are self-adjoint on the $L^2$ space associated with the invariant measure of the dynamics is very natural, the method can still be applied for more general averaging operators with ergodic invariant measures $ \alpha_i $. In such cases, the ranges of the $ P_i $ are contained in RKHSs with reproducing kernels given by left or right symmetrization of $ p_i $ (see Section~\ref{sec:geomHarm}). Moreover, residual bounds analogous to~\eqref{eq:rTLLP} can be derived for the $ L^2( A, \alpha_i ) $ norms. If the $ \alpha_i $ are sufficiently smooth (e.g., absolutely continuous) relative to the invariant measure of the dynamics, then it may be possible to obtain from these bounds useful bounds for the residual norm on $ L^2(A,\alpha) $. Here, we do not study the behavior of the residual in more detail, but we note the result in \cite{FernandezEtAl14} that for complete observations, radial Gaussian kernels, and uniform sampling density, the residual norm decreases with $ l $ faster than any algebraic rate.      

\subsubsection{\label{sec:pyramidsFinite}Laplacian pyramids for finite datasets}

Analog forecasting with Laplacian pyramids and finite datasets is structurally similar to the formulation in the limit of large data, but one has to additionally confront the risk of overfitting. More specifically, in the finite-sample case we replace the probability kernels $  p_i  $ (which involve integrals with respect to the invariant measure; see~\eqref{eq:bistochastic}) with bistochastic kernels $ \hat p_i = \hat \rho_i( \pi( \cdot ), \pi( \cdot ) ) $ on $ L^1( A, \hat \alpha ) $ constructed using the method in Section~\ref{sec:bistochastic}. These kernels satisfy $ \hat  p_i( b, a ) = \hat p_i(a, b ) $ for arbitrary points in $ A$, and moreover $ N^{-1} \sum_{j=0}^{N-1} \hat p_i( b, a_i ) = 1 $. While $ \hat p_i $ is not bistochastic on $ L^1(A,\alpha) $, it converges to the bistochastic kernel $ p_i $ from Section~\ref{sec:pyramidsLarge} as $N \to \infty$.  

Using the same sequence of decreasing bandwidths $ \epsilon_i $ as in Section~\ref{sec:pyramidsLarge}, we construct the sequence of RKHSs $ \{ \hat{\mathcal{H }}_0,  \hat{\mathcal{H}}_1, \ldots, \hat{\mathcal{H}}_l \} $, $ \hat{\mathcal{H}}_i \subset L^2(A,\hat\alpha ) $, associated with $ \hat p_i $, and introduce the positive-semidefinite, self-adjoint averaging operators $ \hat P_i : L^2(A, \hat \alpha ) \mapsto L^2( A, \hat \alpha ) $ such that
\begin{displaymath}
  \hat P_i f( b ) = \int_A \hat p_i( b, a ) f( a ) d\hat \alpha(a ) = \frac{ 1 }{ N} \sum_{j=0}^{N-1}  \hat p_i( b, a_j ) f( a_j ), \quad \hat P_i f \in \hat{\mathcal{H}}_i.
\end{displaymath}
Note that by square-integrability of $ \hat p_i $, the $ \hat{\mathcal{H}}_i $ are also subspaces of $ L^2(A,\alpha) $ (see Section~\ref{sec:geomHarm}).  In what follows, we will assume that $ \hat\alpha $ is an ergodic invariant measure of the Markov semigroups $ \{ \hat P^n_i \}_{n\in\mathbb{N}} $; this condition will be fulfilled so long as the $ N \times N $ Markov matrices $ \hat{\mathbf{ P }}_i = [ \hat p_i( a_j, a_k ) / N  ]$ on the training data are irreducible.  

At the zeroth level of the Laplacian pyramids construction, we evaluate
\begin{displaymath}
  \hat f_{t,0} = \hat P_0 U_t f.
\end{displaymath}
This approximation has a residual in $ L^2(A,\hat \alpha ) $ given by 
\begin{displaymath}
  \hat r_{t,0} = U_t f - \hat P_0 U_t f, 
\end{displaymath}
which we approximate in turn through
\begin{displaymath}
  \hat g_{t,1} =\hat  P_1 \hat r_{t,0}.
\end{displaymath}
 From this point onward, the iterations are repeated $ l $ times analogously to~\eqref{eq:LapApprox} and~\eqref{eq:LapApprox2}, replacing $ P_i $ by $ \hat P_i $,  until the Laplacian pyramids approximation
\begin{displaymath}
  \hat f_{t,l} = \hat f_{t,0} + \sum_{i=1}^l \hat g_{t,l}
\end{displaymath}
is completed. The function $ \hat f_{t,l} $ lies above a unique function $ \hat F_{t,l} $ on $ X $ constructed analogously to $ F_{t,l} $ in Section~\ref{sec:pyramidsLarge}. 

By self-adjointness of the $ \hat P_i $ on $ L^2(A, \hat\alpha ) $ and ergodicity of the associated diffusion semigroups, the residual norm $ \lVert \hat r_{t,l} \rVert_{\hat\alpha} $ on $ L^2( A, \hat \alpha ) $  is a non-increasing function of $ l $. In particular, we have
\begin{equation}
  \label{eq:hatRTL}
  \lVert \hat r_{t,l} \rVert_{\hat \alpha} \leq \lVert \hat r_{t,l-1} \rVert_{\hat\alpha} \leq \cdots \leq   \lVert \hat r_{t,0} \rVert_{\hat\alpha} \leq \lVert U_t f - \hat f_t \rVert_{\hat\alpha},
\end{equation}
where $ \hat f_t = \int_A U_t f \, d \alpha $ the expectation of the time-shifted prediction observable with respect to the sampling measure. Note that unlike the corresponding expectation value with respect to the invariant measure (used, e.g., in~\eqref{eq:rTLLP}),  $ \hat f_t $ is not time-independent (but converges to a time-independent function in the limit of large data). As with~\eqref{eq:rTLLP}, the inequalities in~\eqref{eq:hatRTL} become strict if the $ \hat r_{t,i} $ and $ U_t f $ are not in the nullspaces of the corresponding $ \hat P_i $ operators.          

Of course, that the residual norm on $ L^2( A, \hat\alpha ) $ is non-increasing  does not imply that the residual norm on $ L^2( A, \alpha ) $ is non-increasing too. In fact, at fixed $ N $, we generally expect the discrepancy between $ \lVert \hat r_{t,l} \rVert_{\hat \alpha} $ and $ \lVert \hat r_{t,l} \rVert $ to increase with $ l $. Intuitively, as $ l $ increases $ \hat{\mathcal{H}}_l $ contains functions of progressively smaller scales, which are prone to similar overfitting effects as those discussed in the context of the Nystr\"om extension in Section~\ref{sec:nystrom}. 

In the case of Laplacian pyramids, the occurrence of overfitting is made particularly transparent from the fact that the probability measures induced on $ ( A, \mathcal{ A } ) $ by the $ \hat p_i $ converge as $ i \to \infty $ to pullbacks of $ \delta $-measures in the data space. To examine this in more detail, fix a state $ a_j \in A $ in the training dataset and  consider the probability measure $ \hat \nu_{a_j,i} $ on $ ( A, \mathcal{ A } ) $ such that $ \hat \nu_{a_j,i}( S ) = \int_S \hat p_i( a_j, \cdot ) \, d\hat \alpha $. For typical bandwidth-dependent kernels (including the radial Gaussian kernel in~\eqref{eq:KGauss} and the kernels of Section~\ref{sec:kernel}), as $ i \to \infty $ and $ \epsilon_i \to 0 $, this probability measure converges to $ \Delta_{a_j} = \delta_{\pi(a_j)} $. That is, as $ i \to \infty $, $ \hat P_i $ acts on $ f $ by averaging over its values on the points in $ A $ that are mapped to a single point $ x_j = \pi( a_j )  $ in $ M $. Moreover, if the observations are complete, $ a_j $ is the only point in $ A $ which is sent to $ x_j $ and $ \Delta_{a_j} = \delta_{a_j} $. 

To alleviate this issue, Fern\'andez et al.~\cite{FernandezEtAl14} advocate replacing $ \hat p_i $ by the ``hollow'' transition probability kernels $ \tilde p_i $ such that 
\begin{equation}
  \label{eq:tildeP}
  \tilde p_i( b, a_j ) = \tilde \rho_i( y, x_j ), \quad 
  \tilde \rho_i( y, x_j ) = 
  \begin{cases}
    0, & y = x_j, \\
   \hat \rho_i( y, x_j ) / \zeta( y ), & y \neq x_j.
 \end{cases}
\end{equation}
In the above, $ y = \pi( b ) $, $ x_j = \pi( a_j ) $, and  $ \zeta( y ) = \sum_{x_k \neq y} \hat \rho_i( y, x_k ) / \tilde N( y )$, where $ \tilde N( y ) $ is equal to the number of points in the training dataset $ \{ x_k \} $ different from $ y $. Inspired by leave-one-out cross-validation techniques, this modification of the kernel has a small effect at large $ \epsilon_i $, but prevents the probability measures $ \hat \nu_{a_j,i} $ to degenerate to $ \Delta_{a_j} $ as $ \epsilon_i \to 0 $ (at the expense of introducing a small bias in the invariant measure of the corresponding Markov semigroup). As a result, the residual norm $ \lVert \hat r_{t,l} \rVert_{\hat \alpha} $ may increase with $l $. In the experiments of Section~\ref{sec:results}, we will always use the transition kernels $ \tilde p_i $ when making kernel analog forecasts with Laplacian pyramids. Operationally, we find that as $ l $ grows, $ \lVert r_{t,l} \rVert_{\hat\alpha} $ initially decays, but eventually reaches a minimum and then starts increasing again. Choosing $ l $ as the minimizer of  $ \lVert r_{t,l} \rVert_{\hat \alpha} $ appears to be an effective practical criterion for balancing increased explanatory power with risk of overfitting the training data.

\section{\label{sec:kernel}Choice of kernel}

While the kernel analog forecasting techniques presented in Section~\ref{sec:mathback} place few constraints on the form of the kernel, the choice of kernel will ultimately have a strong impact on forecast skill. In this Section we discuss specific examples of kernels for analog forecasting which we will subsequently employ in the applications of Section~\ref{sec:results}. Here, we focus on two aspects of kernel design, namely (i) the selection of the functional form of the kernel $ K $ in Section~\ref{sec:basicKernel}, and (ii) the normalization and symmetrization of that kernel to construct the various out-of-sample extension and averaging operators in Sections~\ref{sec:nystrom} and~\ref{sec:pyramids}. 

In this discussion, we restrict attention to so-called local kernels \cite{BerrySauer15}, i.e., kernels with an exponential decay controlled by a bandwidth parameter, $ \epsilon $, such as the radial Gaussian kernel $ K_\epsilon $ in~\eqref{eq:KGauss}. Intuitively, $ \epsilon $ controls the ``resolution'' of $ K_\epsilon $ in the phase space $ A $ (e.g., through the associated nested RKHSs of Section~\ref{sec:pyramids}), and one can study the asymptotic behavior of these kernels as $ \epsilon \to 0 $. In several cases of interest (e.g., when $ X $ is a Hilbert space and $ M $ is a smooth manifold embedded in $ X $), the limiting behavior of local kernels is governed by well-studied operators on manifolds such as diffusion operators and vector fields. Thus, one can use the properties of these operators to inform kernel design.          
   
\subsection{\label{sec:delay}Delay-coordinate maps}
 
Delay-coordinate maps \cite{Takens81,BroomheadKing86,SauerEtAl91} is a powerful tool for recovering topological features of attractors of dynamical systems lost through partial observations. Here, we use delay-coordinate maps as a means of actually constructing the data space $ X $ on which subsequent kernel calculations are performed. 

Consider the same dynamical system $ (A,\mathcal{A}, \Phi_t, \alpha ) $ as Section~\ref{sec:notation}, but observed through a different observation map, $ \tilde \pi : A \mapsto Z $, where $ Z $ is a data space. We assume that $ Z $ and $ \tilde \pi $ have similar measurability properties as those required to hold for $ X $ and $ \pi $ in Section~\ref{sec:notation}.  Fixing a positive integer parameter $ q $ (the number of delays), we map the point $ a $ in phase space $ A $ to a point $ x $ in $ X = Z^q $ (hereafter, called delay-coordinate space) via one of the two mappings 
\begin{gather*}
  \begin{aligned}
    a \mapsto x & = \pi_-( a ) = ( \tilde \pi( a ), \tilde \pi( \Phi_\tau^{-1}( a ) ), \ldots, \tilde \pi( \Phi^{-1}_{(q-1)\tau} ( a ) ) ), \\
    a \mapsto x &= \pi_+( a ) = ( \tilde \pi( a ),  \tilde \pi( \Phi_\tau( a ) ), \ldots, \tilde \pi( \Phi_{(q-1)\tau}( a ) ) ), 
  \end{aligned}
\end{gather*}
Note that in the non-invertible (semigroup) case $ \pi_- $ is not defined, and this will have an impact on the utility of delay-coordinate mapping in analog forecasting of these systems. In what follows,  whenever we use $ \pi_{-} $ we will tacitly assume that the dynamics is invertible. Denoting the time series of observations in $ Z $ by $ \{ z_0, z_1, \ldots, \} $, where $ z_i = \tilde \pi( a_i ) $, the action of $ \pi_- $ and $ \pi_+ $ on a state $ a_i $ sampled in the training phase is determined by the forward or backward lagged sequences,
\begin{equation}
  \label{eq:delay}
  x_i = \pi_{-}( a_i ) = ( z_i, z_{i-1}, \ldots, z_{i-(q-1)} ), \quad x_i = \pi_{+}( a_i ) = ( z_i, z_{i+1}, \ldots, z_{i+(q-1)} ),
\end{equation}
respectively. Notice that so long as the time ordering of the training data is available $ \pi_-( a_i ) $ and $ \pi_+( a_i ) $ can be evaluated without explicit knowledge of the underlying state $ a_i $. 
 
According to a theorem of Takens~\cite{Takens81}, which was generalized in \cite{SauerEtAl91} to systems with fractal attractors under prevalent sets of observation functions,
if $ \Phi_t $ is a flow on a compact attractor $ A $ of a dynamical system, then with high probability the points $ x_i = \pi( a_i ) $ lie on a set $ M = \pi( A ) \subseteq X $ which is in one-to-one correspondence with $ A $ (i.e., $ A $ and $ M $ are diffeomorphic manifolds), provided that $ q $  is sufficiently large.  Thus, because knowledge of the position on the attractor is sufficient to determine the future evolution of the system, the time series $ \{ x_i \} $ becomes Markovian even if $ \{ z_i \} $ is non-Markovian. 

Since analog forecasting relies on following the evolution of observables on such time series starting from given initial conditions (analogs), it is natural to identify those initial conditions using the lagged sequences $ x_i $ rather than $ z_i $. Even in situations where $ M $ and $ A $ are not in one-to-one correspondence despite delay-coordinate mapping, we expect $\pi $ to have recovered at least some predictive information which is not contained in $ \tilde \pi  $. For this reason, we will construct our kernels for analog forecasting in delay-coordinate space $ X $, as opposed to the ``snapshot'' space $ Z$. This approach should be beneficial in both traditional analog forecasting with respect to Euclidean distances and the kernel-based methods developed here. Note that the special case $ q = 1 $ corresponds to no embedding, $ X = Z $. 

Before proceeding, it is important to note a key difference between the backward ($\pi_-$) and forward $(\pi_+)$ delay-coordinate maps. Namely, $ \pi_- $ can be evaluated in a ``real-time'' fashion by concatenating previously observed snapshots, whereas $ \pi_+ $ cannot be evaluated without using information from the future. As a result, in analog forecasting scenarios where the snapshots $ z_i $ used in forecast initialization are obtained incrementally, the smallest lead time for which the time interval needed to evaluate $ \pi_+ $ does not overlap with the forecast time interval is $ t_0 = ( q - 1 ) \tau $. In other words, when measuring the skill of forecasts made with $ \pi_+ $, the first $ q -1 $ leads are ``used up'' for forecast initialization. Due to these considerations, when the dynamics are invertible $ \pi_- $ is preferable for analog forecasting (whereas $ \pi_- $ and $ \pi_+ $ perform equally well for topological recovery). When the dynamics is not invertible and $ \pi_- $ cannot be used there is a tradeoff between topological recovery (favored by large $ q $) and ease of forecast initialization (favored by small $ q $).

\subsection{\label{sec:kernels}Local kernels with dynamics-adapted features}

\subsubsection{Local kernels in delay-coordinate space}
 
Hereafter, we assume that the data space $ Z $ is a Hilbert space with inner product $ \langle z, z' \rangle_Z $ and norm $ \lVert z \rVert = \langle z, z' \rangle_Z^{1/2} $. Under this assumption, the delay-coordinate space $ X $ naturally acquires a Hilbert space structure with inner product given by summing the $ \langle \cdot, \cdot \rangle_Z $ inner products at each lag. That is, for $ x = ( \zeta_0, \ldots, \zeta_{q-1} ) $ and $ z' = ( \zeta'_0 , \ldots, \zeta_{q-1}' ) $ with $ \zeta_i, \zeta'_i \in Z $ we define $ \langle x, x' \rangle = \sum_{i=0}^{q-1} \langle \zeta_i, \zeta'_i \rangle_Z $ and $ \lVert x \rVert = \langle x, x \rangle^{1/2} $. An example of a kernel in delay-coordinate space is the radial Gaussian kernel 
\begin{equation}
  \label{eq:KDelay}
  K_\epsilon( y, x ) = e^{-D^2(y,x) / \epsilon } , \quad  D( y, x ) = \lVert y - x \rVert, \quad \epsilon > 0. 
\end{equation}
This kernel is equivalent to~\eqref{eq:KGauss} with the distance function $ D(y, x ) $ set to the norm  $ \lVert y - x \rVert $ of delay-coordinate space.

 As with other kernels encountered in Section~\ref{sec:mathback}, $ K_\epsilon $ induces an integral operator, $ \mathcal{ K }_\epsilon : L^2( A, \alpha ) \mapsto L^2( A, \alpha ) $ acting on square-integrable functions on $ A $ through the formula
\begin{equation}
  \label{eq:KContinuous}
  \mathcal{ K }_\epsilon f( b ) = \int_A K_\epsilon( \pi( b ), \pi( a ) ) f( a ) \, d\alpha(a). 
\end{equation} 
An important property of kernels defined in delay-coordinate space such as~\eqref{eq:KDelay} is that their associated integral operators depend on the dynamical system generating the data.  In particular, because each point in delay-coordinate space corresponds to a segment of dynamical evolution, different dynamical systems will induce different kernel values, and hence different operators, even when observed in the same observation space. We refer to kernels whose properties depend on the dynamical system generating the data as ``dynamics-adapted'' kernels.

In the case that $ M = \pi( A ) \subseteq X $ is a smooth, compact manifold diffeomorphic to $ A $, $ K_\epsilon $ is symmetric, and $ \epsilon $ is small, the influence of the dynamics on $ K_\epsilon $ and $ \mathcal{ K }_\epsilon $ can be studied from the point of view of a Riemannian geometry induced on $ A$ from the ambient geometry of $ X $. Here, the intuition is that as for a fixed reference point $ b $, as $ \epsilon $ becomes small $ K_\epsilon( \pi( b ), \pi( \cdot ) ) $ is only appreciable in an exponential neighborhood of $ b $ (see, e.g.,~\eqref{eq:KAsymp} ahead). In this neighborhood, the behavior of $ \mathcal{ K }_\epsilon $ is governed by the leading integral moments of $ K_\epsilon( \pi( b ), \pi( \cdot ) ) $ on $ A $ \cite{BerrySauer15}, which are governed in turn by the leading derivatives of the distance function. In particular, the second derivative (Hessian) of $D( \pi( b ),\pi( \cdot) )  $ is closely related to the Riemannian metric tensor inherited by $ M $ from the ambient space inner product. 

In practical applications, we are interested in the behavior of integral operators defined for the atomic sampling measure $ \hat \alpha $. That is, instead of~\eqref{eq:KDelay} we consider, $ \hat{\mathcal{ K }}_\epsilon : L^2(A, \hat \alpha ) \mapsto L^2( A, \hat \alpha )$, where
\begin{equation}
  \label{eq:KDiscrete}
  \hat{\mathcal{ K }}_\epsilon f( b ) = \int_A K_\epsilon( \pi( b ), \pi( a ) ) f( a ) \, d\hat\alpha(a). 
\end{equation} 
For the same reasons as those leading to overfitting in Laplacian pyramids (see Section~\ref{sec:pyramids}), as $ \epsilon $ decreases at fixed $ N $, the properties of this operator become increasingly different from those of $ \mathcal{ K }_\epsilon $. Moreover, the smaller $ \epsilon $ is, the larger $N $ must be to achieve good consistency between $ \hat{\mathcal{K}}_\epsilon $ and $ \mathcal{ K}_\epsilon $ \cite{Singer06}. Intuitively, at small $ \epsilon $,  $ \mathcal{ K }_\epsilon $ is sensitive to small-scale features of $ f $, requiring a large number of samples to approximate the integral in~\eqref{eq:KContinuous} by~\eqref{eq:KDiscrete}. In the limit $ \epsilon \to 0 $, infinitely many samples are required. Thus, in practice, the limit $ \epsilon \to 0 $ is realized as a limit of large data. As mentioned in Section~\ref{sec:basicKernel}, additional complications can arise if  $ M $ is noncompact. In such cases, a variable-bandwidth variant of~\eqref{eq:KDelay} can be used to ensure stable behavior as $ \epsilon \to 0 $ and $ N \to \infty $ \cite{BerryHarlim15}.      


In \cite{BerryEtAl13}, Berry et al.\ established that under the assumption that Takens' theorem holds (i.e., $ A $ and $ M $ are diffeomorphic manifolds), increasing the number of delays $ q $ biases the kernel-induced Riemannian metric of $ A $ towards an invariant subspace associated with the most stable Lyapunov direction of the dynamical system (specifically, the most stable Oseledets subspace associated with the Lyapunov metric \cite{Pesin1976}). As has been demonstrated in a number of applications \cite{GiannakisMajda12a,BerryEtAl13,BushukEtAl14,SzekelyEtAl14}, the geometry of the data in delay-coordinate space significantly enhances the ability to extract intrinsic dynamical timescales with kernel eigenfunctions. Thus, it may be desirable to perform delay-coordinate mapping even if the observations are full to take advantage of this property. Below, we will argue that the Riemannian geometry of data in delay-coordinate space is also beneficial for analog forecasting.

\subsubsection{Cone kernels}
  
Besides~\eqref{eq:KGauss}, several other kernels for dynamical systems have been proposed in the literature \cite{SingerEtAl09,GiannakisMajda12a,BerryEtAl13,Talmon2012,Giannakis15,BerryHarlim15} and could be employed in the analog forecasting techniques developed here in different scenarios. In this paper, we work with the family of ``cone kernels'' introduced in \cite{Giannakis15}. A novel aspect of these kernels (which will turn out to be beneficial for analog forecasting) is that they feature an explicit dependence on the time tendency of the data, estimated through finite differences in time. For instance,  $ v = ( x_i - x_{i-1} ) / \tau $ corresponds to a first-order backward finite-difference approximation for the time derivative of the data, but higher-order and/or central or forward schemes can equally be used. In \cite{Giannakis15},  a geometrical interpretation was given for these quantities in terms of the vector field $ V = d \Phi_t / dt \rvert_{t=0} $ on the attractor generating the dynamics (assuming again that $ A $ and  $M $ are diffeomorphic manifolds). As a result, incorporating $ v $ in kernels allows one to bias the Riemannian metric of the data in a way that depends explicitly on the generator. 

In cone kernels, that dependence enters through the norm of $ V $, estimated through the norm $ \lVert v \rVert $ in ambient data space, and the angle between $ V $ and the relative displacement vector $ \omega  = y_i - x_j $ of time-indexed samples in delay coordinate space. The cosine of that angle is estimated through $ \cos \eta = \langle v,  \omega \rangle / ( \lVert v \rVert \lVert \omega \rVert ) $,  and similarly we compute $ w = ( y_i - y_{i-1} ) /  \tau $ and $ \cos \theta = \langle w,  \omega \rangle / (\lVert w \rVert \lVert \omega \rVert ) $. Introducing a parameter $\zeta \in [0, 1) $ (in addition to the kernel bandwidth $ \epsilon $), we define the cone kernel by
\begin{equation}
  \label{eq:conekernel}
  K_{\epsilon,\zeta} ( y_i, x_j ) = e^{-L_\zeta(y_i,x_j )/\epsilon},  \quad L_\zeta(y_i, x_j ) = \frac{ \lVert \omega \rVert^2}{ \lVert w \rVert \lVert v \rVert }\left[(1-\zeta \cos^2 \theta  )(1-\zeta \cos^2 \eta ) \right]^{1/2}.
\end{equation}
Note that as with radial the Gaussian kernels in~\eqref{eq:KDelay}, cone kernels are symmetric, but the function $ L_\zeta $ appearing in the argument of the exponential is anisotropic when $ \zeta > 0$. 


As $\zeta$ approaches $ 1 $, this kernel assigns increasingly higher affinity to data samples whose relative displacement vector is aligned with either $ v $ and/or $ w $; i.e, it assigns strong affinity to pairs of samples whose relative displacement vector lies within a narrow cone aligned with the dynamical vector field. This leads to distance contraction along the dynamical flow, such that the new averaging operators will be well-adapted to extract observables varying on intrinsic slow timescales. In particular, it can be shown \cite{Giannakis15} that in the induced geometry of cone kernels the norm $ \lVert u \rVert_\zeta $ of tangent vectors on the manifold is related to the ambient-space norm $ \lVert u \rVert $ through the expression
\begin{equation}
  \label{eq:coneLength}
  \lVert u \rVert_\zeta^2 = \frac{ 1 }{ \lVert V \rVert^2 } \left( \lVert u \rVert^2 - \zeta  \frac{ \langle V, u \rangle^ 2 }{ \lVert V \rVert^2 } \right). 
 \end{equation}
(We emphasize again that that this results requires that the $ A $ and $ M = \pi( A ) $ are diffeomorphic manifolds; thus, in the case that $ \tilde \pi $ is an incomplete observation map, we also require that Takens' theorem holds.) For $ \zeta > 0 $ the metric preferentially contracts tangent vectors aligned with $ V $ having $ \langle V, u \rangle^2 / \lVert V \rVert^2 \approx \lVert u \rVert^2 $, and the amount of length contraction becomes arbitrarily large as $ \zeta \to 1 $. As a result, in the cone-kernel geometry with $ \zeta \approx 1 $, small spherical balls centered on the initial data $ y $ will preferentially contain samples in the training dataset lying along a tube of integral curves of the dynamical vector field containing the integral curve passing through $ y $. Integral operators constructed from cone kernels (e.g.,~\eqref{eq:KContinuous} and the averaging operators of Section~\ref{sec:normalization} ahead) will therefore have weaker impact on slow observables varying weakly on these neighborhoods aligned with the dynamical flow. On the other hand, fast observables exhibiting strong variability along the orbits of the dynamics will be strongly averaged. Thus, integral operators constructed from cone kernels can be thought of as intrinsic low-pass filters associated with the dynamical system. 

Another important property of the cone kernel construction is that it is stable under changes in observation modality, in the sense that the associated integral operators derived from  datasets related by invertible transformations will be similar even if the corresponding ambient-space metrics (i.e., the dot product and corresponding norm in~\eqref{eq:coneLength}) are significantly different. We also note that as a result of the scaling by $ 1/ ( \lVert w \rVert \lVert v \rVert ) $ cone kernels acquire a scale invariance which is useful for computing  affinities for heterogeneous datasets consisting of measurements with different physical dimension (units) \cite{BushukEtAl14}. The outcome of this scaling factor in the induced geometry is a conformal scaling by the inverse squared norm of the dynamical vector field which can be seen in~\eqref{eq:coneLength}. Heuristically, the result of this conformal factor is to preferentially increase the resolution (i.e., to ``zoom in'') at the regions of the data manifold where the phase space velocity $ \lVert V \rVert $ is small), and decrease the resolution when $ \lVert V \rVert $ is large. Large $ \lVert V \rVert $ (and small sampling density) is expected to occur during metastable regime transitions which will be a prominent feature of the atmospheric model studied in Section~\ref{sec:toyModel}. Therefore, the conformal transformation by $ 1/ ( \lVert w \rVert \lVert v \rVert ) $ should improve the stability of kernel integral operators used to analyze such datasets. In \cite{GiannakisMajda12a}, the atmospheric model of Section~\ref{sec:toyModel} was studied using the  kernel in~\eqref{eq:conekernel} with $ \zeta = 0 $, and it was found that this transformation is highly beneficial for recovering the intermittent regime transitions in Galerkin-reduced models.   
  




\subsection{\label{sec:normalization}Kernel normalization}

As discussed in Section~\ref{sec:mathback}, the raw kernel $ K_\epsilon $ must be normalized prior to use in analog forecasting. In Section~\ref{sec:basicKernel}, we described a simple normalization strategy that converts $ K_\epsilon( y, \cdot ) $ to a density function $ \hat \rho_\epsilon( y, \cdot ) $ on the training dataset through left normalization. Here, we outline two alternative normalization strategies which produce bistochastic kernels in the limit of large data, hence eliminating forecast biases. Throughout this Section, we restrict attention to symmetric kernels with $ K_\epsilon (y, x ) = K_\epsilon( x, y ) $. Most of the results of this Section are valid for general symmetric local kernels, and for notational simplicity we drop the $ \zeta $ subscript from cone kernels unless we are making reference to a result specific to this kernel family. 

As stated in the preamble to Section~\ref{sec:kernel}, we are interested the behavior of these normalization strategies and the associated averaging operators acting on functions on $ A  $ in the limit of large data and small kernel bandwidth. In the case of averaging operators constructed through symmetric kernels, and for complete observations (i.e., $ A $ is diffeomorphic to is image $ M = \pi( A ) $ under the observation map),  the behavior of the averaging operators can be characterized through  second-order self-adjoint differential operators generating a diffusion process on $ A $. In the incomplete observations case, the limiting behavior of the averaging operators is significantly harder to analyze, but nevertheless they remain useful for unbiased analog forecasting.       

\subsubsection{\label{sec:diffMap}Diffusion maps normalization}

The first normalization strategy used in this work was introduced in the diffusion maps algorithm of Coifman and Lafon~\cite{CoifmanLafon06}, and further studied by Berry and Sauer in \cite{BerrySauer15}. Here, instead of left-normalizing $ K_\epsilon $ directly (as done in~\eqref{eq:hatRho}), we first perform a right normalization,
\begin{equation}
  \label{eq:kRight}
  \hat K_\epsilon'( y, x ) = \frac{ K_\epsilon( y, x ) }{ \hat q_\epsilon^{1/2} ( x ) },
\end{equation}
and then left-normalize $ \hat K'_\epsilon $ to produce a row-stochastic kernel:
\begin{equation}
  \label{eq:kLeft}
  \hat \rho_\epsilon( y, x ) = \frac{ \hat K_\epsilon'( y, x ) }{ \hat d_\epsilon( y ) }, \quad \hat d_\epsilon( y ) = \frac{ 1 }{ N } \sum_{i=0}^{N-1} \hat K'_\epsilon( y, x_i ).
\end{equation}
For completeness, we note that in the original formulation of diffusion maps  \cite{CoifmanLafon06}, the right normalization in~\eqref{eq:kRight} is performed using arbitrary powers of $ \hat q_\epsilon $. Here, we have elected to normalize $ K_\epsilon $ by $ \hat q_\epsilon^{1/2} $ as this choice leads to orthogonal eigenfunctions with respect to the invariant measure of the dynamics.  

By the pointwise ergodic theorem, as $ N \to \infty $, the quantities $ \hat q_\epsilon $, $ \hat K_\epsilon' $,  $ \hat d_\epsilon $, and $ \hat \rho_\epsilon $ converge to 
\begin{gather*}
  q_\epsilon( y ) = \int_M K_\epsilon( y, \cdot ) \, d\mu, \quad K'_\epsilon( y, x ) = \frac{ K_\epsilon( y, x ) }{ q^{1/2}_\epsilon( y ) }, \quad d_\epsilon( y ) = \int_M K'_\epsilon( y, \cdot ) \, d\mu, \quad \rho_\epsilon( y, x ) = \frac{ K'_\epsilon(y,x) }{ d_\epsilon( y ) }. 
\end{gather*}
Moreover, $ \hat \rho_\epsilon( y, \cdot ) $ and $ \rho_\epsilon( y, \cdot ) $ are densities on $ L^2( M, \hat \mu ) $ and  $ L^2( M, \mu ) $, respectively, so that  
\begin{displaymath}
  \int_M \hat \rho_\epsilon( y, \cdot ) \, d\hat\mu = \frac{ 1 }{ N } \sum_{i=0}^{N-1} \hat \rho_\epsilon( y, x_i ) = 1, \quad \int_M \rho_\epsilon( y, \cdot ) \, d \mu= 1.
\end{displaymath}
Setting $ \hat p_\epsilon( b, a ) = \hat \rho_\epsilon( \pi( b ), \pi( a ) ) $ and $ p_\epsilon( b, a ) = \rho_\epsilon( b, a ) $, we also have 
\begin{equation}
  \label{eq:rhoNorm}
  \int_A \hat p_\epsilon( b, \cdot ) \, d\hat\alpha = \frac{ 1 }{ N } \sum_{i=0}^{N-1} \hat p_\epsilon( b, a_i ) = 1, \quad \int_A p_\epsilon( b, \cdot ) \, d\alpha = 1,
\end{equation}
and the techniques of Sections~\ref{sec:basicKernel}--\ref{sec:pyramids} can be implemented using these densities. 

Here, we are interested in the behavior of $ \hat \rho_\epsilon $ and $ \rho_\epsilon $ and the associated averaging  operators $ \hat P_\epsilon: L^2(A,\hat\alpha ) \mapsto L^2(A,\hat\alpha ) $ and $ P_\epsilon : L^2(A,\alpha ) \mapsto L^2(A,\alpha ) $, given by
\begin{displaymath}
   \hat P_\epsilon f( b ) = \int_A \hat p_\epsilon(  b,  a  ) f( a ) \, d\hat\alpha( a ), \quad P_\epsilon f( b ) = \int_A p_\epsilon(  b,  a  ) f( a ) \, d\alpha( a ). 
\end{displaymath}
First, note that in general $P_\epsilon $ is not self-adjoint on $ L^2(A,\alpha) $. However, expressing $ P_\epsilon f( b ) $ as 
\begin{displaymath}
  P_\epsilon f( b ) = \int_A s_\epsilon( b, a ) f( a) \frac{ d_\epsilon( x ) }{ q_\epsilon( x ) } \, d\alpha( a ), \quad s_\epsilon( b, a ) = \frac{ K_\epsilon( b, a ) }{  d_\epsilon( \pi( b ) ) d_\epsilon( \pi( a ) ) },
\end{displaymath}
where $ y = \pi( b ) $, and $ x = \pi( a ) $, it is evident that $ P_\epsilon $ is positive-semidefinite, and self-adjoint on $ L^2( A, \alpha_\epsilon ) $ for the measure $ \alpha_\epsilon $ with $ d \alpha_\epsilon / d\alpha = d_\epsilon / q_\epsilon $. By the properties of $ K_\epsilon $ listed in Section~\ref{sec:basicKernel}, $ \alpha_\epsilon $ is absolutely continuous with respect to $ \alpha $. Similarly, $ \hat P_\epsilon $ is not self-adjoint on $ L^2(A, \hat \alpha ) $, but it can be expressed as
\begin{displaymath}
  \hat P_\epsilon f( b ) = \int_A \hat s_\epsilon( b, a ) f( a) \frac{ \hat d_\epsilon( x ) }{ \hat q_\epsilon( x ) } \, d\hat\alpha( a ), \quad \hat s_\epsilon( b, a ) = \frac{ K_\epsilon( b, a ) }{  \hat d_\epsilon( \pi( b ) ) \hat d_\epsilon( \pi( a ) ) }.
\end{displaymath}
Thus, $ \hat P_\epsilon $ is self-adjoint on $ L^2(A,\hat \alpha_\epsilon ) $ for the atomic measure $ \hat \alpha_\epsilon $ with density $ \hat d_\epsilon/ \hat q_\epsilon $ relative to $ \hat \alpha $.  

Consider now the behavior of $ P_\epsilon$ as $ \epsilon \to 0 $ assuming that $ M = \pi( A ) $ is diffeomorphic to $ A $.  Under this condition, due to the exponential decay of the kernel it is possible to approximate kernel integrals using 
\begin{equation}
  \label{eq:KAsymp}
  \frac{ 1 }{ \epsilon^{m/2} } \int_A K_\epsilon( \pi( b ), \pi( a ) ) f( a ) \, d\alpha(a) = \gamma( b ) f( b ) + O( \epsilon ),
\end{equation}
where $ m $ is the dimension of  $A $ and $ \gamma( b ) $ is a smooth positive function that does not depend on $ f  $ or $ \epsilon $ (e.g., \cite{CoifmanLafon06,BerrySauer15,Giannakis15}). (More specifically, $ \gamma( b ) $ depends on the density of $ \alpha $ relative to the volume form of the Riemannian metric induced by the kernel.) Note that the completeness of the observations is essential to the validity of this approximation. In particular, if $ \pi $ is not invertible on its image, the kernel will fail to localize to an $ m $-dimensional ball of arbitrarily small radius as  $ \epsilon \to 0 $, invalidating~\eqref{eq:KAsymp}. 

Using~\eqref{eq:KAsymp} and the fact that $ K_\epsilon( y, x ) = K_\epsilon( x, y ) $, we find
\begin{displaymath}
  \int_A p_\epsilon( b, a ) \, d \alpha( b ) = 1 + O( \epsilon ). 
\end{displaymath}
This means that with the diffusion maps normalization, $ p_\epsilon( b,a ) $ becomes asymptotically bistochastic as $ \epsilon \to 0 $. Moreover, $ d \alpha_ \epsilon / d \alpha \to 1 $,  and $ P_\epsilon $ becomes self-adjoint on $ L^2( A, \alpha )$. It is straightforward to verify that $ p_\epsilon( b, a ) $ is also bistochastic in the limit $ \epsilon \to \infty $ where $ K_\epsilon $ attains the constant value 1. By the pointwise ergodic theorem, $ \lim_{N\to\infty} \hat P_\epsilon f( a ) = P_\epsilon f( a )$ for $ \alpha $-almost every $ a $ and starting state $ a_0 $ is the training data, so in the limit of large data $ \hat P_\epsilon $ acquires the self-adjointness properties of $ P_\epsilon $ on $ L^2(A,\alpha)$. 

It follows from these results that, with the diffusion maps normalization, the basic kernel analog forecast through~\eqref{eq:basicAnalogLarge} becomes unbiased in the limit of large data and for $ \epsilon \to 0 $ and $ \epsilon \to \infty $, in the sense that the expected forecast error in~\eqref{eq:barE} over initial conditions drawn from the invariant measure on $ A $ vanishes. (In fact, that would also be the case for the simpler kernel normalization via~\eqref{eq:rho}.) However, at intermediate values of $ \epsilon $, $ \int_A p_\epsilon( b, a ) \, d\alpha(a) \neq 1 $ and $ p_\epsilon $ is not bistochastic (though it is always row-stochastic in accordance with~\eqref{eq:rhoNorm}), and the expected error can be nonzero even as $ N \to \infty $.  


Of course, the fact that basic kernel analog forecasting with large-bandwidth kernels becomes unbiased does not have high practical significance, since such forecasts are likely to be highly uninformative to begin with. That property, however, turns out to be important in the case of analog forecasting with Laplacian pyramids where one sweeps through a broad range of bandwidths. Here, we have seen that the diffusion maps normalization improves the forecast biases via Laplacian pyramids over the basic left normalization at small $ \epsilon $ and when the observations are complete (i.e., $ M $ and $ A $ are diffeomorphic manifolds and~\eqref{eq:KAsymp} holds), but biases are likely to remain at intermediate $ \epsilon $ values (and at both intermediate and small $ \epsilon $ values if the observations are incomplete). In Section~\ref{sec:bistochastic}, we will discuss an alternative normalization strategy \cite{CoifmanHirn13} leading to unbiased forecast as $ N \to \infty $ for arbitrary $ \epsilon $ and observation map $ \pi $. 

\subsubsection{Diffusion eigenfunctions and data-driven observables}

We now examine the impact of the diffusion maps normalization to the eigenfunctions of the averaging operators $ P_\epsilon $ and $ \hat P_\epsilon $ and the Nystr\"om-extension based forecasting scheme in Section~\ref{sec:nystrom}. 

First, assuming that the observations are complete, it is possible to show \cite{CoifmanLafon06,BerrySauer15} that, uniformly on $ A $,
\begin{equation}
  \label{eq:pEps}
  P_\epsilon f( b ) = f( b ) - \epsilon \mathcal{ L } f( b ) + O( \epsilon^2 ),
\end{equation}
where $ \mathcal{ L } : L^2( A, \alpha ) \mapsto L^2( A, \alpha ) $ is the (positive semidefinite) generator of a stochastic gradient flow on $ A $ for a Riemannian metric $ g $ (that depends on the kernel) and a potential determined by the density $ \theta $ of the invariant measure of the dynamics relative to the volume form of $ g $. Specifically, we have
\begin{equation}
  \label{eq:LDiff}
  \mathcal{ L } f = \upDelta f - \frac{ \upDelta \theta^{1/2} }{ \theta^{1/2} } f,
\end{equation} 
where $ \upDelta f = - \divr_g \grad_g f$ is the Laplace-Beltrami operator associated with  $ g $. 

Equation~\eqref{eq:pEps} establishes the pointwise convergence of $ \epsilon^{-1} ( I - P_\epsilon ) f $ to $ \mathcal{ L }f $ as $ \epsilon \to 0 $. By ergodicity we can conclude that $ \hat{\mathcal{ L }}f( a ) = \lim_{\epsilon\to 0 } \epsilon^{-1} \lim_{N\to\infty} ( I - \hat P_\epsilon ) f( a ) $ is $ \alpha $ -almost surely equal to $ \mathcal{ L }f( a ) $ too.  However, spectral convergence, i.e., convergence of the eigenvalues and eigenfunctions of $ \epsilon^{-1} ( I - P_\epsilon ) $ and  $ \epsilon^{-1} ( I - \hat P_\epsilon ) $ to those of $ \mathcal{ L } $ is a more subtle question which, to our knowledge, has not been addressed in the literature for the diffusion maps normalization used here to construct $ P_\epsilon $ and $ \hat P_\epsilon $. However, a number of related results suggest that spectral convergence may hold for the class of averaging operators employed in this work. In particular, in \cite{BelkinNiyogi07} Belkin and Niyogi establish spectral convergence of the graph Laplacian (an operator analogous to $ \hat{\mathcal{ L }} $ in our notation) to the Laplace-Beltrami operator for radial Gaussian kernels and uniform sampling density relative to the Riemannian measure. Moreover, in \cite{VonLuxburgEtAl04} von Luxburg et.\ al establish spectral convergence of normalized kernel integral operators which have structural similarities with the operators $ P_\epsilon $ and $ \hat P_\epsilon $ studied here. On the basis of these results, it appears plausible that spectral convergence holds for the operators employed in this work too. In what follows, we provide a geometrical interpretation of the eigenfunctions of $ P_\epsilon $ and $ \hat P_\epsilon $, and the associated geometric harmonics, assuming spectral convergence of $ \epsilon^{-1} ( I - P_\epsilon ) $ and $ \epsilon^{-1} ( I - \hat P_\epsilon ) $ to  $ \mathcal{ L } $. 

In particular, an important property of the eigenfunctions of $ \mathcal{ L } $ is that  they  are the extrema of the ratio 
\begin{equation}
  \label{eq:dirichletR}
  r( f ) = \frac{ E_g( f ) }{ \lVert f \rVert^2 }, \quad E_g( f ) = \int_A \lVert \grad_g f \rVert^2 \, d\alpha,
\end{equation}
 where $ E_g( f ) $ is a weighted Dirichlet energy measuring the expected roughness of $ f $ with respect to the Riemannian metric $ g $ and the invariant measure of the dynamics. Denoting the eigenvalues and eigenfunctions of $ \mathcal{ L } $ by $ ( \kappa_i, \varphi_i ) $, we have $ 0 = \kappa_0 < \kappa_1 \leq \kappa_2 \leq \cdots \uparrow \infty $, $ \langle \varphi_i, \varphi_j \rangle = \delta_{ij} $, and $ \kappa_i = E_g( \varphi_i ) $. Moreover, $ P_\epsilon \varphi_i = \lambda_{\epsilon,i} \varphi_i + O( \epsilon^2 )  $, where $ \lambda_{\epsilon, i } = e^{-\epsilon \kappa_i } $. It therefore follows that as $ \epsilon \to 0 $, the leading $ l $ eigenfunctions of $ P_\epsilon $ have a geometrical interpretation as the least lough orthonormal set of $ l $ functions in $ L^2(A,\alpha ) $  in the sense of~\eqref{eq:dirichletR}. By ergodicity, the eigenfunctions of $ \hat P_\epsilon $ acquire the same property in the limits $ \epsilon \to 0 $ and $ N \to \infty $. Below, we will see that for the choice of kernels in Section~\ref{sec:kernels} this geometrical interpretation also has a dynamical interpretation which motivates the use of the $ \varphi_i $ as data-driven prediction observables.

A similar geometrical interpretation can be made for the geometric harmonics employed in kernel analog forecasting with Nystr\"om extension. In particular, recall that in the geometric harmonics construction as formulated in Section~\ref{sec:geomHarm} one computes the eigenvalues and eigenfunctions, $ ( \lambda_{\epsilon, i}, \phi_{\epsilon,i} ) $, of a symmetric kernel integral operator $ G_\epsilon : L^2( A, \alpha ) \mapsto L^2(A,\alpha ) $ constructed from left or right symmetrization of $ p_\epsilon $. Noticing that in the case of left symmetrization $ G_{L,\epsilon} = P_\epsilon^*P_\epsilon $ and in the case of right symmetrization $ G_{R,\epsilon} = P_\epsilon P^*_\epsilon $, where the adjoints are taken on $ L^2(A, \alpha ) $, the eigenfunctions of $ G_{L,\epsilon} $  ($ G_{R,\epsilon} $) are given by the right (left) singular vectors of $ P_\epsilon $, and the eigenvalues are given by the squares of the corresponding singular values. As $ \epsilon \to 0 $, $ P_\epsilon $ becomes self-adjoint on $ L^2(A,\alpha ) $, and the $ \phi_{\epsilon,i} $ converge to $ \varphi_i $ for both left and right symmetrization. Moreover, the eigenvalues become $ \lambda_{\epsilon,i} = e^{-2\epsilon \kappa_i } + O( \epsilon^2 ) $. Thus, as $\epsilon $ becomes small, the geometric harmonic $ \psi_{\epsilon,i} = S_{\epsilon,i} \phi_{\epsilon,i} $ corresponding to $ \phi_i $ is approximated by $ \psi_{\epsilon,i} \approx e^{-\epsilon\kappa_i} \varphi_i $. It therefore follows that for small $ \epsilon $ (and thus large $ N $),  the spaces of bandlimited observables $ \hat B_l $ used in Section~\ref{sec:nystromAnalog} for analog forecasting have a geometrical interpretation as the $ l $-dimensional subspaces of $ L^2(A,\alpha) $ consisting of the least rough observables in the sense of~\eqref{eq:dirichletR}. Note also that for two functions $ f = \sum_{i=0}^\infty c_i \phi_{\epsilon,i} $ and $ f' = \sum_{i=0}^\infty c' \phi_{\epsilon,i} $ in the RKHS $ \mathcal{ H }_\epsilon $ associated with $ p_\epsilon $ we have $ \langle f, f' \rangle_{\mathcal{H}_\epsilon} \approx \sum_{i=0}^\infty c_i c'_i / e^{-2\epsilon\kappa_i } $.



%

Consider now the geometrical properties of the eigenfunctions of $ P_\epsilon $ for the dynamics-adapted kernels of Section~\ref{sec:kernels}. First, for kernels constructed in delay-coordinate space, including~\eqref{eq:KDelay} and~\eqref{eq:conekernel}, as more delays are added  the data-driven observables constructed from the eigenfunctions are expected to become increasingly biased towards stable Lyapunov directions of the system \cite{BerryEtAl13}. That is, given an orthonormal basis $ \{ e_1, \ldots, e_m \} $ of the tangent space at $ a \in A $ with respect to a Lyapunov metric on $ A $, ordered in order of increasing Lyapunov exponent, we heuristically expect that $ f \in B_l = \spn \{ \psi_0, \ldots, \psi_{l-1} \} $ will vary predominantly along $ e_1, \ldots e_{m_s} $, where $m_s $ is the dimension of the most stable Oseledets subspace. That is, we expect that  $ \lVert \grad_{s} f  \rVert \gg \lVert \grad_{s\perp} ( f ) \rVert $ where $ \grad_s $ and $ \grad_{s\perp} $ denote the orthogonal projection of the gradient along the subspaces spanned by $ \{ e_1, \ldots, e_{m_s} \} $ and $ \{ e_{m_s+1}, \ldots, e_m \}  $, respectively. By invariance of the Oseledets subspaces, we also expect this  property to hold for a reasonably long time under dynamical evolution. As a result, prediction observables initially lying in $ B_l $ are expected to remain in this space at least over moderately long times, alleviating the risk of bias in analog forecasting with Nystr\"om extension (see Section~\ref{sec:nystromAnalog}). 

Next, consider the eigenfunctions from the cone kernels in~\eqref{eq:conekernel}. In this case, it is possible to show that a consequence of the length contraction along the dynamical flow as $ \zeta $ approaches 1, the corresponding Laplace-Beltrami operator admits the asymptotic expansion \cite{Giannakis15},
\begin{equation}
  \label{eq:LBCone}
  \upDelta f = - \frac{ \divr_g( V( f ) V ) }{ 1 - \zeta } + O( ( 1 - \zeta )^0 ),
\end{equation}
where in the above equation $ V( f ) $ is the directional derivative of $ f $ along the dynamical vector field. Note that a general Laplace-Beltrami operator $ \upDelta f = - \divr \grad f $ (approximated, e.g., through a radial Gaussian kernel) depends on the full gradient of $ f $, which generally depends on the observation map for the dynamical system through the ambient-space induced metric. On the other hand, the  directional derivative $ V( f ) $ is intrinsic to the dynamical system, and is independent of the observation map $ \pi$. It also follows from~\eqref{eq:LBCone} that as $ \zeta \to 1 $ the Dirichlet energy in~\eqref{eq:dirichletR} reduces to $ E_g( f ) = ( 1 - \zeta )^{-1} \int_A ( V( f ) )^2 \, d \alpha + O( ( 1 - \zeta)^0 ) $. This functional assigns large roughness to functions with large directional derivative $ V ( f ) $, i.e., functions that exhibit strong variability along the dynamical flow. Equivalently, observables which are strongly bandlimited with respect to cone kernels are expected to vary slowly in the course of dynamical evolution. Such observables are good candidates for revealing intrinsic slow dynamical timescales, and are also expected to have favorable predictability properties. 


\subsubsection{\label{sec:bistochastic}Bistochastic kernels}

The diffusion maps normalization in Section~\ref{sec:diffMap} eliminates analog forecast biases with averaging operators at small and large bandwidths, but does not address biases at intermediate bandwidths. Here, we present an alternative normalization technique that eliminates biases at arbitrary bandwidths (in fact, the method is applicable for general kernels without bandwidth parameters). Our approach is based on the bistochastic kernel construction of Coifman and Hirn \cite{CoifmanHirn13}, which we adapt here to the case of finite datasets consisting of time-ordered observations of ergodic dynamical systems. 

Recall that the diffusion maps normalization consists of the right normalization of the raw kernel in \eqref{eq:kRight}, followed by a left normalization in~\eqref{eq:kLeft}. Here, we carry out this procedure in the opposite sense. Specifically, we first left-normalize $ K_\epsilon $ to construct the density function $ \hat \rho_\epsilon( y, x )$  via~\eqref{eq:hatRho}, and then make a right normalization as follows:
\begin{displaymath}
  \hat \beta_\epsilon( y, x ) = \frac{ \hat \rho_\epsilon( y, x ) }{ \hat \omega_\epsilon^{1/2}( x ) }, \quad \hat \omega_\epsilon( x ) = \frac{ 1 }{ N } \sum_{ i=0}^{N-1} \hat \rho_\epsilon( x_i, x ).
\end{displaymath}     
In the limit of large data, these quantities converge to
\begin{displaymath}
  \beta_\epsilon( y, x ) = \frac{ \rho_\epsilon( y, x ) }{ \omega^{1/2}( x ) }, \quad \omega_\epsilon( x ) = \int_M \rho( y, x ) \, d \mu( x ).
\end{displaymath}
To construct bistochastic kernels from $ \hat \beta_\epsilon $ and $ \beta_\epsilon $, we symmetrize these kernels using right symmetrization as in Section~\ref{sec:geomHarm}. That is, we set
\begin{equation}
  \label{eq:bistochastic}
  \hat \sigma_\epsilon( y, x ) = \frac{ 1 }{ N } \sum_{i=0}^{N-1} \hat \beta_\epsilon( y, x_i ) \hat \beta_\epsilon( x, x_i ), \quad \sigma_\epsilon( y, x ) = \int_M \beta_\epsilon( y, x' ) \beta_\epsilon( x, x' ) \, d\mu( x' ), 
\end{equation} 
and define the corresponding symmetric kernels on $ A \times A $ through $ \hat s_\epsilon( b, a ) = \hat \sigma_\epsilon( \pi( b ), \pi( a ) ) $ and $ s_\epsilon( b, a ) = \sigma_\epsilon( \pi( b ), \pi( a ) ) $. Clearly, $ \hat s_\epsilon $ and $ s_\epsilon $ are bistochastic on $ L^1(A,\alpha) $ and $ L^1( A, \hat \alpha ) $  since $ \hat \beta_\epsilon $ and $ \beta_\epsilon $ are column stochastic. Of course $ \hat s_\epsilon $ is neither row- nor column-stochastic on $ L^1( A, \alpha ) $ but it converges to a bistochastic kernel in the limit of large data. By virtue of the fact that this result holds for arbitrary bandwidths, the averaging operators $ \hat P_\epsilon $ constructed from $ \hat s_\epsilon $ yield asymptotically unbiased analog forecasts when used in conjunction with the Laplacian pyramids algorithm for both complete and incomplete observations.

\section{\label{sec:err}Error metrics and uncertainty quantification}

\subsection{Error metrics}

In Section~\ref{sec:results}, we will assess the methods described in Sections~\ref{sec:mathback} and~\ref{sec:kernel} in hindcast experiments against test datasets consisting of sequences of initial data $ \{ y_0, y_1, \ldots, y_{\tilde N - 1 + k } \}$ independent of the training data. If the test data are generated by the same dynamical system $ ( A, \mathcal{ A }, \Phi_t, \alpha ) $ as the training data (i.e., in a perfect-model scenario) we have $ y_i = \pi( b_i )$, where $ b_i $ are states in $ A $ such that $ b_i = \Phi_{t_i}(  b_0 ) $, $ t_i = ( i -1 ) \tau $. In the presence of model error, the $ y_i $ are generated by a different dynamical system, $ (\tilde A, \tilde{ \mathcal{ A}}, \tilde \Phi_t, \tilde \alpha ) $, and we have $ y_i = \tilde \pi( \tilde b_i ) $ with $ \tilde b_i =  \tilde \Phi_t ( \tilde b_0 ) $, $ \tilde b_i \in \tilde A $, and $ \tilde \pi $ an observation map from $ \tilde A $ to $ X $. 

 As stated in Section~\ref{sec:notation}, we perform these tests using two distinct types of forecast observables. Observables of the first type are those objectively defined on the test dataset independently of a data analysis algorithm. That is, in the perfect-model case we have the ground-truth values $ G_i = f( b_i ) $ of the forecast observable $ f $. In the imperfect-model case, the ground-truth values are $ G_i = \tilde f( \tilde b_i ) $ where $ \tilde f $ is the observable map for the dynamical system generating the test data. Examples of objectively defined observables are the components of the state vector, as it will be the case in Section~\ref{sec:toyModel}. Observables of the second type are data-driven observables, constructed through a data analysis algorithm applied to the training data. For example, in Section~\ref{sec:ccsm}, the forecast observables will be the leading low-frequency eigenfunctions of North Pacific SST recovered by cone kernels. For data-driven observables, there is no objective notion of ground truth on the test data, but a natural reference value is given by the out-of-sample extension of $ f $, i.e., $ G_i = \hat F_0( y_i ) $ for the Nystr\"om-based forecast in~\eqref{eq:fNystrom}), or  $ G_i = \hat F_{0,l}( y_i ) $ for the Laplacian pyramids forecast in Section~\ref{sec:pyramidsFinite}. 
In all cases, we compute the forecast error at lead time $ t = k \tau  $ for initial data $ y_i $ through the difference $ R_t(y_i ) = \hat{F}_t( y_i ) - G_{i+k} $. 

Based on these definitions for the ground truth and the corresponding forecast errors, we compute two time-averaged skill scores, namely the root mean squared error (RMSE) and pattern correlation (PC) scores. These are given by  
\begin{equation}
  \label{eq:err}
  \text{RMSE}( t )  =\sqrt{ \frac{1}{\tilde N}\sum_{i=0}^{\tilde N-1} \lvert R_t(y_i)\rvert^2 }, \quad \text{PC}(t)  = \frac{1}{\tilde N} \sum_{i = 0}^{\tilde N-1} \frac{(\hat{F}_t(y_i) - \bar F_t )(G_{i+k} -\bar G_k )}{\sigma_F \sigma_G },
\end{equation}
where $ \bar F_t = \sum_{i=0}^{\tilde N-1} \hat{F}_t( y_i  ) / \tilde N $, $ \bar G_k = \sum_{i=0}^{\tilde N -1} G_{i+k} / \tilde N $, $ \sigma_F^2 = \sum_{i=0}^{\tilde N-1} ( \hat{F}_t( y_i  ) - \bar F_t )^2 / \tilde N $, and $ \sigma_G^2 =  \sum_{i=0}^{\tilde N-1} ( G_{i+k} - \bar G_k )^2 / \tilde N $. 

\subsection{Uncertainty quantification}

Uncertainty quantification (UQ) is an important part of any statistical forecasting technique, but it appears to have received limited attention in the case of analog forecasting. Here, as a basic form of UQ we estimate the forecast error by local averaging the residual of the training data. In particular, the residual can be evaluated within the training dataset using $ R_t(x_i )  = \hat{F}_t(x_i ) - f_{i+k} $. Using a normalized local kernel $ \hat \rho( y, x_i ) $ on the training dataset (computed, e.g., via the methods of Section~\ref{sec:kernel}), we compute
\begin{equation}
\label{eq:UQ}
\varepsilon_t^2(y) = \frac{ 1 }{ N } \sum_{i = 0 }^{N-1} \hat \rho (y, x_i) \lvert R_t(x_i)\rvert^2,
\end{equation}
and use $ \pm \varepsilon_t( y ) $ to place two-sided error bars about our forecasts $ \hat F_t( y ) $. 

Clearly, this approach is rather rudimentary, and carries a risk of underestimating the forecast uncertainty if $ \hat F_t $ overfits the training data. However, if steps are taken to prevent overfitting in the training phase (e.g., through the use of the transition probability kernels in~\eqref{eq:tildeP}), it appears that~\eqref{eq:UQ} provides a reasonable UQ. We defer a more rigorous study of forecast uncertainty in kernel analog forecasting to future work.


\section{\label{sec:diffusion}Comparison with diffusion forecast}

Recently, Berry et al.~\cite{BerryEtAl15} developed a technique for nonparametric forecasting of dynamical systems called diffusion forecasting, which has some common aspects with the kernel analog forecasting methods developed here. In this Section, we discuss the similarities and differences of these techniques and comment on their relative strengths and weaknesses. 

Similarly to kernel analog forecasts, diffusion forecasts do not assume a parametric structure for the forecast model, and rely instead on observations of the system state in the past (analogous to the training samples $ \{ x_i  \} $ used here) to approximate operators (analogous to $U_t$) governing the evolution of observables. Moreover, as with kernel analog forecasting, kernel techniques also play an important role in diffusion forecasts. There, the eigenfunctions from a variable-bandwidth kernel \cite{BerryHarlim15} are used to build a data-driven orthonormal basis $ \{ \phi_0, \phi_1, \ldots \} $ for the $L^2$ space of the dynamical system, and this basis is used to create a matrix representation of the forward operator of the dynamics. More specifically, the dynamical system in diffusion forecasting is assumed to evolve on a manifold $M $, and the $ \{ \phi_i \} $ basis is orthonormal on $ L^2( M, \mu ) $ space associated with the system's invariant measure $ \mu $. Diffusion forecasting can handle both deterministic and stochastic systems; in the former case it approximates the Koopman operator, and in the latter case the Kolmogorov operator. These operators are represented by $ l \times l $ matrices with elements $ A(t)_{ij} = \langle \phi_i, U_t \phi_j \rangle $. Given an observable $ f = \sum_{i=0}^{l-1} c_i \phi_i $, and an initial probability measure with density $ \rho = \sum_{i=0}^{l-1} b_i \phi_i \in L^2( M, \mu ) $ relative to $ \mu $, the diffusion forecast at lead time $ t $ is given by the expectation value 
\begin{displaymath}
  \int_M U_t f \, \rho \, d \mu \approx \sum_{i,j=0}^{l-1} A(t)_{ij} b_i c_j. 
\end{displaymath}

An advantage of diffusion forecasting over kernel analog forecasting is its greater flexibility with regards to the initial probability measure. In kernel analog forecasting, the initial probability density is a kernel-dependent function of the initial data which is cumbersome to explicitly control. On the other hand, the initial density in diffusion forecasts is entirely up to the user, provided of course that it can be well represented in the eigenfunction basis. Another advantage of diffusion forecasting is that it is applicable to both deterministic and stochastic systems. It is plausible that kernel analog forecasting can be generalized to the stochastic case using similar error estimation techniques as in~\cite{BerryEtAl15}.

An advantage of kernel analog forecasting over diffusion forecasting is its ability to handle partially observed systems. Applying diffusion forecasting to the partially observed systems studied here would produce an operator of the form $ \Pi_l U_t \Pi_l $ where $ \Pi_l $ is the orthogonal projector to the $ l $-dimensional subspace of $ L^2( A, \alpha ) $ spanned by $ \{ \phi_0 \circ \pi, \phi_1 \circ \pi, \ldots, \phi_{l-1} \circ \pi \}  $. Unless that subspace happens to be an invariant subspace of the dynamics, forecasting with $ \Pi_l U_t \Pi_l $ will lead to biases. The analog forecasting with Nystr\"om extension in Section~\ref{sec:nystrom} suffers from similar biases, but the techniques in Sections~\ref{sec:analog} and~\ref{sec:pyramids} which are based on averaging operators will produce more uncertain, rather than biased, forecasts when faced with partial observations. Kernel analog forecasting also appears to make fewer assumptions than diffusion forecasting about the smoothness of the dynamical system and the manifold, though it is plausible that some of the smoothness assumptions in diffusion forecasting could be relaxed. In particular, the theoretical development of diffusion forecasting in \cite{BerryEtAl15} is for dynamical systems on smooth manifolds, but it is possible that a similar framework could be developed for more general measurable spaces (in fact, in \cite{BerryEtAl15} it was shown that diffusion forecasting can be successfully applied to the Lorenz 63 system, where the attractor clearly violates the smooth manifold assumption).   
 
Overall, while kernel analog forecasting and diffusion forecasting have their individual strengths and weaknesses, both methods are useful in a variety of forecasting scenarios where skillful parametric models are hard to construct. Even in situations where parametric models are available, these techniques can provide useful benchmarks for the forecast skill that should at a minimum be expected from parametric models. 

\section{\label{sec:results}Experimental results}

\subsection{\label{sec:toyModel}A chaotic intermittent low-order atmosphere model}
\begin{figure}
  \centering \includegraphics[width=0.5\textwidth]{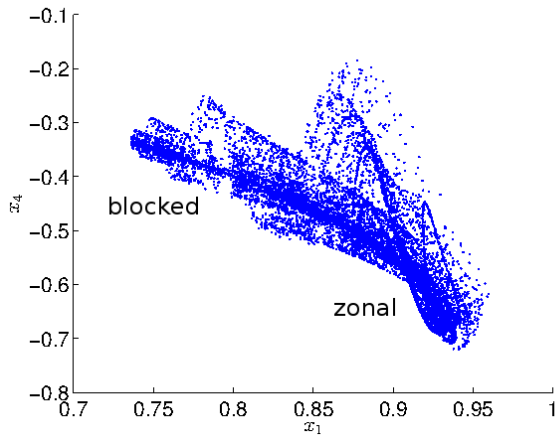}
  \caption{\label{fig:trajectories}Scatter plot of the state vector components $ ( u_1, u_4 ) $ from a 10,000-day simulation of the low-order atmospheric model, showing approximate locations of the zonal and blocked regimes.}
\end{figure}

We illustrate several key aspects of the forecasting methods described in Sections~\ref{sec:mathback} and~\ref{sec:kernel}, including kernel analog forecasting with Laplacian pyramids{eq:LapApprox}, lagged sequences in~\eqref{eq:delay}, and cone kernels in~\eqref{eq:conekernel}, through  experiments on a six-dimensional low-order model of the atmosphere featuring chaotic metastability \cite{CrommelinEtAl04,CrommelinMajda04}. Specifically, setting $u( t ) = (u_1( t ), \dots, u_6( t ) ) \in \mathbb{R}^6$, we consider a deterministic dynamical system of the form $\dot{u}( t ) = V(u(t))$, where the vector field, $ V(u) = (V_1(u), \ldots, V_6(u) ) $, is given by  
\begin{gather*}
  V_1( u ) =  \gamma_1^* u_3 - C( u_1 - u_1^* ), \quad V_2( u ) = - ( \alpha_1 u_1 - \beta_1 ) u_3 - C u_2 - \delta_1 u_4 u_6, \\
  V_3( u ) = ( \alpha_1 u_1 - \beta_1 ) u_2 - \gamma_1 u_1 - C u_3 + \delta_1 u_4 u_5, \; V_4( u ) =  \gamma_2^* u_6 - C( u_4 - u_4^*) + \epsilon ( u_2 u_6 - u_3 u_5 ), \\
  V_5( u ) = - ( \alpha_2 u_1 - \beta_2 ) u_6 - C u_5 - \delta_2 u_4 u_3, \quad V_6( u ) = ( \alpha_2 u_1 - \beta_2 ) u_5 - \gamma_2 u_4 - C u_6 + \delta_2 u_4 u_2.
\end{gather*}
Here, $ \alpha_i $, $ \beta_i $, $ \gamma_i $, $ \gamma^*_i $, $ \epsilon $, $ C $, and $ u^* $ are model parameters. This model is derived by a low-order truncation of the streamfunction in the barotropic vorticity equation in a channel in the presence of topography and a zonal (east--west) forcing profile. In particular, the state-vector components $u_i$ are expansion coefficients in the spatial Fourier basis for the channel. Physically, this model provides a coarse approximation of atmospheric dynamics in the presence of orography and a zonal background flow---it is a simplified model for the polar jet stream interacting with continental mountain ranges, such as the mountain ranges of North America. Observationally, it is known that such flows exhibit metastable transitions between so-called zonal and blocked states, where the jet propagates relatively unimpeded, or is significantly blocked by the presence of topography, respectively. With an appropriate choice of parameters~\cite{CrommelinMajda04}, the six-dimensional model exhibits a qualitatively similar behavior, and the two regimes are manifested by distinct regions in the $ ( u_1, u_4 ) $ plane (see Figure~\ref{fig:trajectories}). In the spatial domain, these components of the state vector represent purely zonal flow, with no variation in the meridional direction, reflecting the interesting regime switching between zonal and blocked states.

In what follows, we apply the techniques of Sections~\ref{sec:mathback} and~\ref{sec:kernel} to predict the evolution of the observable $ f = ( u_1, u_4 ) $ representing the regimes, using as initial data either the full state vector, $ y = u $, or a vector formed by applying delay-coordinate maps to the projected state vector $ z = ( u_1, u_4 ) $ in accordance with~\eqref{eq:delay}. We note that forecasting with reduced parametric models is particularly challenging in this example. For instance, in \cite{CrommelinMajda04} it was found that even a modest dimension reduction of the full model to four degrees of freedom by Galerkin projection onto the PCA basis produces models that either decay to fixed points, or become locked on orbits of unrealistically high temporal regularity, failing to reproduce chaotic regime transitions. On the other hand, Galerkin projection onto a basis derived by eigenfunctions from the kernel in~\eqref{eq:conekernel} with $ \zeta = 0 $ was able to exhibit chaotic regime transitions using only three dimensions \cite{GiannakisMajda12a}. This suggests that the patterns derived from Laplace-Beltrami eigenfunctions are more intrinsic to the dynamical system than the linear PCA modes.    

Our training dataset, $\hat{M}$, 
 consists of $N = 10^4$ samples taken every $ \tau = 1 $ day (d) from an equilibrated simulation of the full dynamical system performed using a Runge-Kutta method. A scatter plot of the components  $ ( u_1, u_4 ) $ of the state vector in the training data is depicted in Figure~\ref{fig:trajectories}. To generate the test dataset, we changed the initial conditions and ran another equilibrated simulation with $\tilde N = 5000$ data points. 

\begin{figure}
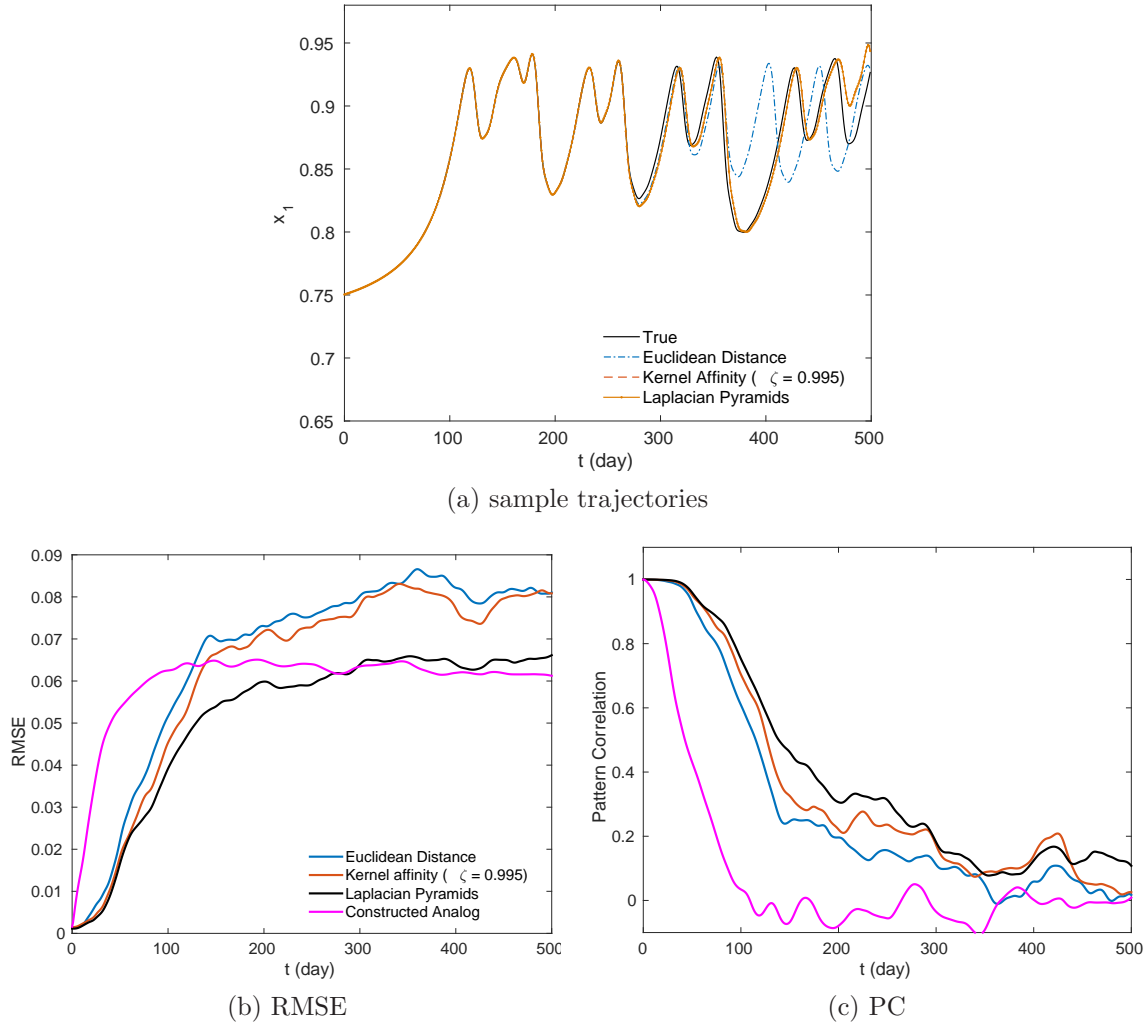

  \centering
  \subfloat[sample trajectories]{
    \includegraphics[width=0.5\textwidth]{./trajectory_full_id3202.eps}
    \label{fig:cdv_fullmodel_traj}
  }\\
  \subfloat[RMSE]{
    \includegraphics[width=0.45\textwidth]{./rmse_cdv_x1_full_wCA.eps}
    \label{fig:cdv_fullmodel_rmse}
  }
  \subfloat[PC]{
    \includegraphics[width=0.45\textwidth]{./pc_cdv_x1_full_wCA.eps}
    \label{fig:cdv_fullmodel_pc}
  }
\caption{\label{fig:cdv_fullmodel_pred}Hindcast experiments for the low-order atmospheric model with complete initial data. (a) Ground-truth and predicted time series for $u_1 $ via conventional Euclidean-distance analogs, Kernel affinity-based analogs with $ \zeta = 0.995$, and kernel analog forecasting with Laplacian pyramids. (b,c) Root mean squared error (RMSE) and pattern correlation (PC) scores, respectively, for the forecast models in (a).}
\end{figure}

First, we discuss hindcast experiments using the full state vector as initial data. In these experiments we did not perform delay-coordinate mapping as the initial data are complete [i.e., we set $ q = 1 $ in~\eqref{eq:delay}]. Therefore, the only dependence of the kernel \eqref{eq:conekernel} in the dynamics is through the finite-difference estimates, $ v $ and $ w $, of the dynamical vector field $ V $, which were evaluated here using a first-order backward scheme. Figure~\ref{fig:cdv_fullmodel_pred} shows sample forecast trajectories and the RMSE and PC skill scores from~\eqref{eq:err} obtained via the empirical forecasting methods described in Sections~\ref{sec:mathback} and~\ref{sec:kernel}; namely, conventional analog forecasting with Euclidean distances, single analog forecasting with dynamics-adapted kernel affinity, kernel analy forecasting with Laplacian pyramids, and constructed analog method introduced in~\cite{vandenDoolEtAl03}. The constructed analog method makes prediction based on a linear combination of the historical data, and the weights for averaging are determined by fitting the initial data with least squares. Note that we do not use the Nystr\"om approach in~\eqref{eq:fNystrom} as $ f $ is not a bandlimited observable in the eigenfunction basis.  

For the affinity-based analog forecasts, we used the cone kernel in~\eqref{eq:conekernel} with $\zeta = 0.995$, and also examined the case with $ \zeta = 0 $ corresponding to no influence of the angular terms. We find that the analogs chosen with respect to the $ \zeta = 0 $ kernel affinities via~\eqref{eq:LapApprox} have essentially identical skill to the conventional analog forecast, but there is a marked improvement for the cone kernel with $ \zeta = 0.995 $. The improvement of skill is reflected in the improved RMSE and PC results in Figures~\ref{fig:cdv_fullmodel_pred}(b) and~\ref{fig:cdv_fullmodel_pred}(c), and is more prominent in the forecast time series in Figure~\ref{fig:cdv_fullmodel_pred}(a). There, the analog based on Euclidean distance deviates from the ground truth for $ u_1( t ) $ after $ t \approx 360 $ d, but the affinity-based analog with $ \zeta = 0.995 $ tracks the true trajectory up to $ t \approx 480 $ d. We therefore see that choosing analogs with respect to an affinity measure that preferentially selects samples with similar time tendency (as $ \zeta \approx 1 $ cone kernels do) leads to improvement of skill. For constructed analog method, finding the linear combination weights is challenging. Since we have more samples in time than the number of spatial points, we need to solve an under-determined system to get the weights. The least squares solution might not be the optimal solution. Therefore, the forecast skill of the constructed analogue method is low compared to other analog forecasting methods and the sample trajectory is not shown in Figure~\ref{fig:cdv_fullmodel_traj}.  


The most significant improvement of skill takes place for the kernel analog forecasting with Laplacian pyramids in~\eqref{eq:LapApprox}. Here, we used $ \zeta = 0.995 $, and truncated the kernel $ \hat p_{\epsilon_l} $ with 10 nearest neighbors. The initial bandwidth parameter $ \epsilon_0 $ was set to  the median of the distances between the 10 nearest neighbors. The iteration of the Laplacian pyramids stopped at $ l = 2 $ iterations. In Figure~\ref{fig:cdv_fullmodel_pred}, the kernel analog forecasting accurately tracks the true trajectory to at least 480 d, the RMSE is lower at both intermediate and long times ($ t \gtrsim 60$ d), and the PC values exceed those from other methods for up to 300 d leads. 
To summarize, in these experiments the cone-kernel geometry with $ \zeta \approx 1 $ (which produces neighborhoods aligned with the dynamical flow), improves the identification of single analogs, and moreover there are further benefits from using multiple samples in the training data through Laplacian pyramids.       

\begin{figure}
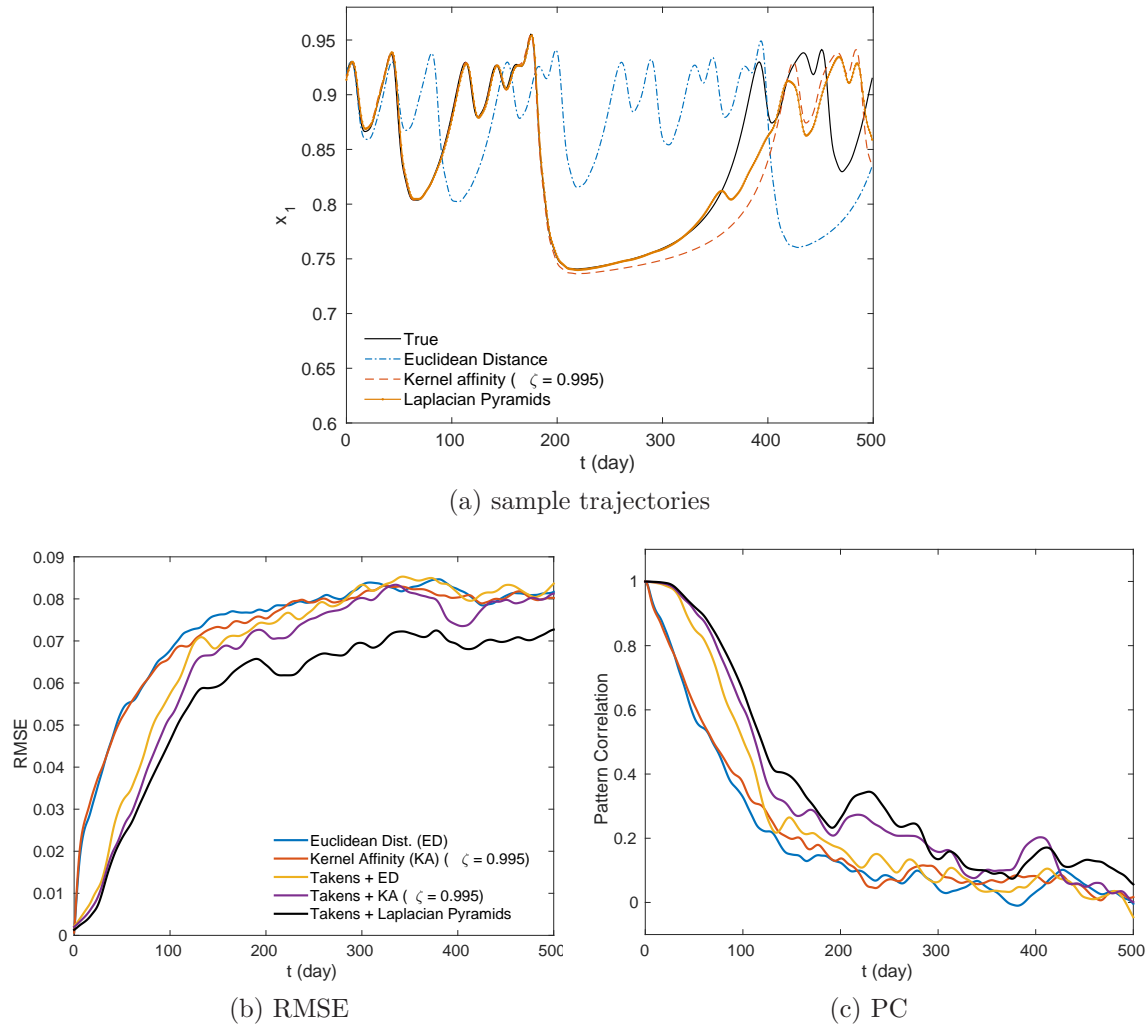

  \centering
  \subfloat[sample trajectories]{
    \includegraphics[width=0.5\textwidth]{./trajectory_partial_id1575.eps}
    \label{fig:cdv_partial_traj}
  }\\
  \subfloat[RMSE]{
    \includegraphics[width=0.45\textwidth]{./rmse_cdv_x1_partial.eps}
    \label{fig:cdv_partial_rmse}
  }
  \subfloat[PC]{
    \includegraphics[width=0.45\textwidth]{./pc_cdv_x1_partial.eps}
    \label{fig:cdv_partial_pc}
  }
  \caption{\label{fig:cdv_partial_pred}Same experiments as in Figure~\ref{fig:cdv_fullmodel_pred}, but for incomplete initial data, $ z = ( u_1, u_4 ) $, reconstructed using the delay-coordinate mapping in~\eqref{eq:delay} with $ q $ lags. The experiments shown here are for $ q= 1 $ and $ q = 22 $, corresponding to no delay-coordinate mapping and delay coordinate mapping over a physical time window of 21 days, respectively.}
\end{figure}

Next, to mimic a real-world scenario where only partial observations are available, we perform an experiment where only two coordinates, $ z = ( u_1, u_4 ) $, of the full state vector are observed. Following~\ref{eq:delay}, we form the vectors of initial data $ x_i $ by applying delay-coordinate mapping to $ z_i $ with $ q $ lags. In an operational setting, this would correspond to predicting the state of the atmosphere given its observed evolution over $ q $ days in the past. The results from hindcast experiments with $ q = 1 $ and $ q = 22 $ [i.e., no delay coordinate mapping and delay coordinate mapping over a physical time window $ ( q - 1 ) t = 21 $ d, respectively] and the same forecasting techniques used in the fully-observed case are shown in Figure~\ref{fig:cdv_partial_pred}. 

Without Takens' time-lagged embedding, the analog forecasts with partial observations perform significantly worse than the same analog forecasts with complete observations for all forecasting methods. This is consistent with the two-dimensional initial data with $ q = 1 $ being insufficient to determine neighborhoods on the attractor, and there is evidence that the intrinsic dimension of the attractor in this class of systems is greater than two \cite{CrommelinMajda04}. On the other hand, using Takens' time-lagged embedding with $ q = 22 $ delays leads to nearly equal skill as the case with full observations. Thus, in this relatively low-dimensional example, delay-coordinate mapping is a highly effective way of recovering the necessary dynamical information for empirical forecasting. Similarly to the complete-observation experiments, the $ \zeta = 0.995 $ cone-kernel affinity measure gave the best single-analog forecast, and the kernel analog forecasting with Laplacian pyramids provides the highest skill across the board. 

\subsection{\label{sec:ccsm}Long-range forecasting in the North Pacific sector of a comprehensive climate model}

As a high-dimensional application, we study low-frequency SST variability in the North Pacific sector of the coupled climate model CCSM3 \cite{CCSM3}. The datasets used in this study are two long equilibrated control integrations available in the public domain (\url{https://www.earthsystemgrid.org}) with designations b30.004 and b30.009. Integration b30.004 spans 900 years (y) using a 2.8$^\circ $ nominal resolution for the atmospheric component of the model; b30.009 spans 500 years, but employs a higher, 1.4$^\circ$, atmosphere resolution. Both integrations employ the same $1^\circ $ ocean resolutions. Two sets of experiments are discussed below, the first of which is a perfect-model experiment using  the first 400 years of integration b30.004 as the training dataset and the second half of b30.004 as the test dataset. The second set of experiments introduces model error and we use the data from b30.009 as test data from ``nature'' and the first 800 years of b30.004 as training. The training datasets are also used to construct data-driven low-frequency observables for prediction. Elsewhere \cite{ComeauEtAl15}, we perform hindcast experiments against actual observational data acquired via remote sensing, involving multivariate datasets of SST, regional arctic sea ice concentration and sea ice volume. The b30.004 and b30.009 datasets were also used in \cite{GiannakisMajda12b}, where the dominant spatiotemporal modes of North Pacific SST variability as extracted by NLSA algorithms were compared across different GCMs. 

\subsubsection{Recovering low-frequency observables.}
Following \cite{GiannakisMajda12b,Giannakis15}, we extract low-frequency SST modes using $ q = 24 $ delay-coordinate lags (corresponding to a two-year temporal embedding window) to induce timescale separation. Throughout, we work with cone kernels with $ \zeta = 0.995 $ as these kernels provide better timescale separation in the eigenfunctions \cite{Giannakis15} than the $\zeta = 0$ kernels used in \cite{GiannakisMajda12b}. The SST snapshots have dimension $ d = 6671 $ (equal to the number of North Pacific ocean gridpoints in CCSM3), and the dimension of the data vectors $ x_i $ after delay-coordinate mapping via~\eqref{eq:delay} is $ m = 24 \times 6671 = \text{160,104} $. Throughout, we work with a kernel bandwidth parameter $ \epsilon = 2.25 $, but our results are qualitatively robust for bandwidths in the interval 0.5--5. 

\begin{figure}
  \centering
  \includegraphics[width=0.8\textwidth]{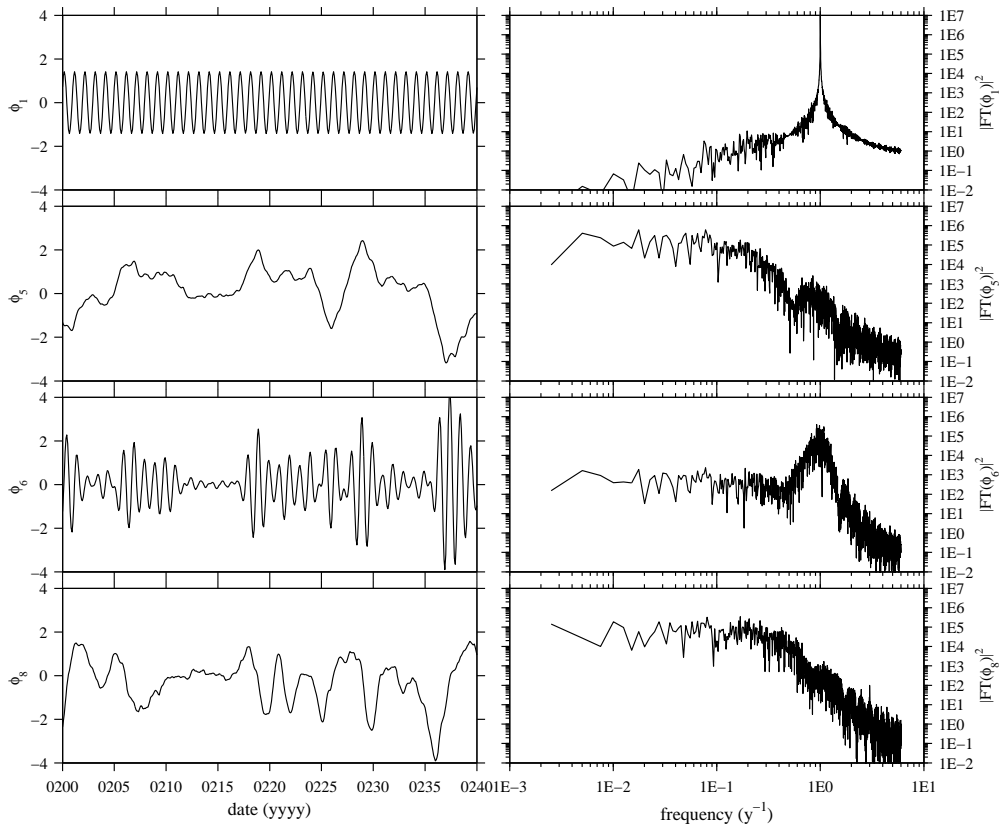}
  \caption{Representative eigenfunctions of North Pacific SST in the CCSM3 model. The eigenfunctions, $ \phi_i( x_j ) $, are shown as time series with respect to the timestamps $ t_j = ( j - 1 ) \tau $ for a 40-year segment of the 400-year training dataset. Frequency spectra, computed by applying the discrete Fourier transform of the 400-year time series are also shown. (a) Annual periodic mode; (b) $L_1 = \phi_5$, first low frequency mode, Pacific decadal oscillation (PDO); (c) annual intermittent mode corresponding to the PDO; (d) $L_2 = \phi_8$, second low frequency mode.}
\label{fig:modes}
\end{figure}

Representative eigenfunctions computed from the training data with these delay-coordinate and kernel parameters are shown in Figure~\ref{fig:modes}. The eigenfunctions separate the temporal variability of the data into qualitatively distinct families, namely periodic [Figure~\ref{fig:modes}(a)], low-frequency [Figures~\ref{fig:modes}(b,d)], and intermittent [Figure~\ref{fig:modes}(c)] modes. The periodic modes closely resemble sinusoids with frequencies given by integer multiples of 1 y$^{-1}$. These modes represent the harmonics of the seasonal cycle (a prominent source of variability in North Pacific SST), and form doubly-degenerate pairs with a phase offset of $\pi/2 $. The low-frequency modes describe oscillations taking place at interannual to decadal timescales, and are characterized by red-noise-like power spectra. The intermittent modes consist of periodic signals at the seasonal-cycle harmonics modulated by low-frequency envelopes. We refer the interested reader to \cite{GiannakisMajda12b,BushukEtAl14,Giannakis15} for detailed discussions on the properties and physical significance of these mode families. 

In what follows, our focus will be on forecasts of the leading-two low-frequency eigenfunctions, i.e., $ L_1 = \phi_5 $ and $ L_2 = \phi_8 $. These eigenfunctions carry significant power on interannual to decadal timescales, making them good candidate observables for long-range forecasting, but in addition they are associated with physically meaningful patterns in spatiotemporal reconstructions (e.g., see the online supporting animation in \cite{Giannakis15}). In particular, $ L_1 $ and $ L_2 $ display the salient features of two prominent low-frequency climate patterns in North Pacific. The first low frequency mode represent PDO~\cite{MantuaHare02} and the second mode is a manifestation of ENSO, which is most prominent in tropical Pacific~\cite{Trenberth1997} 
, respectively. The former is characterized by a horseshoe-like temperature anomaly pattern, developing east of Japan, together with an anomaly of the opposite sign along the west coast of North America. Based on these observations, we select eigenfunctions $ L_1 $ and $ L_2 $ as data-driven observables for empirical forecasting.  It is important to note that our kernel construction is essential to achieve timescale separation; in particular, if no Takens embedding is used ($q=1$) the temporal character of the eigenfunctions becomes corrupted, mixing low-frequency and periodic variability. However, despite their low-frequency character, $L_1$ and $L_2$ are particularly challenging to predict with parametric models, and first-order autoregressive models often fail to beat the persistence forecast \cite{ComeauEtAl15}. In the predictive skill results discussed below we include the persistence forecast as a reference for the forecast skill that can be achieved in these time series with simple autoregressive models.  We used constructed analog method to predict the evolution of the leading two low-frequency modes. The monthly climatology is removed from both the training and test datasets to find the coefficients for the weighted averaging. However, the pattern correlation scores for this method on the first two low-frequency modes are below 0.5 at $t = 0$. Therefore, these results are not included in the following subsections. 

\subsubsection{\label{sec:exp_ccsm3_perfect}Forecasting in a perfect-model environment.}
\begin{figure}
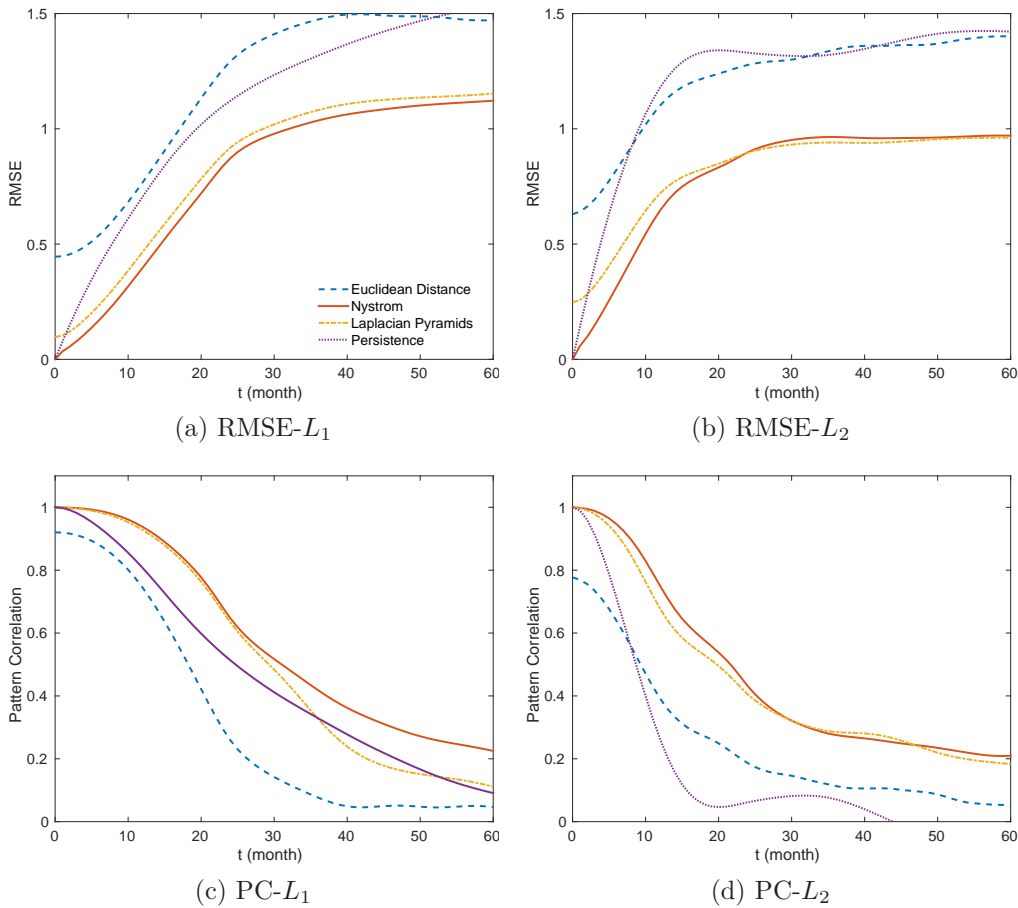

  \centering
  \subfloat[RMSE-$L_1$]{
    \includegraphics[width=0.4\textwidth]{./rmse_ccsm3_phi6}
    \label{fig:PDO_emb24_rmseb}
  }
  \subfloat[RMSE-$L_2$]{
    \includegraphics[width=0.4\textwidth]{./rmse_ccsm3_phi9}
    \label{fig:ENSO_emb24_rmseb}
  }\\
  \subfloat[PC-$L_1$]{
    \includegraphics[width=0.4\textwidth]{./pc_ccsm3_phi6}
    \label{fig:PDO_emb24_pcb}
  }
  \subfloat[PC-$L_2$]{
    \includegraphics[width=0.4\textwidth]{./pc_ccsm3_phi9}
    \label{fig:ENSO_emb24_pcb}
  }
  \caption{Forecast skill for the low-frequency eigenfunctions $ L_1 = \phi_5 $ and $ L_2 = \phi_8 $ in a perfect-model scenario. The RMSE and PC scores from~\eqref{eq:err} where evaluated for the persistence forecast, single-analog forecast based on Euclidean distances, and kernel analog forecasting with Nystr\"om extension and Laplacian pyramids.}
\label{fig:emb24b}
\end{figure}

We begin with an application of the techniques of Sections~\ref{sec:mathback} and~\ref{sec:kernel} in hindcasts of a test dataset generated by the same model as the model generating the training data---as stated earlier, the test dataset in this set of experiments consists of the 400 years of CCSM3 integration b30.004 following the 400 years of the training data. Figure~\ref{fig:emb24b} display predictive skill results for $L_1$ and $L_2$ obtained with the methods described in Section~\ref{sec:mathback}, as well as with the persistence forecast. The prediction observables in these experiments are pure eigenfunctions, so we performed kernel analog forecasting using both the Nystr\"om method and Laplacian pyramids via~\eqref{eq:fNystrom} and~\eqref{eq:LapApprox}, respectively. Moreover, because the prediction observables are data driven (i.e., they are not objectively defined on the test dataset), the skill scores were evaluated treating the Nystr\"om out-of-sample extension of $ L_1 $ and $ L_2 $ (Figure~\ref{fig:emb24b}) as ground truth. The persistent forecast follows the Nystr\"om out-of sample extension of the functions at lead time $t = 0$.  

\begin{figure}
\begin{center}
\includegraphics[width=0.45\textwidth]{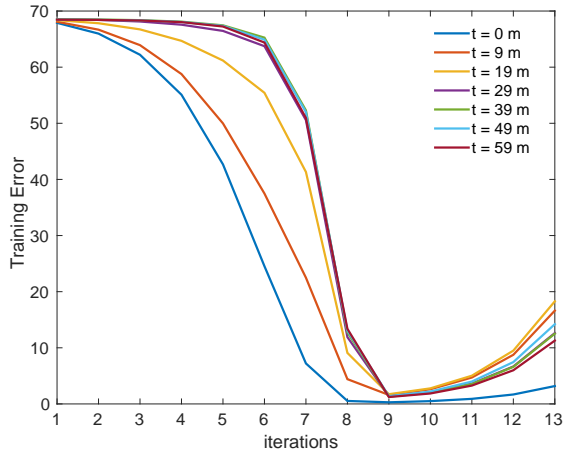}
\end{center}
\caption{Training errors with respect to the number of iterations in Laplacian pyramids for $L_1 = \phi_5$ at different lead time $t$. The initial bandwidth $\epsilon_0$ is chosen to be the median of all pairwise distances.}
\label{fig:LP_err}
\end{figure}

Using Laplacian pyramids for predicting the first two low frequency modes $L_1$ and $L_2$, we chose the initial kernel bandwidth to be the median of all pairwise distances. The training error in Laplacian pyramids first decreases with the increasing number of iterations, and then increases (see Figure~\ref{fig:LP_err}). The iterations stop at the minimum training error. For this dataset, 8 or 9 iterations are needed. Because of the choice of the ground truth for the computation of error metrics, the Laplacian pyramids estimation has small initial error at $t=0$.
 
The main finding from these experiments is that the kernel analog forecasting, implemented either with the Nystr\"om extension in~\eqref{eq:fNystrom} or with Laplacian pyramids in~\eqref{eq:LapApprox}, are the best performers across the board, substantially improving over single-analog forecasting approaches as well as persistence forecasts. In the case of the PDO, PC scores exceeding 0.5 (a popular lower threshold  for ``useful'' forecasts) persist for 32-month lead times using Nystr{\"o}m method and 30-month lead times for Laplacian pyramids. In contrast, the persistence forecast crosses the $\text{PC}=0.5$ threshold at 25-month leads, and has decayed to 0.38 at $ t = 32 $ months. Note that for this dataset the persistence forecast performs better than linear autoregressive models \cite{ComeauEtAl15}, which are popular parametric models for the PDO \cite{NewmanEtAl03}. Improvements over the persistence forecast are even more significant for the second low frequency mode $L_2$, which is a more rapidly decorrelating variable than the PDO. In this case, the $ \text{PC}=0.5 $ crossing time for the kernel-weighted forecasts is 22 months, which is more than a factor of two improvement compared to the 9-month crossing time for the persistence forecast. 

Consider now the forecasts made by single-analog methods. In this example with a high-dimensional ambient data space, the analogs selected via the cone-kernel affinity measure using~\eqref{eq:conekernel} have nearly identical skill as the conventional Euclidean-distance analogs (cf.\ the low-dimensional example in Section~\ref{sec:toyModel}). Therefore, the affinity-based results are not included in Figures~\ref{fig:emb24b}. 
Note that in the literature it is customary to calibrate the forecast error vs.\ lead time curves such that the error at $ \tau = 0 $ vanishes (e.g., \cite{Branstator2012}), and our skill scores with single analogs will generally appear lower in those approaches. 

\begin{figure}
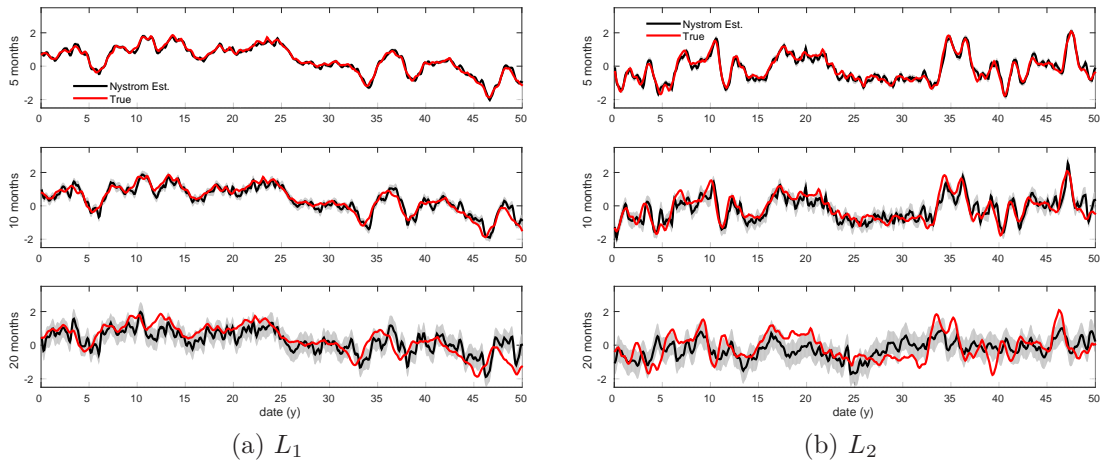

\begin{center}
\subfloat[$L_1$]{
\includegraphics[width=0.42\textwidth]{./ccsm3_prediction_PDO}
\label{fig:4cast_phi6}
}\quad
\subfloat[$L_2$]{
\includegraphics[width=0.42\textwidth]{./ccsm3_prediction_ENSO}
\label{fig:4cast_phi9}
}
\end{center}
\caption{Prediction of (a) $L_1 $ and (b) $L_2 $ at 5, 10, and 20 months lead. The red line shows the true signal and the black line shows the kernel analog forecasting with Nystr\"om extension. The gray shaded areas indicate uncertainty with $\pm \varepsilon_t$ defined in~\eqref{eq:UQ}.}
\label{fig:4cast}
\end{figure}

The prediction results of kernel analog forecast with Nystr\"om extension for $L_1$ and $L_2$ at 5, 10, 20 months lead are illustrated in Figure~\ref{fig:4cast}. At 5 months lead, the predicted time series agree very well with the true signals. The estimated uncertainty is also very small.  When lead time $t$ increases, the prediction error increases and the uncertainty $\varepsilon_t$ in~\eqref{eq:UQ} becomes larger (see the gray shaded area in Figure~\ref{fig:4cast}). 

It is evident from Figure~\ref{fig:emb24b} that the single-analog method is affected by large error at short lead times. This error is caused by poor reconstruction accuracy using a single analog. That is, unless the sampling density on the attractor is large, the ground truth at initialization time ($ t = 0 $) may deviate significantly from the estimated value $ \hat{F}( y, 0 ) $ from the analog. Substantial errors at initialization time typically persist at least for the short term, and in such cases single-analog methods will have lower short-term skill than methods which are able to accurately reconstruct the forecast observable at initialization, including the persistence forecast. Indeed, as shown in Figures~\ref{fig:emb24b}(a,c), the single-analog forecast for $L_1$ (a slowly-decorrelating observable) fail to beat persistence for both short and long times. Single-analog forecast of $L_2$, which is a more rapidly decorrelating variable, does outperform persistence at moderate to long lead times [$t \gtrsim 10 $ months in Figures~\ref{fig:emb24b}(b,d)], but the short-term skill for this observable is also poor. Note that poor short-term skill with single-analog method was not an issue in the experiments in Section~\ref{sec:toyModel} with the low-order atmospheric model (see Figures~\ref{fig:cdv_fullmodel_pred} and~\ref{fig:cdv_partial_pred}). There, the dimension of both the ambient data space and the attractor was low, and the sampling density was sufficiently  for accurate reconstruction at short leads even with a single analog. In contrast, by appropriately weighing multiple samples in the training data, the kernel analog forecasting performs well in both low- and high- dimensional ambient spaces, and for both short and long lead times. 
  
The computational complexity for Nystr\"om extension is $O(N^3 + l_1N)$, where the first term, $O(N^3)$, is the complexity for computing the eigendecomposition of the $N \times N$ kernel matrix and the second term is for extending the eigenfunctions and reconstructing the signal with the top $l_1$ eigenfunctions. The computational complexity for Laplacian pyramids is $O(l_2N^2)$, where $l_2$ is the number of iterations. In this example, the eigendecomposition of the appropriate kernel matrix was performed to derive the data-driven observables, $L_1$ and $L_2$, and therefore, the kernel analog forecasting with Nystr\"om extension only takes $O(l_1N)$ operations. The running time is 85 seconds for Nystr\"om extension and 276 seconds for Laplacian pyramids implemented with MATLAB on a machine 16 cores with 500 GB RAM. Although Laplacian pyramids is slower than Nystr\"om extension, in a more general problem, where the observables are not Laplace-Beltrami eigenfunctions, it can be more efficient than Nystr\"om method.  
    
\subsubsection{\label{sec:exp_ccsm3_modelerror}Forecasting with model error in the training data}

To test the robustness of our techniques in the presence of model error in the training data, we now study the forecast skill of the first two low frequency modes $L_1$ and $L_2$ in an experiment where the training data come from the first 800 years of the b30.004 CCSM integration, but the test data come from 500 years of the b30.009 integration. As mentioned earlier, these two integrations differ in the resolution of the atmospheric component of CCSM. These two datasets therefore have different atmospheric dynamics, which lead in turn to differences in SST variability through the associated atmosphere--ocean momentum and heat fluxes. Because b30.009 has higher atmospheric resolution than b30.004, in these experiments we view b30.009 as ``nature'' and b30.004 as an imperfect model. More broadly, this experimental setup is representative of many empirical modeling applications in the physical sciences, where long-term observations of nature are not available for training. 

\begin{figure}
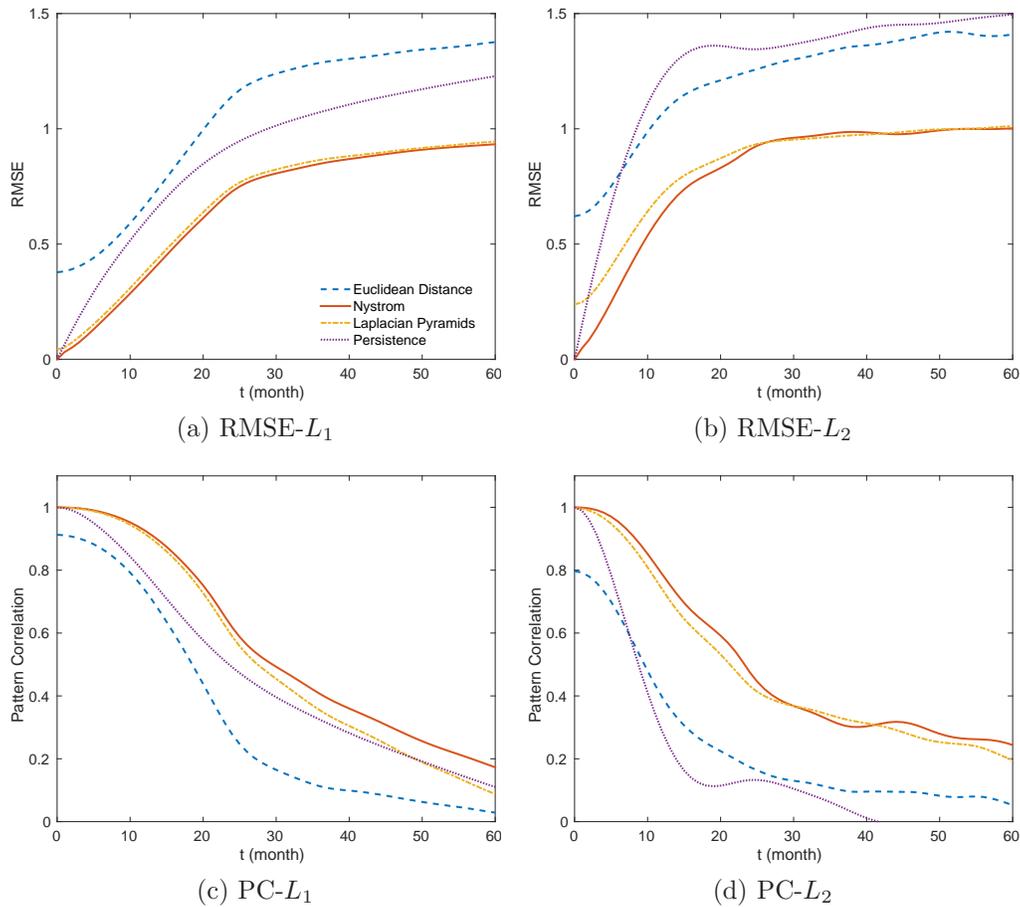

\begin{center}
  \subfloat[RMSE-$L_1$]{
    \includegraphics[width=0.4\textwidth]{./rmse_ccsm3_modelerr_phi6}
    \label{fig:PDO_ME_rmseb}
  }
  \subfloat[RMSE-$L_2$]{
    \includegraphics[width=0.4\textwidth]{./rmse_ccsm3_modelerr_phi9}
    \label{fig:ENSO_ME_rmseb}
  }\\
  \subfloat[PC-$L_1$]{
    \includegraphics[width=0.4\textwidth]{./pc_ccsm3_modelerr_phi6}
    \label{fig:PDO_ME_pcb}
  }
  \subfloat[PC-$L_2$]{
    \includegraphics[width=0.4\textwidth]{./pc_ccsm3_modelerr_phi9}
    \label{fig:ENSO_ME_pcb}
}
\end{center}
\caption{Forecast skill for the low-frequency eigenfunctions $L1 = \phi_5 $ and $ L_2 = \phi_8 $ respectively, for training data with model error. Similarly to Figure~\ref{fig:emb24b}, the RMSE and PC scores were computed for the persistence forecast, single-analog forecast based on Euclidean distances, and kernel analog forecasting using the Nystr\"om extension and Laplacian pyramids.} 
\label{fig:emb24_model_errorb}
\end{figure}
\begin{figure}
\begin{center}
\subfloat[$L_1$]{
\includegraphics[width=0.42\textwidth]{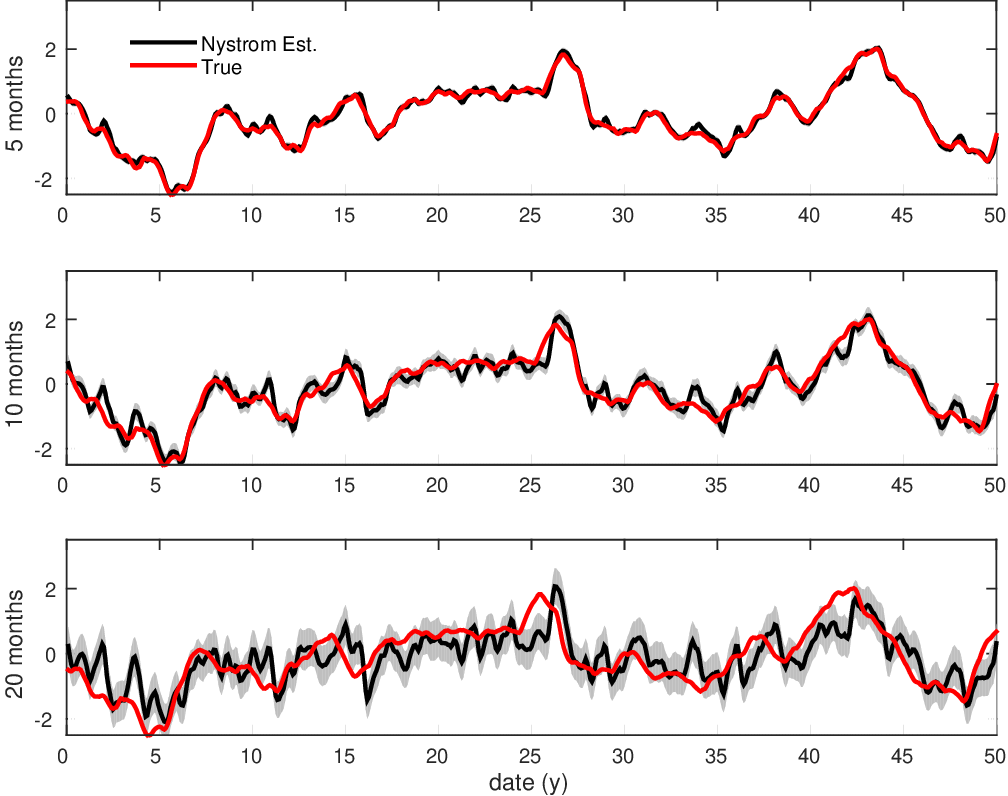}
\label{fig:4cast_phi6_modelerr}
}\quad
\subfloat[$L_2$]{
\includegraphics[width=0.42\textwidth]{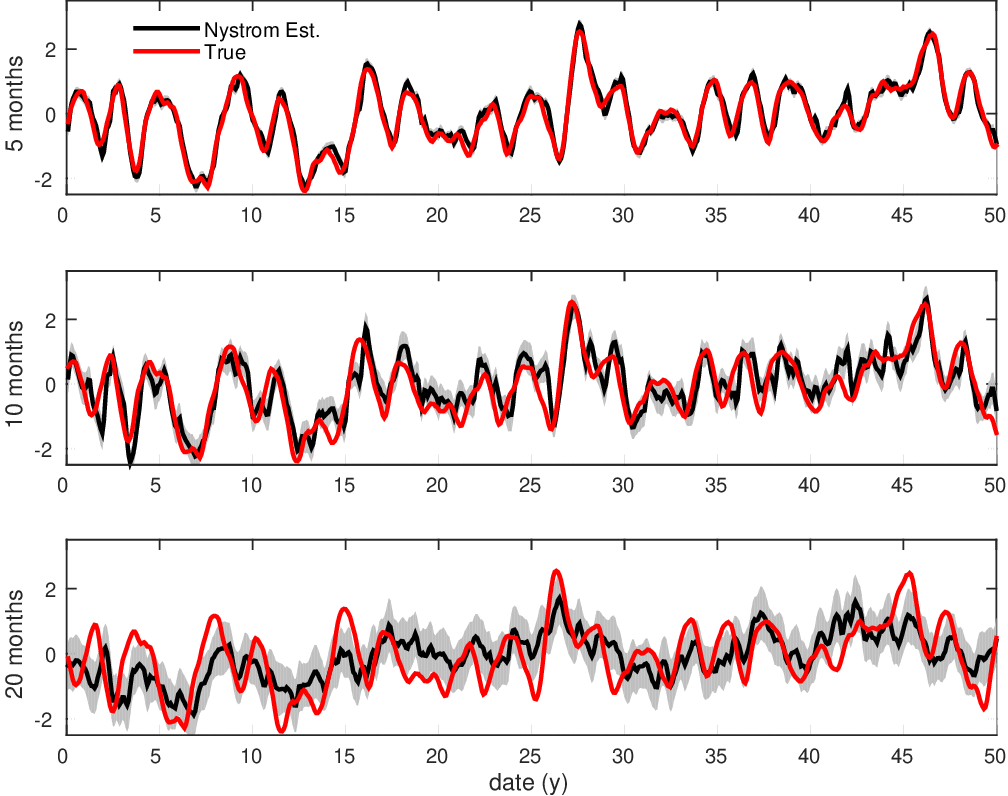}
\label{fig:4cast_phi9_modelerr}
}
\end{center}
\caption{Prediction of (a) $L_1 $ and (b) $L_2$ with model error at 5, 10, and 20 months lead. The red line shows the true signal and the black line shows the kernel analog forecasting with Nystr\"om extension. Light gray shaded area indicate uncertainty with $\pm \varepsilon$ in defined in~\eqref{eq:UQ}.}
\label{fig:4cast_modelerr}
\end{figure}

The skill scores in Figure~\ref{fig:emb24_model_errorb} were evaluated treating the Nystr\"om out-of-sample extension of $L_1$ and $L_2$ from b30.004 to b30.009 as ground truth, since the prediction observables are pure eigenfunctions. The persistence forecast follows the Nystr\"om out-of-sample extension at time $t = 0$. As expected, the skill scores in these experiments with model error in the training data are generally lower than their perfect-model counterpart in Figures~\ref{fig:emb24b}. In the case of the first low frequency eigenfunction $L_1$, PC scores exceeding 0.5 persist for 30-month lead times, i.e., $2$ months less than in the perfect model experiment. In contrast, the persistence forecast crosses the $\text{PC}=0.5$ threshold at 24-month leads, and has decayed to 0.40 at $ \tau = 30 $ months. The kernel analog forecasting methods still provide significantly higher skill than persistence, and also outperform single-analog forecasting using Euclidean distances. In the case of the second low frequency mode $L_2$, the $\mathrm{PC} = 0.5$ crossing time for the kernel analog forecasts remains more than a factor of two higher than the persistence forecast [see Figure~\ref{fig:emb24_model_errorb}(d)]. Figure~\ref{fig:4cast_modelerr} shows the Nystr\"om kernel analog forecasting results for $L_1$ and $L_2$ at 5, 10, and 20 months lead time.        

\section{\label{sec:conclusions} Conclusions}

In this paper, we proposed a family of data-driven, nonparametric forecasting methods for observables of dynamical systems. Inspired by Lorenz's analog forecasting technique \cite{Lorenz1969}, these methods perform forecasting through local averaging operators that predict the value of the forecast observable by a weighted average over multiple states in the training data that closely resemble the initial data. These operators are constructed through suitably normalized kernels which measure the pairwise similarity of data taking dynamics into account. In particular, the kernels used here incorporate empirically accessible aspects of the dynamical system through Takens delay-coordinate maps \cite{GiannakisMajda12a,BerryEtAl13} and finite-difference approximations of the dynamical vector field on the attractor \cite{Giannakis15}, enhancing timescale separation and robustness to changes of observation modality. Due to these features, the properties of the associated averaging operators can be understood in terms of a modified Riemannian geometry favoring samples in the training data with similar dynamical evolution to the observed data at forecast initialization. 

Mathematically, kernel analog forecasting has strong connections with kernel methods for out-of-sample extension of functions. Here, we studied two such approaches based on Nystr\"om extension \cite{CoifmanLafon06b} and the Laplacian pyramids technique \cite{Rabin2012,FernandezEtAl14}. The Nystr\"om-based approach naturally lends itself to the identification and forecasting of data-driven observables constructed from kernel eigenfunctions with high smoothness and favorable predictability. On the other hand, kernel analog forecasting with Laplacian pyramids performs well with more general classes of forecast observables. Used in conjunction with a bistochastic kernel normalization \cite{CoifmanHirn13}, this method is also able to make unbiased forecasts from incomplete initial data.  




We demonstrated the efficacy of these schemes in hindcast experiments involving a low-order deterministic model for the atmosphere with chaotic regime metastability, as well as interannual to decadal variability of sea surface temperature (SST) in the North Pacific sector of a comprehensive climate model. We also studied a North Pacific SST experiment with model error in the training data. In all experiments, our proposed weighted-ensemble methods led to significant improvement of forecast skill compared to conventional analog forecasting with Euclidean distances. The kernel analog forecasting also significantly outperformed forecasts made with single analogs identified using Euclidean distances. 


In the North Pacific SST applications, we constructed data-driven forecast observables $L_1$ and $L_2$ through kernel eigenfunctions. The timescale separation capabilities of the kernels used here contributed significantly to the clean low-frequency character and favorable predictability properties of the observables, extending the range of useful forecasts [$>0.5$ pattern correlation score (PC)] to 2.6 years for the PDO and 1.8 years for the the second low frequency eigenfunction that is associated with ENSO. As a benchmark, the persistence forecast for these variables (which is more skillful than simple autoregressive models) decays below the 0.5 PC threshold at 2 years ($L_1$) and 0.8 years ($L_2$). Again, kernel analog forecasting with Nystr\"om extension and Laplacian pyramids (as opposed to single analog) were crucial to obtain these results.

Stemming from this work are several future research directions. Here, we focused on forecasting with the kernels developed in \cite{GiannakisMajda12a,GiannakisMajda14}, but other kernels for dynamical systems have been proposed in the literature \cite{SingerEtAl09,Talmon2012,BerryHarlim15} and could be employed in our forecasting schemes. Moreover, we have focused on Nystr\"om-type extensions and Laplacian pyramids, but other interpolation and extrapolation techniques, such as kriging, can be explored. Finally, we have studied the asymptotic behavior of our algorithms in the limit of large data, but have not addressed the rate of convergence to the asymptotic limit. In particular,  it should be fruitful to study the convergence properties of our methods in different classes of dynamical systems (e.g., geometrically ergodic systems), and related implications to risk of overfitting and uncertainty quantification. 


\section*{Acknowledgments}

D. Giannakis and Z. Zhao wish to acknowledge support from ONR MURI Grant 25-74200-F7112 and NSF Grant DMS---1521775. The research of D. Giannakis is partially supported by ONR DRI grant N00014-14-0150. The authors wish to thank Darin Comeau and Andrew Majda for stimulating conversations.

\section*{References}
\bibliographystyle{plain}
\bibliography{prediction}

\end{document}